\documentclass[pre,aps,twocolumn,notitlepage,showpacs,floatfix,superscriptaddress]{revtex4-2}
\usepackage{amsmath}
\usepackage{amssymb}
\usepackage{physics}
\usepackage{bm}
\usepackage{booktabs}
\usepackage{graphicx}
\usepackage{hyperref}
\usepackage[margin=1in]{geometry}
\usepackage[dvipsnames]{xcolor}
\usepackage[T1]{fontenc}

\usepackage{booktabs}
\usepackage{rotating}

\newcounter{appsection}
\newcounter{appsubsection}[appsection]
\renewcommand{\theappsection}{\Alph{appsection}}
\renewcommand{\theappsubsection}{\Alph{appsection}\arabic{appsubsection}}

\newcommand{\appsec}[1]{%
  \refstepcounter{appsection}%
  \section*{Appendix \theappsection: #1}%
}
\newcommand{\appsubsec}[1]{%
  \refstepcounter{appsubsection}%
  \subsection*{\theappsubsection: #1}%
}
\usepackage{xcolor}
\begin{document}
\title{Boundary shape controls the stress distribution in two-component elastic
materials}

\author{Kyungeun Kim}
\email{kkim@math.ubc.ca}
\affiliation{Department of Mathematics, University of British Columbia, Vancouver, BC Canada}
\author{Adam Clare}
\email{aclare@mech.ubc.ca}
\affiliation{Department of Mechanical Engineering,  University of British Columbia, Vancouver, BC Canada}
\author{Miranda Holmes-Cerfon}
\email{holmescerfon@math.ubc.ca}
\affiliation{Department of Mathematics, University of British Columbia, Vancouver, BC Canada}

\date{\today}
\begin{abstract}
We investigate how the shape and arrangement of high-stiffness regions govern stress redistribution in two-component triangular elastic spring networks subjected to spatially correlated and uncorrelated random prestress. We compare boundary roughness, partitioning into disconnected regions, and graded transitions between regions of different stiffness. Our simulations identify interface bonds as channels for energy transfer: increasing their number provides additional transfer paths and reduces stress concentration in mechanically rigid regimes, while offering little benefit in compliant subnetworks. This is analogous to defects in naturally occurring crystal structures, where the defects themselves isolate stress. At a fixed fraction of high-stiffness material, partitioning reduces stress concentration for both correlated and uncorrelated prestress. The benefit diminishes with further subdivision as the number of interface bonds becomes comparable to the number of interior nodes. Stress reduction, however, depends on the spatial correlations of the imposed prestress: random prestress generally yields a stronger effect, though the difference diminishes in some regimes. Graded interfaces further reduce peak stress by distributing the stiffness transition over a broader region. Together, these results suggest design strategies based on increasing interface area, matching subdivision scale to prestress correlations, and grading stiffness transitions. This minimal framework identifies how internal structure controls energy transfer and stress redistribution, suggesting design principles for mechanical robustness under both internal and external loading.

\end{abstract}

\maketitle

\section{Introduction}
\begin{figure}
    \centering
    \begin{tabular}{cc}
        \includegraphics[width=0.45\linewidth]{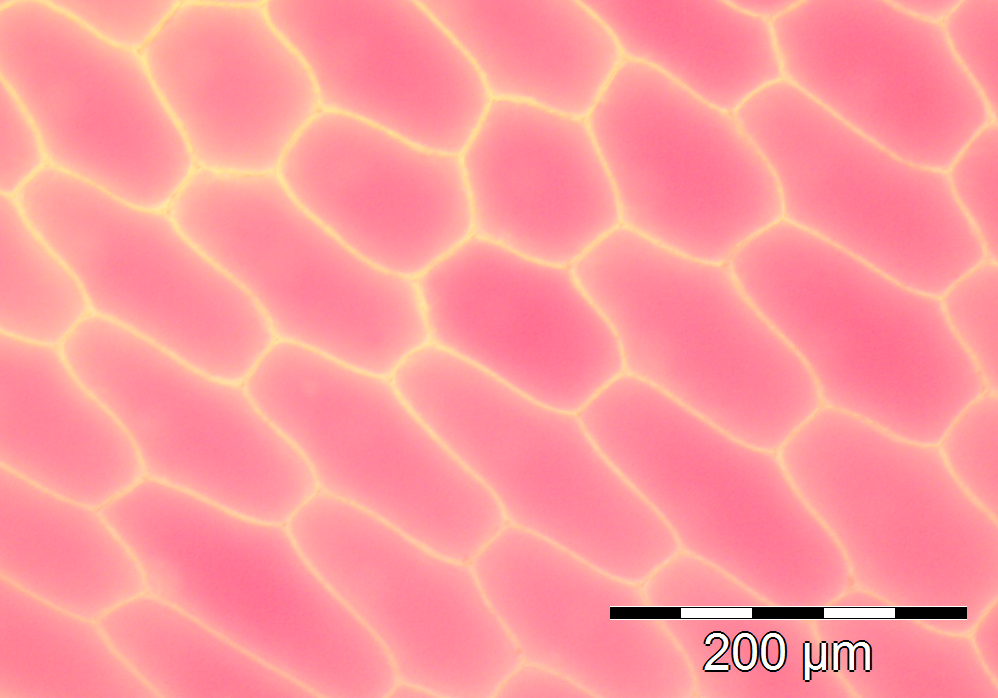} &
        \includegraphics[width=0.45\linewidth]{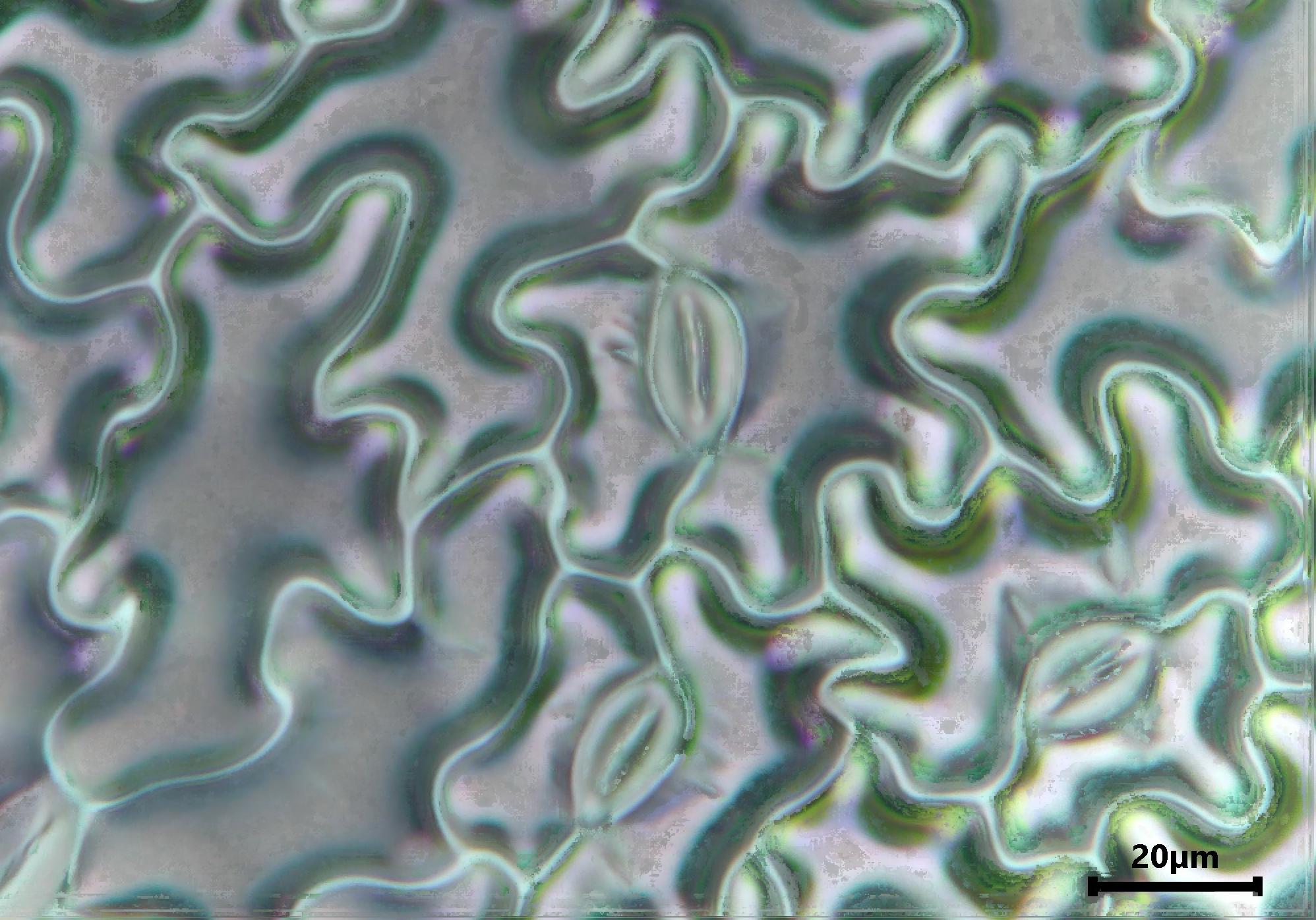} \\
        (a) & (b) \\[6pt]
        \includegraphics[width=0.45\linewidth]{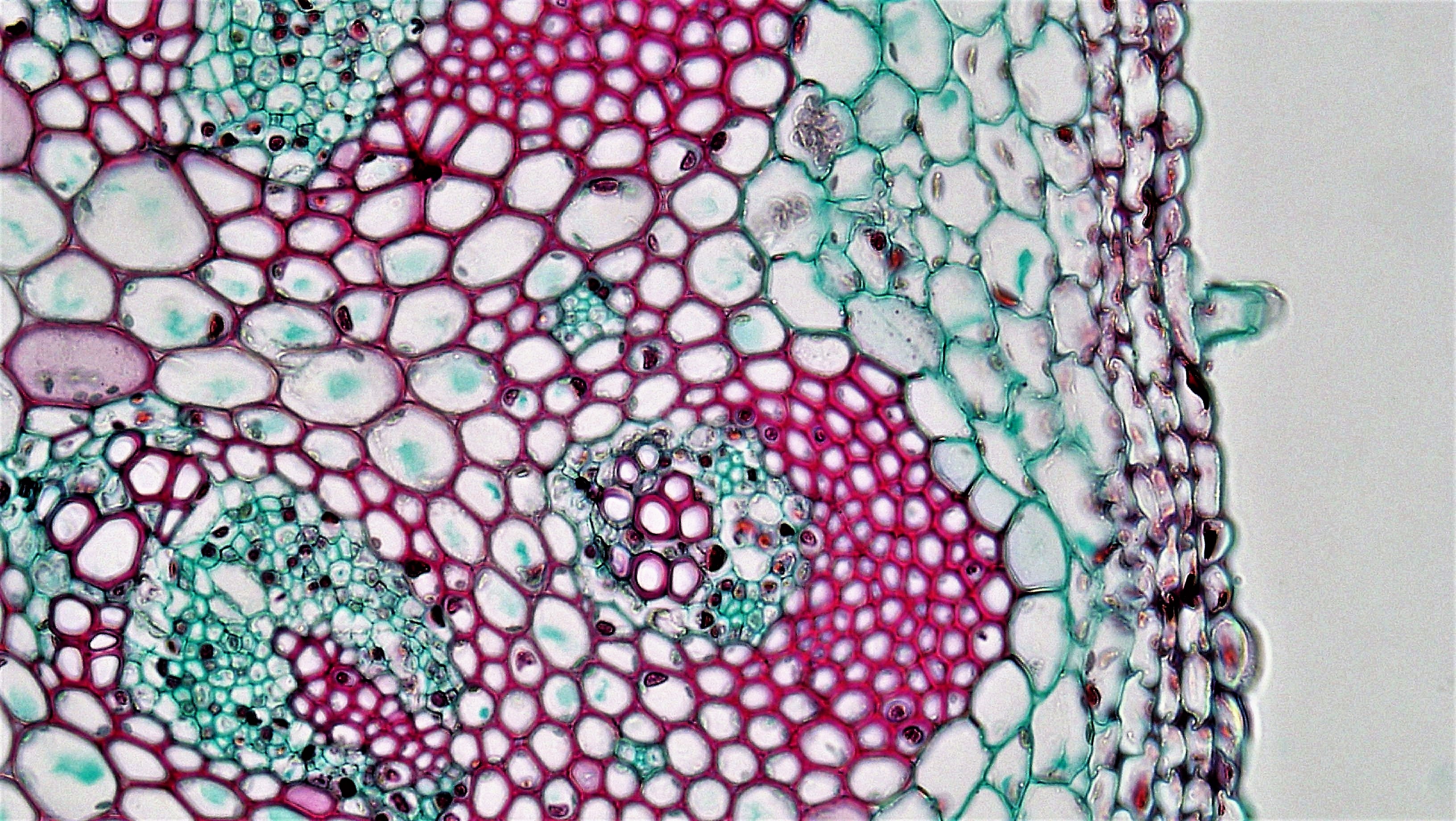} &
        \includegraphics[width=0.45\linewidth]{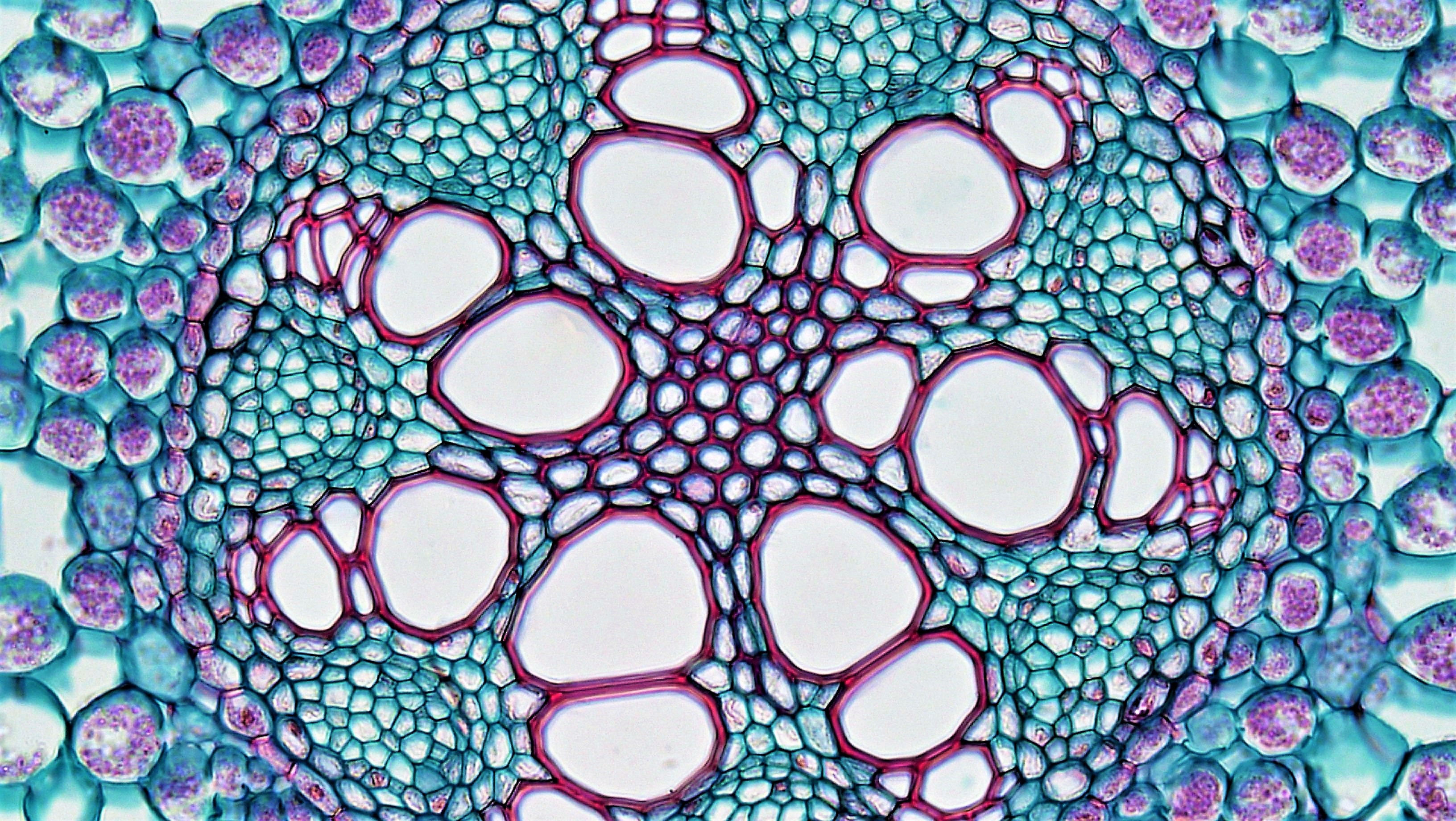} \\
        (c) & (d)
    \end{tabular}
    \caption{Representative plant cell structures.
        (a) Epidermal peel from red onion (\textit{Allium cepa}) showing regular packed pigmented cells~\cite{wikimedia_onionepidermis} (CC BY-SA 3.0, Adrian J. Hunter).
        (b) Cropped image of Leaf epidermis of \textit{Phaseolus coccineus} with stomata, showing irregular tiling~\cite{wikimedia_leafepidermis} (CC BY-SA 4.0, ChrisInMilton).
        (c) Vascular bundle in older \textit{Helianthus} dicot stem, showing hierarchical multi-scale packing~\cite{wikimedia_Stem_Vascular_Bundle} (CC0 1.0, Meisam; Berkshire Community College Bioscience Image Library).
        (d) Monocot root vascular bundle in \textit{Acorus} showing Casparian strip~\cite{wikimedia_Monocot_Root} (CC0 1.0, Netha Hussain; Berkshire Community College Bioscience Image Library).}
    \label{fig:bio}
\end{figure}

Biological tissues exhibit a remarkable range of mechanical states, from ordered solid-like structures to disordered liquid-like arrangements. Plant tissues in particular span this spectrum: stiff structural elements such as bark and woody stems contrast with softer compartments such as epidermal skin cells, and their mechanical properties depend sensitively on both the material composition and the spatial arrangement of cells. Figure~\ref{fig:bio} illustrates this diversity --- Fig.~\ref{fig:bio}(a) regular close-packed arrays (red onion epidermal cells), Fig.~\ref{fig:bio}(b) irregular tilings (leaf epidermis with stomata), and Figs.~\ref{fig:bio}(c)--(d) hierarchical multi-scale packings (dicot stem and monocot root vascular bundles) --- yet all of these structures share a common feature: they are mechanically robust meaning they are able to resist complex environmental loading cases such as wind, humidity fluctuations, and osmotic pressure changes. Plant cell walls are themselves multi-component structures --- cellulose, hemicellulose, lignin, and pectin --- so mechanical function is set by how components of differing stiffness are spatially arranged~\cite{Gibson2012}.

This raises fundamental questions: what determines the degree of regularity in biological cell packings and why? What topological features of the inclusion arrangement confer mechanical robustness --- specifically, the ability to dissipate or redistribute internal prestress without catastrophic failure? and How do different combinations of soft and stiff components affect overall mechanical stability?

Internal prestress arises naturally during growth---cells can experience compression when their expansion is constrained by neighboring cells and the surrounding tissue. Unlike animal cells, which can migrate and rearrange within tissues, plant cells are generally fixed relative to their neighbors by shared cell walls, and their internal stress distributions depend on turgor pressure, cell wall anisotropy, and differential growth or swelling. Mechanical stresses can influence tissue morphogenesis and amplify differences in growth rate between neighboring plant cells~\cite{Hamant2008,Uyttewaal2012}. Cell geometry, in turn, affects how these stresses are accommodated. At the cellular scale, Sapala et al.~\cite{Sapala2018} showed how lobed cell shapes can limit the formation of large unsupported wall regions susceptible to bulging under turgor pressure. Simplified elastic-network models have been used to investigate how elastic interactions and network architecture determine the mechanical response of biological fibre networks~\cite{Broedersz2011,Sharma2016}. While previous studies have examined stress redistribution under specific geometries and loading conditions, here we develop a general framework using a two-component triangular spring network with high-stiffness regions embedded in a low-stiffness region and subjected to random internal prestress. We investigate how the shape and arrangement of these regions influence energy transfer between regions, thereby governing prestress redistribution and the resulting stress concentrations. We compare two geometric mechanisms --- boundary roughness of a single region of high-stiffness and topological partitioning into multiple disconnected regions --- in relaxed regimes, and identify the key design principle for robustness in terms of crack propagation (or meandering) and stress distribution. The aim of this study is to identify and generalize the core mechanical properties observed in nature by geometrical and topological characteristics, and to translate them into design principles for engineering robust structures --- specifically, how to design different material regions and boundaries and allocate stiffness density to minimize internal stress concentration.
\section{Simulation Model}
\begin{figure}
    \centering
    \begin{tabular}{c}
        \includegraphics[width=0.7\linewidth]{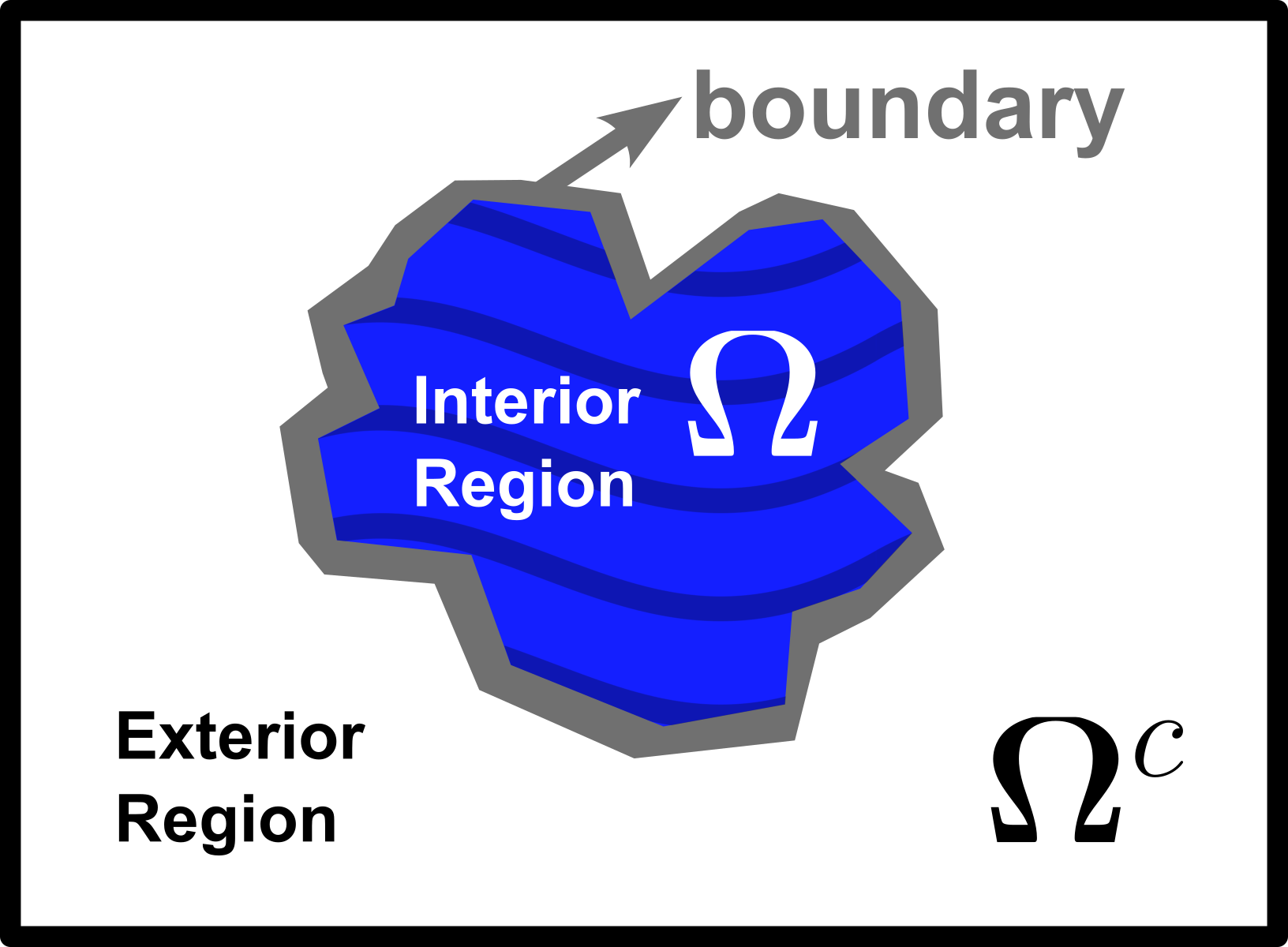} \\
\end{tabular}
\caption{Definition of regions relative to stress redistribution and energy propagation. The wavy blue region represents $\Omega$ (interior region), the white space represents $\Omega^c$ (exterior region), and the boundary represents a layer connecting them.}
    \label{fig:boundaryregion}
\end{figure}

We model the elastic material as a two-dimensional triangular spring network comprising a regular array of $N_x \times N_y$ nodes with reference positions $\{R_i\}$. Internal prestress is introduced through perturbations $\delta \ell_{ij}$ to the bond rest lengths. We consider stress propagating from $\Omega$ to $\Omega^c$ through the boundary layer via optimization (relaxation) when stress is applied only in the region $\Omega$, as represented in Fig.~\ref{fig:boundaryregion}.
. 
\subsection{Node and Bond Classification}
We define the whole domain as a disjoint union $\Omega \sqcup \Omega^c$, which represents a set of nodes $R_i$. Additionally, we define a boundary layer of nodes $\partial\Omega$ derived from the edge set $\mathcal{E}$ such that $\partial\Omega = \{\, i \in \Omega^c: \exists\, j \in \Omega \text{ with } \langle i,j\rangle \in \mathcal{E} \,\} \subset \Omega^c$ where these edges connect interior nodes to exterior nodes. A bond is assigned the high stiffness $k_{\mathrm{in}}$ only if both endpoints are interior nodes. Nodes and bonds are classified as follows:
\begin{itemize}
    \item \emph{Interior nodes}: nodes located strictly inside the prescribed
    boundary (dashed line in Fig.~\ref{fig:boundary}), $i \in \Omega$.
    \item \emph{Boundary nodes}: nodes outside the interior with at least one
    interior node as a nearest neighbor, $i \in \partial\Omega$.
    \item \emph{Exterior nodes}: nodes classified as neither interior nor
    boundary nodes, $i \in \Omega^c \setminus \partial\Omega$.
    \item \emph{Interior bonds} ($\mathcal{E}_\text{int}$): bonds connecting two interior nodes,
    $i,j \in \Omega$, assigned stiffness $k_{\mathrm{in}}$.
    \item \emph{Boundary bonds} ($\mathcal{E}_\text{bnd}$): bonds with at least one boundary-node
    endpoint, $\{i,j\} \cap \partial\Omega \neq \emptyset$, assigned
    stiffness $k_{\mathrm{out}}$.
    \item \emph{Exterior bonds} ($\mathcal{E}_\text{ext}$): bonds connecting two exterior nodes,
    $i,j \in \Omega^c \setminus \partial\Omega$, assigned stiffness
    $k_{\mathrm{out}}$.
\end{itemize}
We denote $|\mathcal{E}_\text{bnd}$| as the number of boundary bonds (crossing between interior and exterior), $N_\text{bnd}$ as the number of boundary nodes, and $N_\text{int}$ as the number of interior nodes. Fig.~\ref{fig:boundary} shows the lattice classification. Boundary nodes are defined by connectivity: nodes connecting interior to exterior nodes, giving exactly one layer at the interface, which is aimed to measure stress transfer from interior to exterior nodes. The apparent thickness in Fig.~\ref{fig:boundary} is due to the orange bonds (boundary bonds), which span from boundary nodes to their neighbors.

\subsection{Prestress Models}
\begin{figure}
    \centering
    \includegraphics[width=0.95\linewidth]{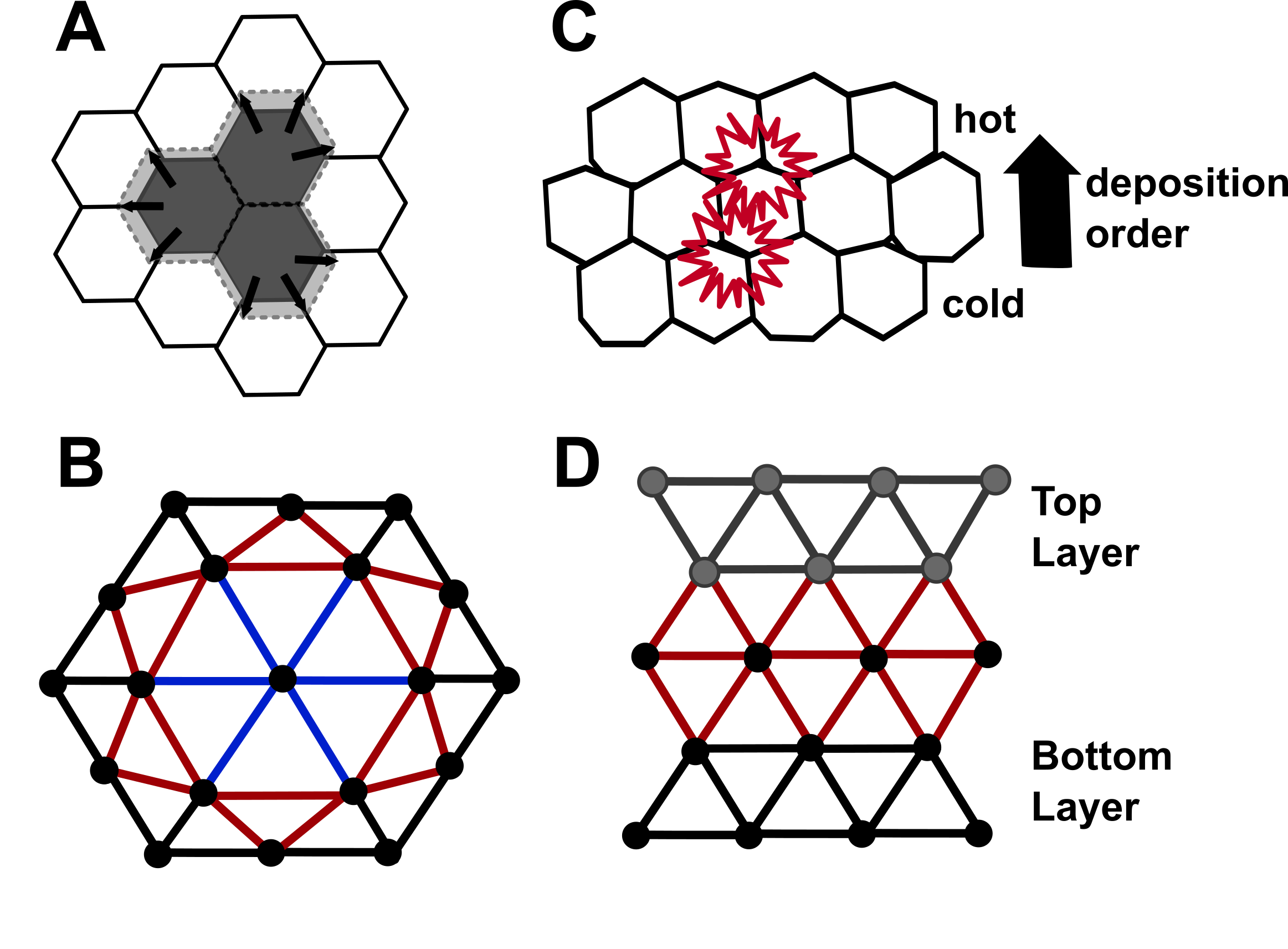}
\caption{Two stress scenarios. A: cell growth from the center; arrows represent the growing direction. B: lattice representation of A. Red denotes compressive stress applied to the region boundary and blue denotes growing bonds. C: layer-by-layer mismatch followed by volume (length) change due to dehydration (plants) or cooling (metal or crystal growth, or 3D printing process); the gray stars highlight the mismatch points. D: triangular lattice representation of C. Gray bonds represent top layer and black bonds represent bottom layer. Red denotes bonds under stress due to layer-by-layer mismatch.}
    \label{fig:stressmodel}
\end{figure}

In biological growth, stress originates from localized growth centers whose mechanical influence decays with distance (Fig.~\ref{fig:stressmodel}(a)--(b)). Similarly, in additive manufacturing, thermal gradients during layer deposition produce spatially correlated residual strains (Fig.~\ref{fig:stressmodel}(c)--(d)). We represent this as prestress by changing the original bond rest lengths. The elastic energy with rest-length perturbation $\delta \ell_{ij}$ is:\begin{equation}
	E = \frac{1}{2}\sum_{\langle i,j\rangle} k_{ij}
	\left[|r_i - r_j| - (\ell_0^{ij} + \delta \ell_{ij})\right]^2,
	\label{eq:energy}
\end{equation}
where $r_i = R_i + u_i$ and $k_{ij} = k_\text{in}$ for interior bonds and
$k_{ij} = k_\text{out}$ elsewhere.

\begin{figure}
    \centering
\begin{tabular}{c c}
        \includegraphics[width=0.5\linewidth]{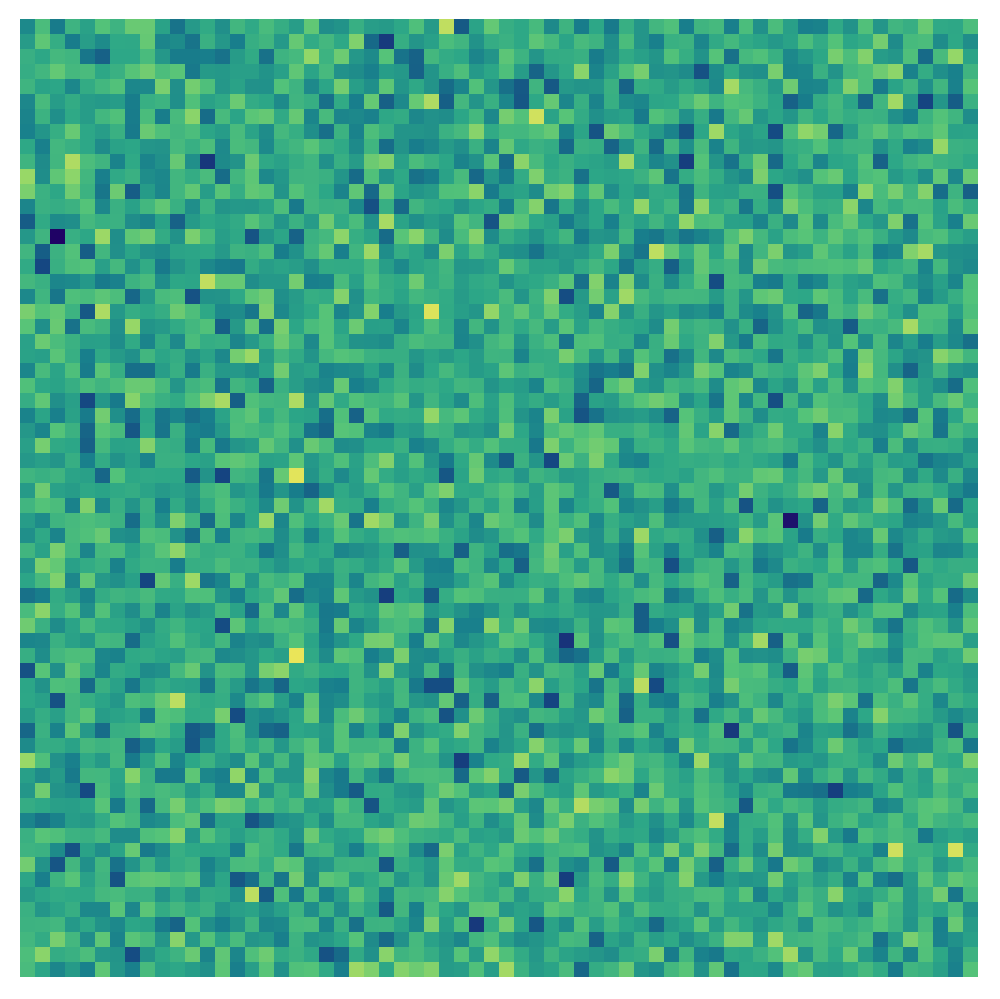} &     \includegraphics[width=0.5\linewidth]{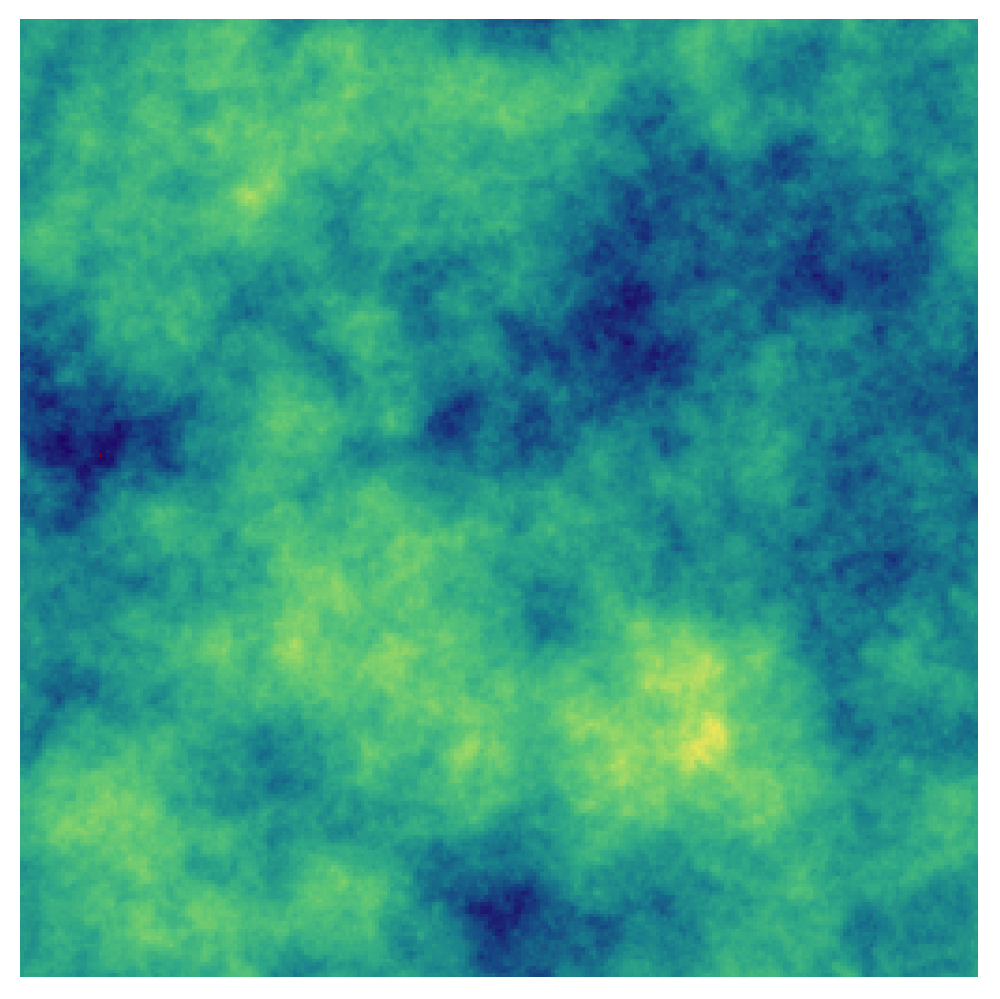} \\
        (a) & (b) \\
\end{tabular}\caption{Prestress model: (a) Gaussian Random Field to represent correlated stress (b) Randomized stress model,}
\label{fig:prestress}
\end{figure}

\paragraph{Correlated.}
To represent correlated stress, we model it as a Gaussian random field (GRF) applied to the inside bonds with an exponential covariance:
\begin{equation}
    \langle\delta \ell_{ij}\,\delta \ell_{kl}\rangle = \beta^2\exp\!\left(-\frac{|R_{ij}-R_{kl}|}{\xi_c}\right),
\end{equation}
where $R_{ij}=(R_i+R_j)/2$ is the bond midpoint and $\xi_c$ is the correlation
length. Samples are drawn via Cholesky decomposition of the covariance matrix.
Fig.~\ref{fig:prestress}(a) shows representative realizations: blue regions have
$\delta \ell_{ij}>0$ (elongated bonds, compressive prestress) and yellow regions have
$\delta \ell_{ij}<0$ (compressed bonds, tensile prestress).
\paragraph{Random (uncorrelated).}
In many manufacturing processes and biological growth, material heterogeneity arises
from local fluctuations with no spatial memory --- each bond misfits independently
of its neighbors. We model this as $\delta \ell_{ij} \sim \mathcal{N}(0,\beta^2)$
drawn independently for each interior bond, with covariance
$\langle \delta \ell_{ij}\,\delta \ell_{kl}\rangle = \beta^2\delta_{(ij),(kl)}$.
Fig.~\ref{fig:prestress}(b) shows a representative realization: blue and yellow
regions are spatially uncorrelated with no visible structure.

\paragraph{Constant (comparison)}
The constant prestress has no physical background. However, it is conducted to demonstrate how residual energy generated by prestress propagates through the elastic energy relaxation.
\subsection{Two Physical Regimes}
Assuming two materials with stiffness $k_\text{in}$ and $k_\text{out}$, we separate Eq.~\eqref{eq:energy} into contributions from interior nodes belonging to region $\Omega$, boundary nodes in $\partial\Omega$, and exterior nodes in $\Omega^c\setminus \partial \Omega$:
\begin{align}\label{eq:energy_split}
E &= \frac{1}{2}\sum_{\langle ij\rangle \in \mathcal{E}_{\text{int}}} k_\text{in}\left[|r_i-r_j|-(\ell_0^{ij}+\delta \ell_{ij})\right]^2 \notag \\
&+ \frac{1}{2}\sum_{\langle ij\rangle \in \mathcal{E}_{\text{bnd}}}  k_\text{out}\left[|r_i-r_j|-\ell_0^{ij}\right]^2\notag \\
&+ \frac{1}{2}\sum_{\langle ij\rangle \in  \mathcal{E}_{\text{ext}}} k_\text{out}\left[|r_i-r_j|-\ell_0^{ij}\right]^2 
\end{align}
Practically, in real lattice structures, the second term may have $k_\text{in}$ instead of $k_\text{out}$ in addition to boundary mismatch (prestress) if we assume exterior nodes do not experience much prestress compared to the interior nodes. In the above equation, the second term can be modified to include prestress: increasing the number of such bonds amplifies the total prestress applied to the system when the number of interior nodes is the same for all cases, making the structure more fragile. To isolate the effect of varying the number of nodes experiencing prestress, in this study, prestress is restricted to interior bonds only.

In a totally inextensible system (no displacement allowed from the initial configuration), the total energy from $\Omega$ is unchanged by boundary shape (if the number of interior nodes is fixed) --- prestress is fully trapped inside. Once displacement is permitted, the number of boundary bonds controls how much prestress energy propagates into $\partial\Omega$ and $\Omega^c$. We relax the network by optimization. This lets us compare different stiffness ratios, boundary shapes, and partitioning, and measure how stress redistributes globally as a function of the prestress applied to the interior region. We set $k_\text{in}>k_\text{out}$ to capture the boundary-region effect on stress propagation from higher to lower energy; in the opposite case, stress may not relax sufficiently, and both cases show similar effects. For consistent evaluation including soft or biological materials, we considered two cases.
\begin{itemize}
    \item Unconstrained: the full nonlinear energy Eq.~\eqref{eq:energy_split} is minimized via FindMinimum (Mathematica) or LBFGS (Julia).
    \item Constrained: node displacements beyond $0.1\,\ell_0$ are penalized by $\mu(|u_i| - 0.1\,\ell_0)^2$ with $\mu = 10^4$, where $\ell_0$ is the bond rest length, to mimic partially extendable materials.

\end{itemize}
\section{Simulation Results}
\subsection{Simple Prestress Test}
\begin{figure}
    \centering
    \begin{tabular}{c}
        \includegraphics[width=0.95\linewidth]{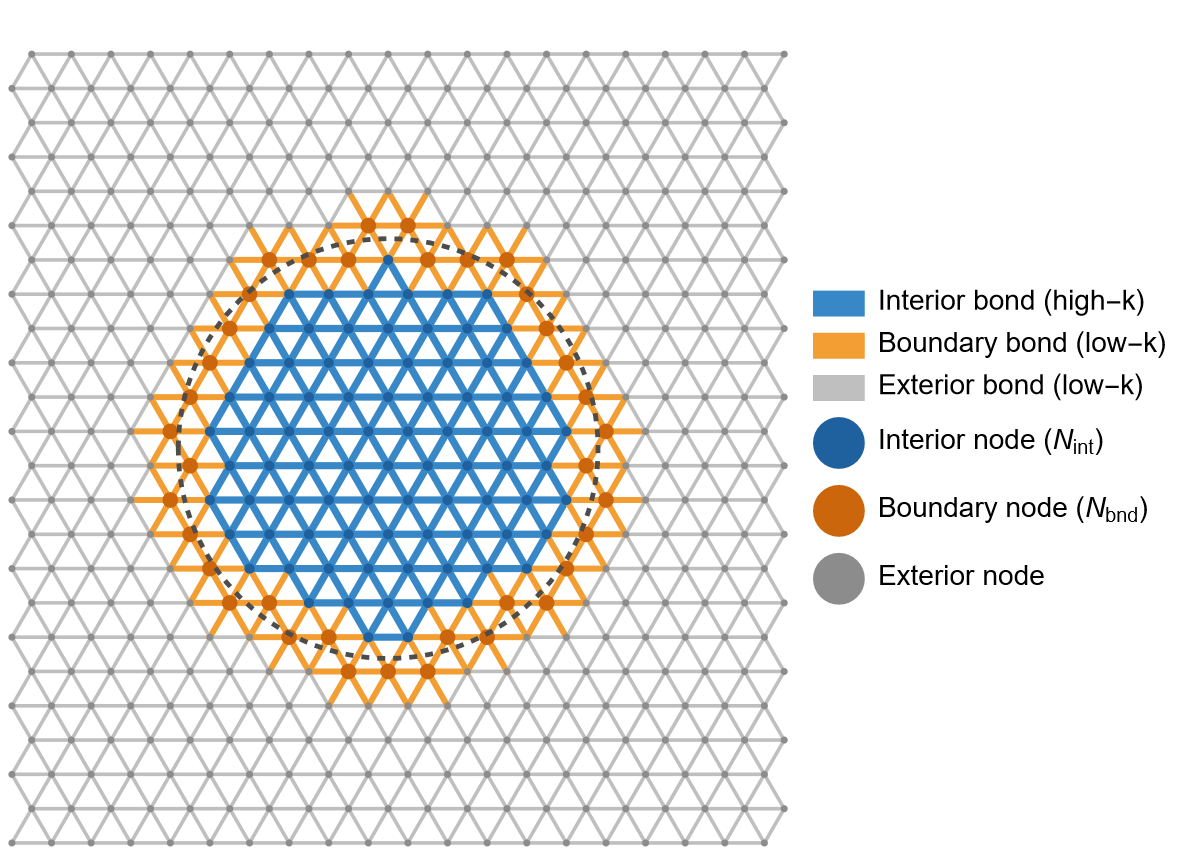} \\
        (a)\\
        \includegraphics[width=0.95\linewidth]{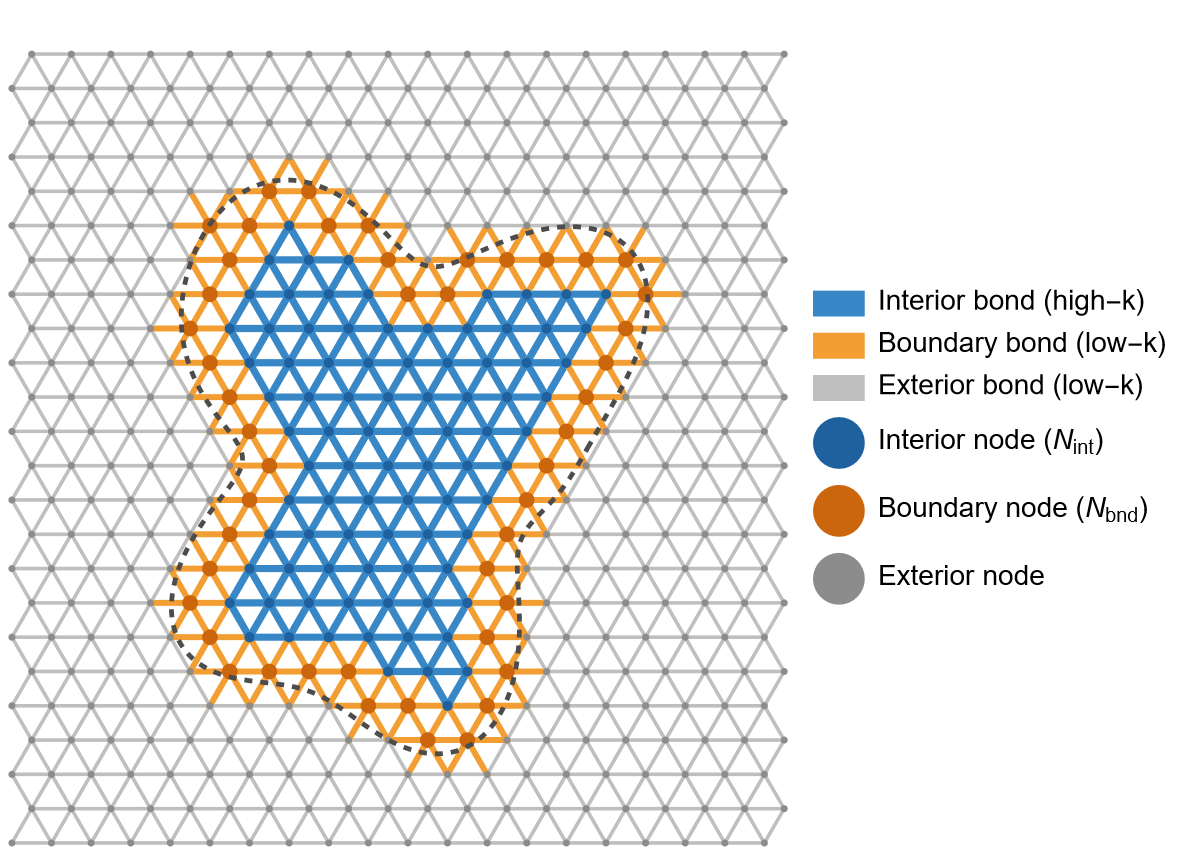}\\
        (b)\\
\end{tabular}
 \caption{Node and bond classification on a 2D triangular lattice ($20\times24$) for two inclusion geometries: (a) circular disk, (b) rough boundary ($\sigma = 0.3,l_c = 1, n_\text{min} = 2, n_\text{max} = 6$). Blue bonds: interior bonds (both endpoints interior, high-$k$). Orange bonds: boundary bonds. Gray bonds: exterior bonds (low-$k$). Filled circles denote interior nodes ($N_\text{int}$, dark blue), boundary nodes ($N_\text{bnd}$, dark orange), and exterior nodes (gray).}
    \label{fig:boundary}
\end{figure}

We conduct a simple prestress test to analyze the relationship between boundary shape and stress distribution, inspired by the plant epidermal cell geometries shown in Fig.~\ref{fig:bio}(a)--(b). Two boundary shapes are compared as shown in Fig.~\ref{fig:boundary}(a)--(b): a circular disk and a rough boundary, under bounded conditions for the $20\times24$ lattice with constant prestress, where the outermost exterior nodes are fixed. For the larger $40\times40$ lattice, we used periodic boundary conditions by wrapping the boundaries. Both the bounded lattice (Mathematica) and the periodic lattice (Julia) apply the prestress quasi-statically, in five equal increments with full relaxation at each step. A single minimization carries no notion of loading history, whereas incremental loading represents slow, continuous deformation such as growth or cooling. The rough boundary $\partial\Omega$ is parameterized by a random Fourier series:
\begin{equation}
    R(\theta) = R_0\!\left[1 + \sigma\sum_{n=n_\text{min}}^{n_\text{max}}
    \frac{a_n\cos n\theta + b_n\sin n\theta}{n^{l_c}}\right],
\end{equation}
where $a_n,b_n\sim\mathcal{N}(0,1)$ and $\sigma$ controls the roughness amplitude.
The power-law weight $n^{-l_c}$ controls the spectral decay: larger $l_c$ suppresses
high-frequency modes, giving smoother boundaries; $l_c=0$ gives white-noise roughness. Note that $R_0$ is adjusted to keep interior bonds comparable. Throughout, $l_c = 1$, $n_\text{min} = 2$, and $n_\text{max} = 6$.
\subsubsection{Constant Prestress}
A uniform prestress $\delta \ell_{ij}=0.2$ is applied to all interior bonds on a $20\times24$ triangular lattice. To eliminate statistical effects from variations in node count, $|\mathcal{E}_\text{int}|=218$ is fixed for both boundary shapes, with prestress applied only to interior bonds. Energy minimization is performed via \texttt{FindMinimum} with 5 continuation steps using $k_\text{in}=10$, $k_\text{out}=1$, and fixed boundary conditions (exterior boundary nodes are held fixed). Bond stress is measured as $s=|\ell_{ij}-(\ell_0^{ij}+\delta \ell_{ij})|$ after optimization, and results are rescaled by the same factor in both cases for a fair comparison.

\begin{figure*}
    \centering
\begin{tabular}{c c}
        \includegraphics[width=0.3\linewidth]{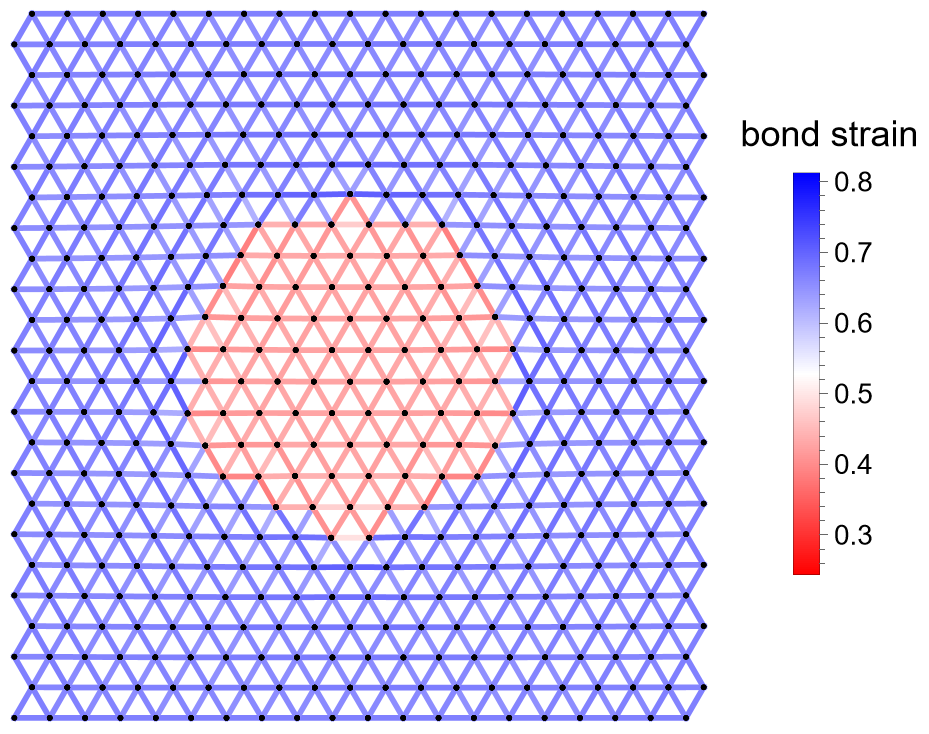} &
        \includegraphics[width=0.3\linewidth]{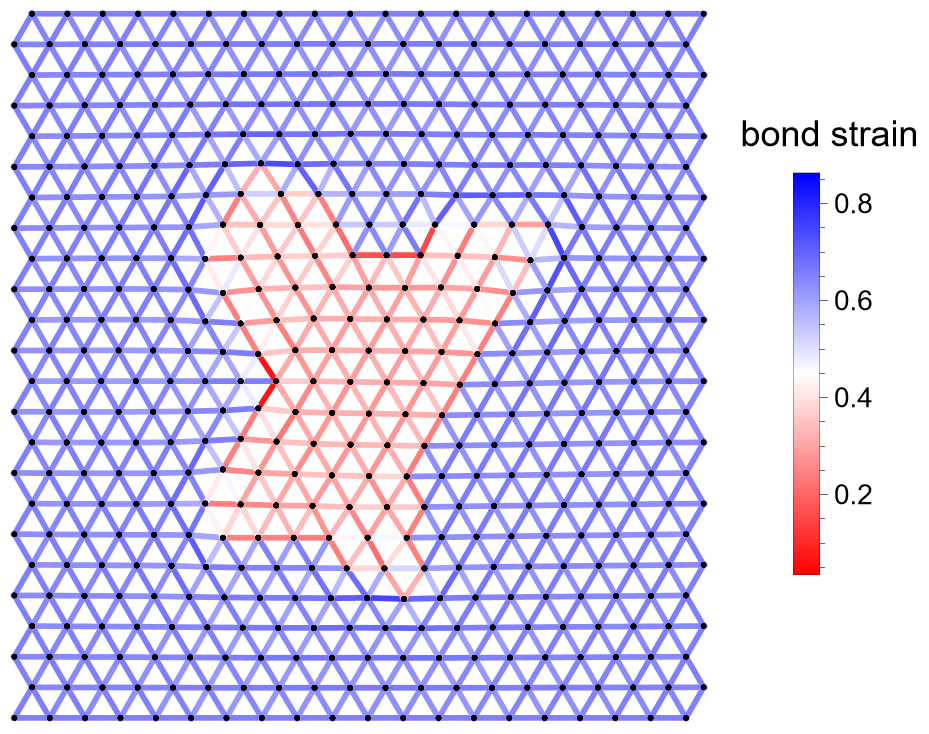} \\ 
             (a) Disk (constrained) &(b) rough boundary (constrained)\\
        & 
\end{tabular}
\begin{tabular}{c c c}
        \includegraphics[width=0.33\linewidth]{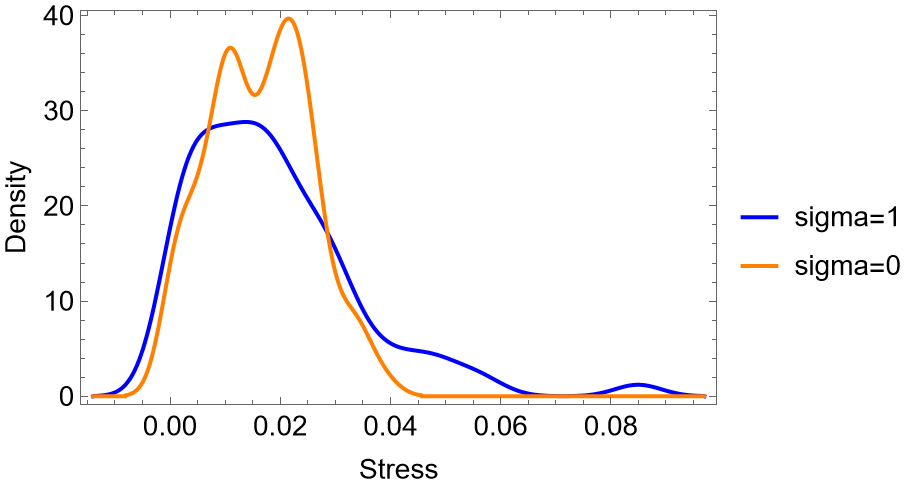} &       \includegraphics[width=0.33\linewidth]{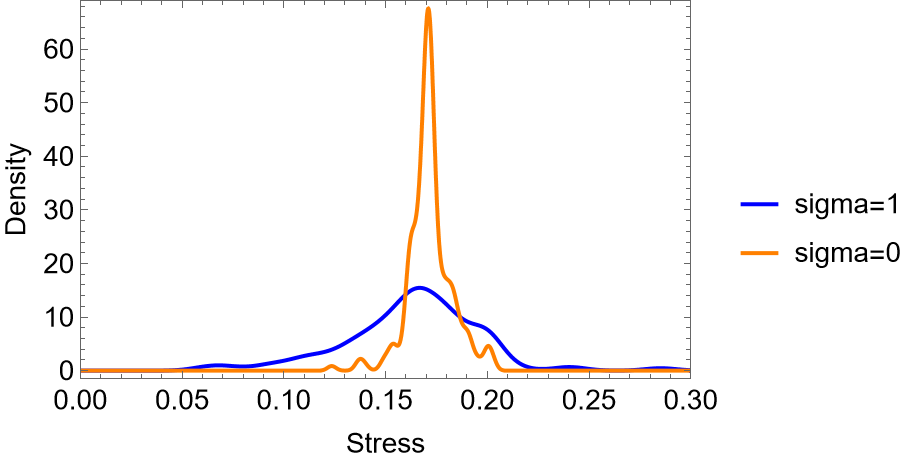}&
         \includegraphics[width=0.33\linewidth]{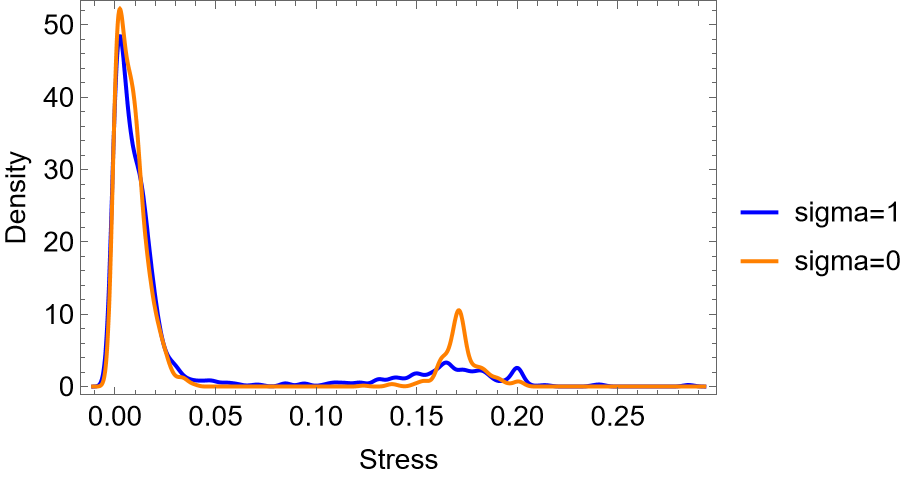}\\
 (c) Boundary bonds (constrained) & (d) Interior bonds (constrained) & (e) All bonds (constrained)\\
        \includegraphics[width=0.33\linewidth]{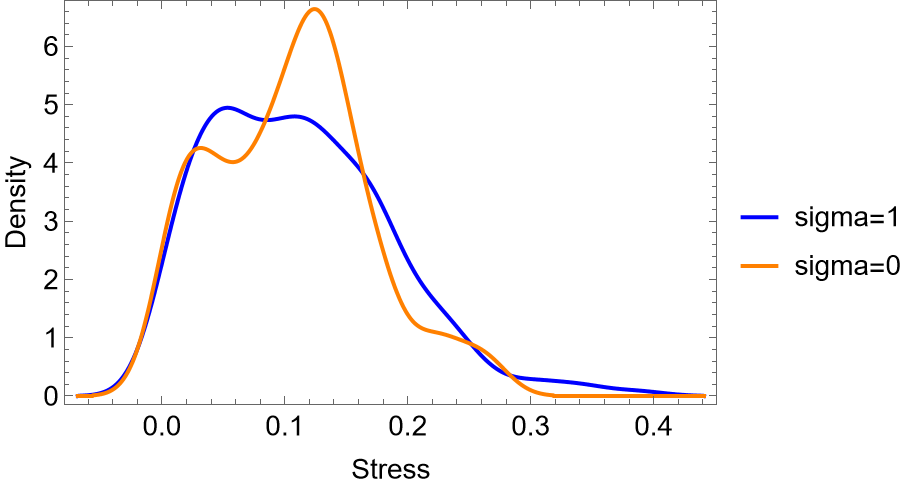} &       \includegraphics[width=0.33\linewidth]{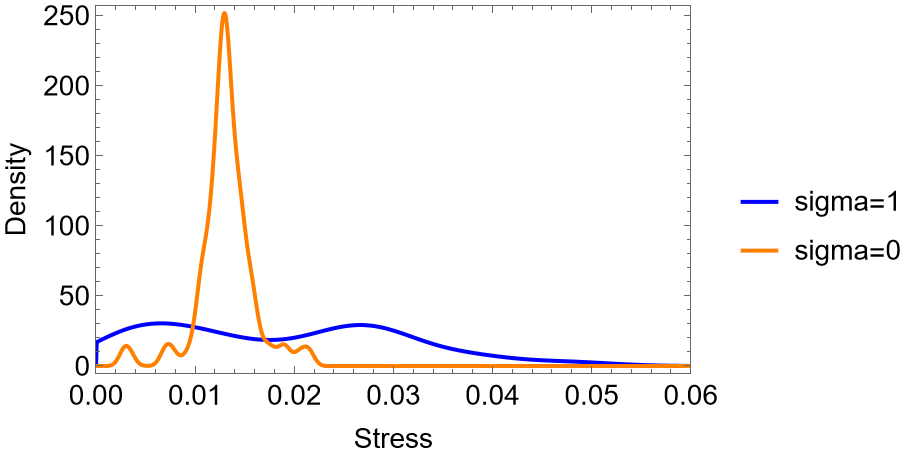}&
         \includegraphics[width=0.33\linewidth]{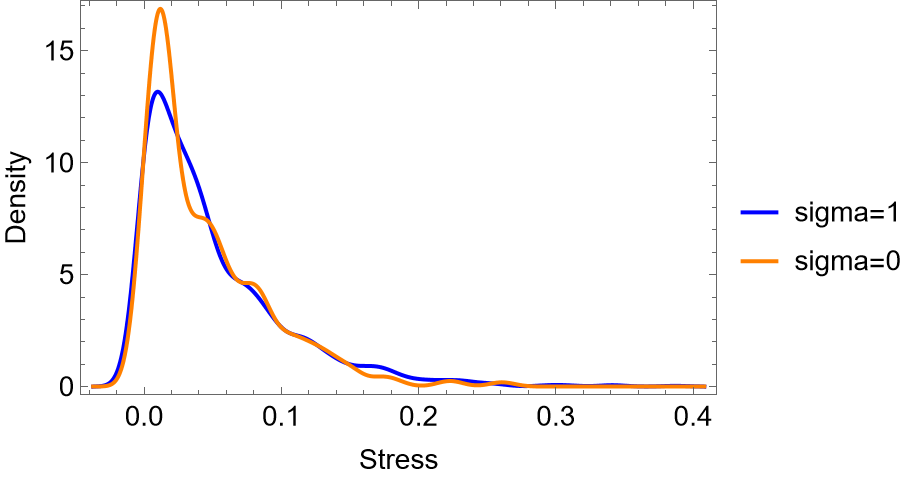}\\
 (f) Boundary bonds (unconstrained) & (g) Interior bonds (unconstrained) & (h) All bonds (unconstrained)\\
 \end{tabular}
\caption{{\it Stress distribution under constrained and unconstrained relaxation.}
(a)--(b) Rescaled stress distributions on the disk boundary ($\sigma=0$) and rough boundary ($\sigma=0.3$, $R_0 = 7.2$), respectively, with red and blue indicating relative stress magnitude.
(c)--(e) Bond stress distributions for the constrained case, comparing rough boundary $\sigma=0.3$ (blue) and disk $\sigma=0$ (orange).
(f)--(h) Same as (c)--(e) for the unconstrained case. Increasing the number of boundary bonds lowers the peak boundary-bond stress and broadens the interior stress distribution, in both the constrained and unconstrained cases.}
    \label{fig:Cconstrained}
\end{figure*}

Figs.~\ref{fig:Cconstrained}(a)--(b) illustrate the geometric differences between boundary shapes; (c)--(e) and (f)--(h) show the corresponding stress distributions for the constrained and unconstrained cases, respectively. The rough boundary ($\sigma=0.3$, butterfly shape) exhibits a broader stress distribution than the circular boundary ($\sigma=0$) for both interior and boundary bonds. Moreover, the rough boundary exhibits lower boundary bond stress and a broader interior stress distribution, which can be partially attributed to its larger number of boundary bonds ($\sigma=0$ 185 vs.\ $\sigma=0.3$ 234), providing more parallel pathways for stress dissipation into the exterior region.

\begin{figure}
    \centering
    \begin{tabular}{c c}
        \includegraphics[width=0.5\linewidth]{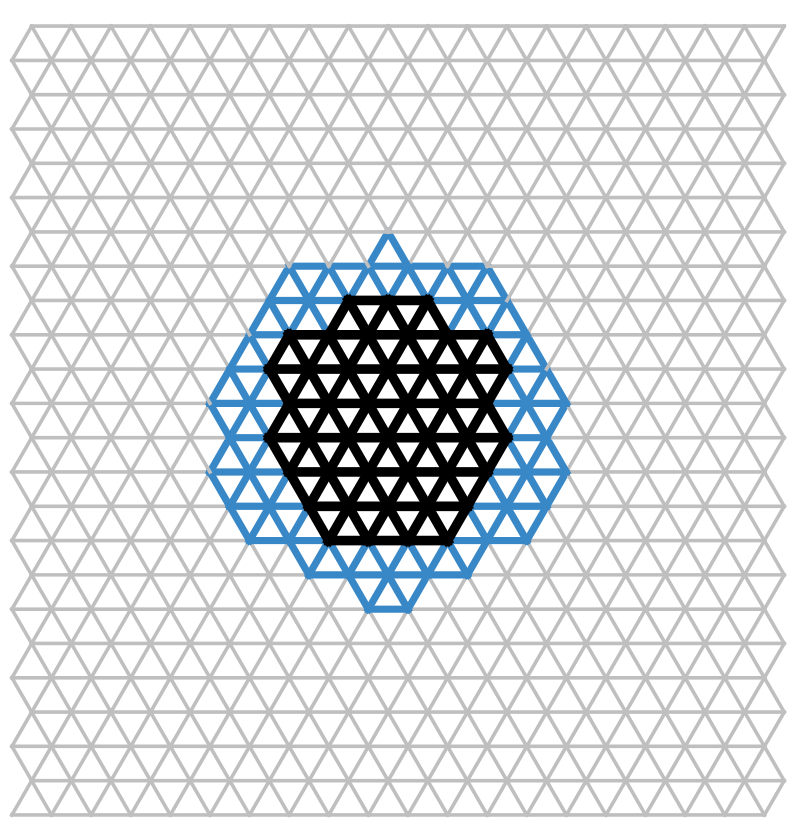} &
        \includegraphics[width=0.5\linewidth]{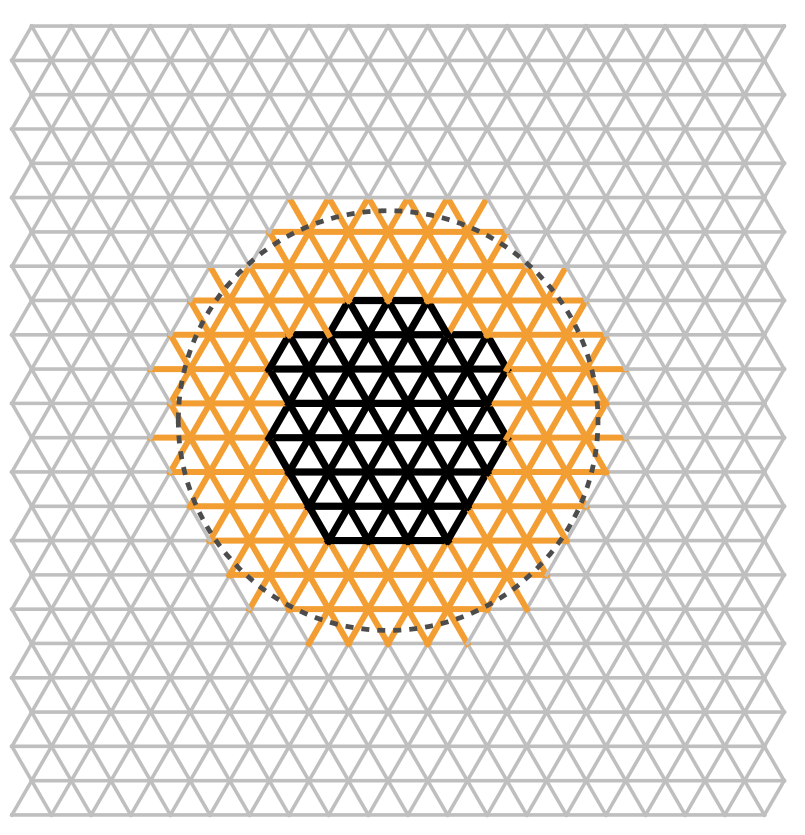} \\
        (a) & (b) \\[6pt]
    \end{tabular}
    \begin{tabular}{c}
        \includegraphics[width=0.95\linewidth]{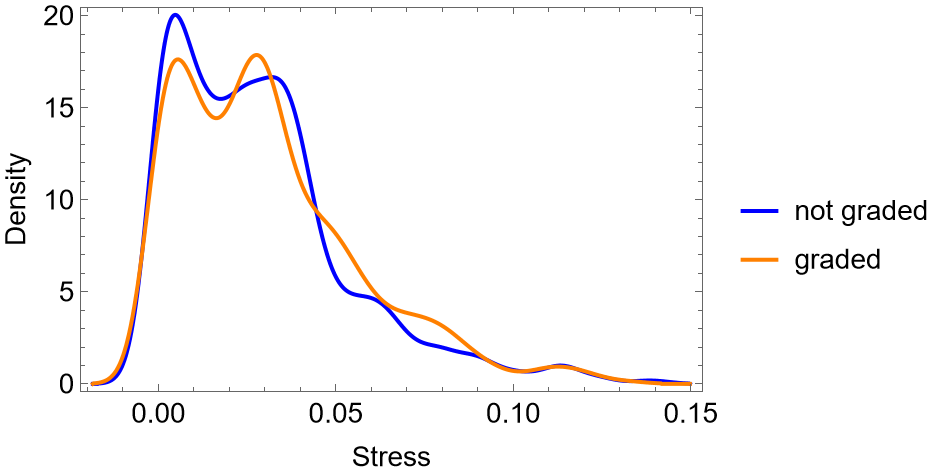} \\
        (c)\\
        \includegraphics[width=0.95\linewidth]{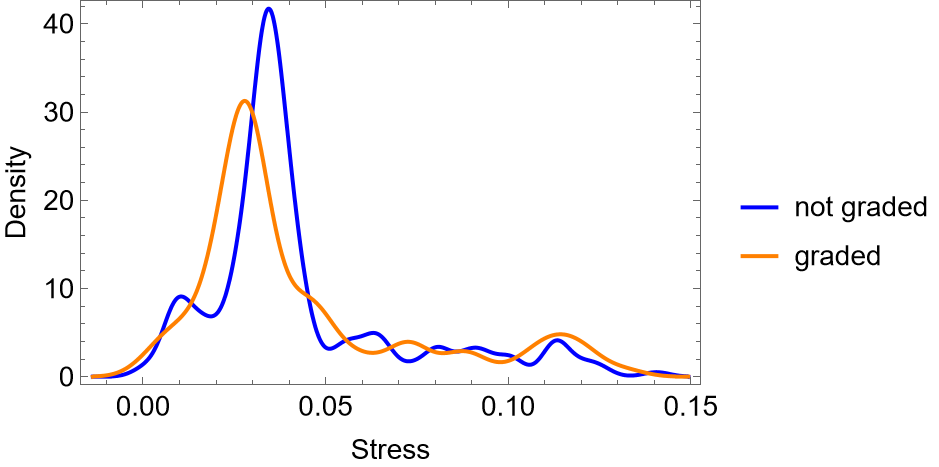} \\
         (d) \\
    \end{tabular}
    \caption{
(a) Original interior region with $k=10$ (blue), where prestress is applied to the entire interior.
(b) Graded boundary configuration: the blue region retains $k=10$ with prestress applied only to the black bonds, while the orange region represents an extended intermediate layer with $k=5$, twice the width of the boundary region in (a).
(c) Stress distribution of all bonds.
(d) Stress distribution of interior bonds only (blue shaded bonds in both cases). Distributing the stiffness transition over a wider region relieves interior stress concentration relative to an abrupt interface.}
\label{fig:Graded}
\end{figure}

We also tested a graded transition region by introducing a buffer layer. The graded boundary is constructed so that half the bonds are drawn from the interior and half from the exterior, with equal contributions from boundary bonds (Fig.~\ref{fig:Graded}(a), blue vs.\ (c) orange shaded bonds). The stiffness of the graded region is set to half the interior stiffness ($k=5$). Prestress is then applied to the same set of bonds (black shaded) to isolate the graded effect. While the full bond distribution in Fig.~\ref{fig:Graded}(c) is comparable between the two configurations, restricting to interior bonds in Fig.~\ref{fig:Graded}(d) reveals reduced stress concentration in the graded case, confirming that the graded transition is effective. The effect is modest due to the relatively small stiffness ratio between interior and exterior ($k_\text{in}/k_\text{out}=10$) and the limited extent of the graded region imposed by the lattice size ($20\times24$).

\subsubsection{Correlated vs.\ Uncorrelated Prestress}
\begin{figure}
    \centering
        \includegraphics[width=0.95\linewidth]{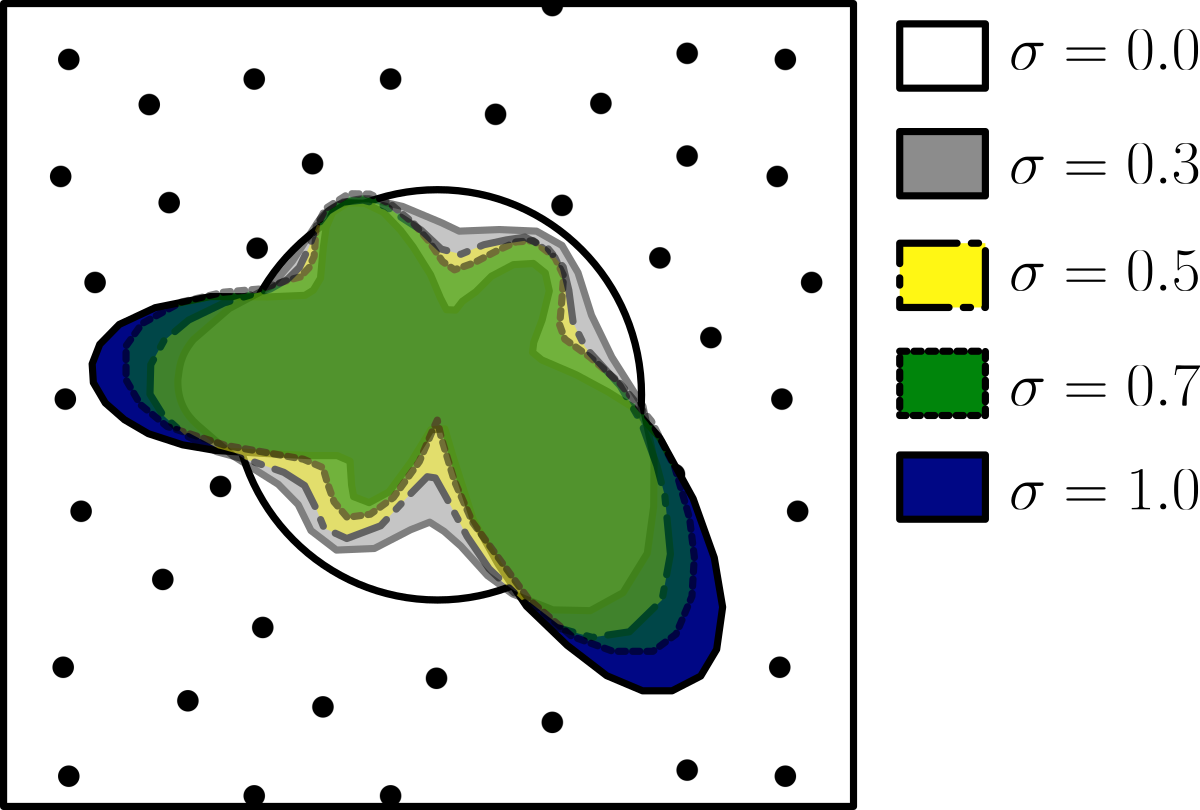} 
\caption{Rough boundary shapes on a $40\times40$ triangular lattice. Increasing $\sigma$ produces rougher boundaries. Colors and border patterns distinguish different values of $\sigma$. The black dot-filled region represents the exterior.}
    \label{fig:FourierBD}
\end{figure}

To obtain broader statistics, we increased the lattice size to $40\times40$ with $k_\text{out}=1$, $\xi_c=2$, $\beta\in\{0.05,0.1,0.2\}$, $R_0 = 15$, and fixed $N_\text{inside}$ where $N_\text{inside} \equiv N_\text{int} + N_\text{bnd}$, under periodic boundary conditions. Each parameter set was averaged over $100$ independent realizations. As a result, $|\mathcal{E}_\text{int}|$ varies with $\sigma$ (a $\sim$6\% decrease from $\sigma=0$ to $\sigma=1.0$), while $|\mathcal{E}_\text{bnd}|$ increases by $\sim$52\%, making the boundary geometry the dominant variable. We tested five boundary shapes $\sigma\in\{0,0.3,0.5,0.7,1\}$ as shown in Fig.~\ref{fig:FourierBD}. Since $s_\text{int}$ increases monotonically with $\beta$, we present $\beta=0.2$ as the representative case; results for $\beta\in\{0.05,0.1\}$ are qualitatively consistent and omitted for brevity (numerical results for all $\beta$ values are provided as supplementary data in \ref{sec:supp_table}). We measured displacement magnitude and bond stress per region $r$:
\begin{equation}
    \xi_r = \langle|u_i|\rangle_{i\in r}, \quad
    \text{Var}_r = \text{var}(|u_i|)_{i\in r}
\end{equation}
\begin{equation}
    s_b = k_b\,|\ell_{ij} - (\ell_0^{ij} + \delta \ell_{ij})|
\end{equation}
\begin{equation}
    s_r = \langle s_b\rangle_{b\in B_r}, \quad
    \text{Var}^s_r = \text{var}(s_b)_{b\in B_r}
\end{equation}
where $r \in \{\text{int}, \text{bnd}, \text{tot}\}$ and $B_r$ is the corresponding bond set. The primary quantity of interest is $s_\text{int}$, the mean bond stress inside $\Omega$, which measures residual stress concentration in the interior after prestress is applied; we therefore focus mainly on $s_\text{int}$ and $s_\text{bnd}$ (see Appendix Sec. \ref{sec:supp_table} for full data of selected sets). All quantities are averaged over all realizations. Prior to applying prestress, the network is relaxed to equilibrium using unit rest lengths over 500 LBFGS iterations. Prestress is then applied and the system is re-minimized via LBFGS with 200 iterations for the constrained case and 100 iterations for the unconstrained case.
\begin{equation}
	\frac{\partial E}{\partial u_i^\alpha} = -\sum_{j\sim i} k_{ij}
	\bigl(|r_i-r_j| - \ell_0^{ij} - \delta \ell_{ij}\bigr)\,
	\frac{\partial |r_i-r_j|}{\partial u_i^\alpha} = 0.
	\label{eq:relaxed}
\end{equation}

\begin{figure*}
    \centering
    \begin{tabular}{c c}
        \includegraphics[width=0.5\linewidth]{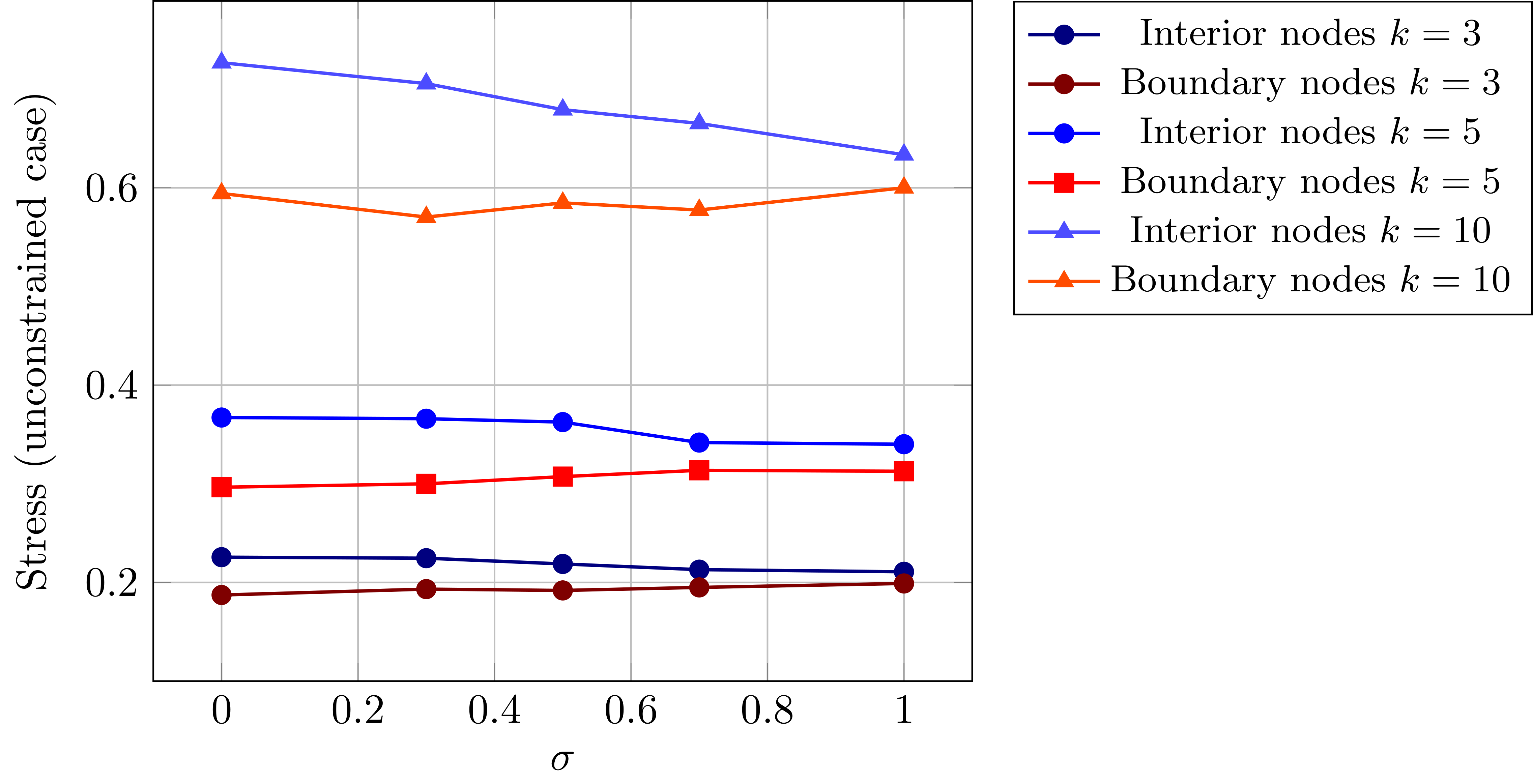} &
        \includegraphics[width=0.5\linewidth]{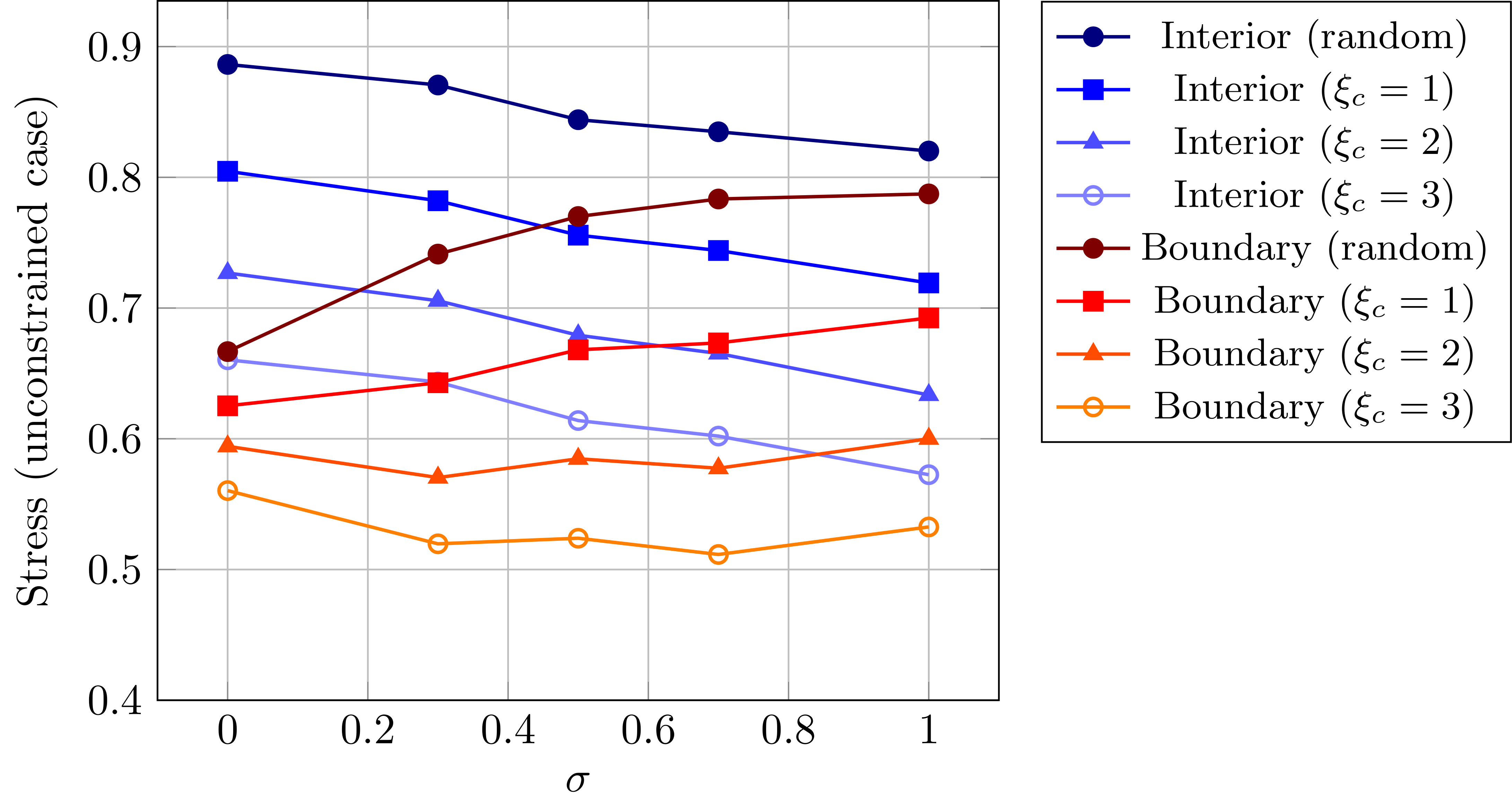} \\
        (a) Interior stiffness ($p_\text{in}=p_\text{out}=1$)& (b) Prestress correlation ($p_\text{in}=p_\text{out}=1$)\\
        \includegraphics[width=0.5\linewidth]{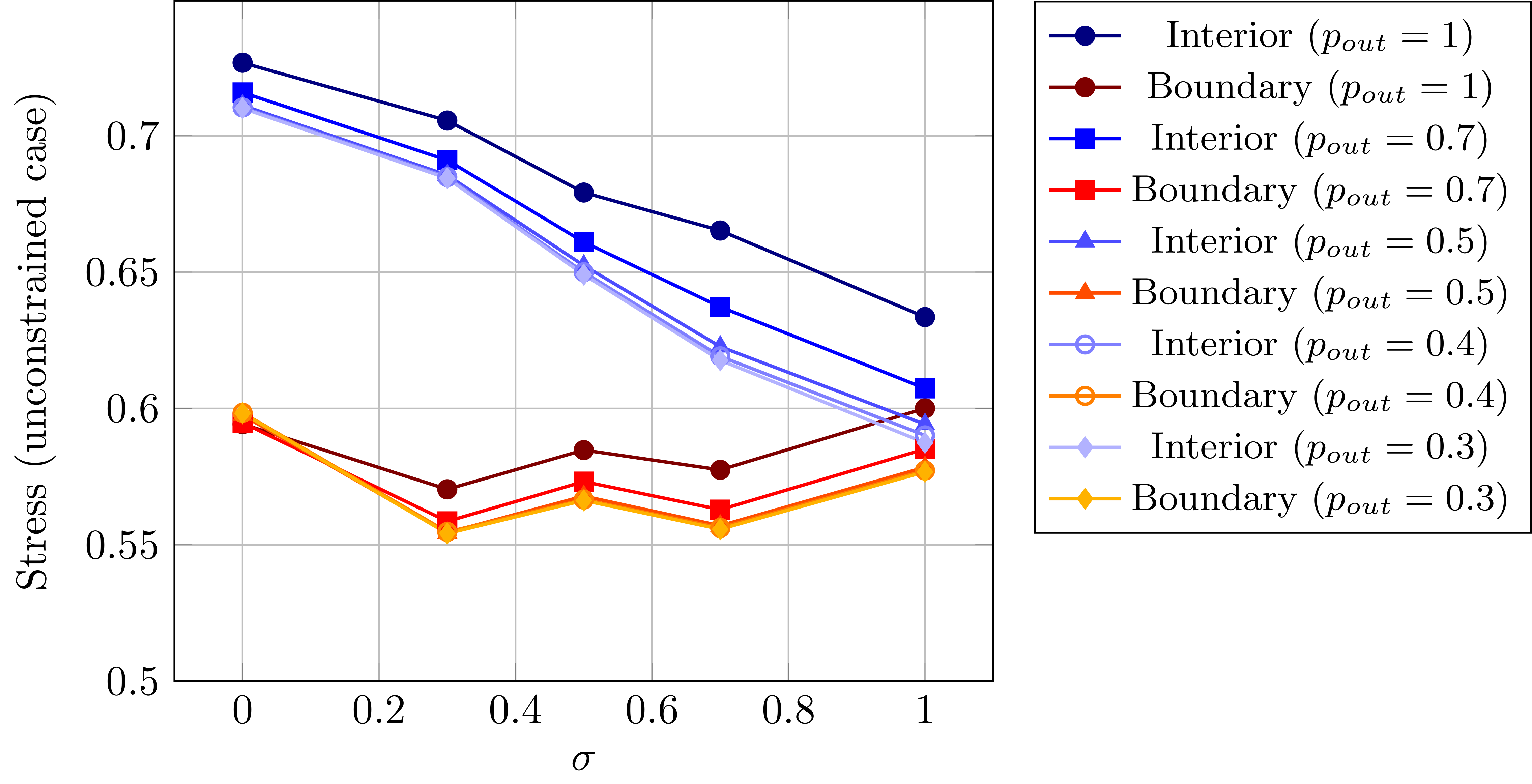} &
        \includegraphics[width=0.5\linewidth]{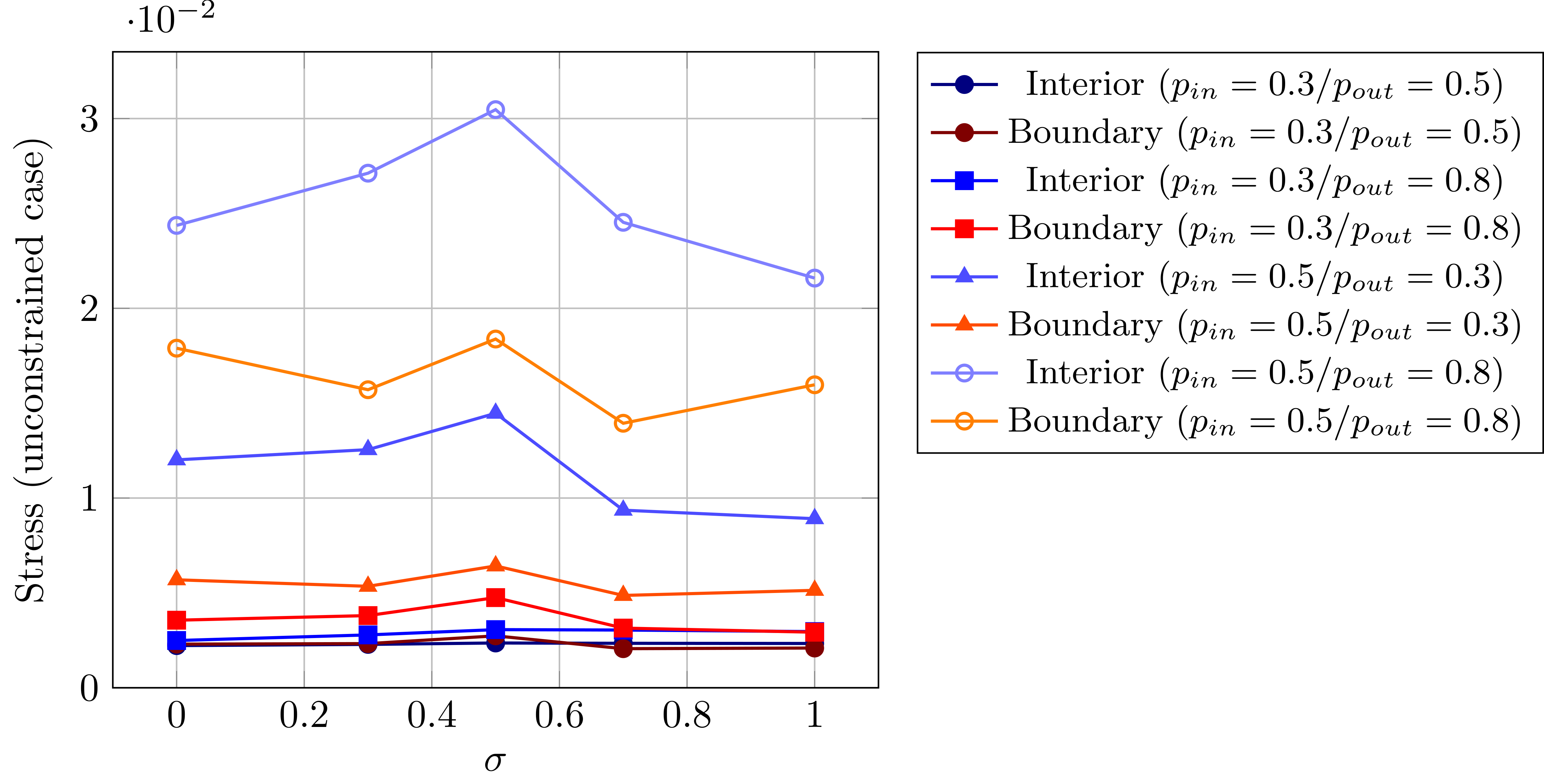} \\
(c) Interior region fraction with $p_\text{out}=1$ & (d) Floppy interior region\\[6pt]
        \includegraphics[width=0.5\linewidth]{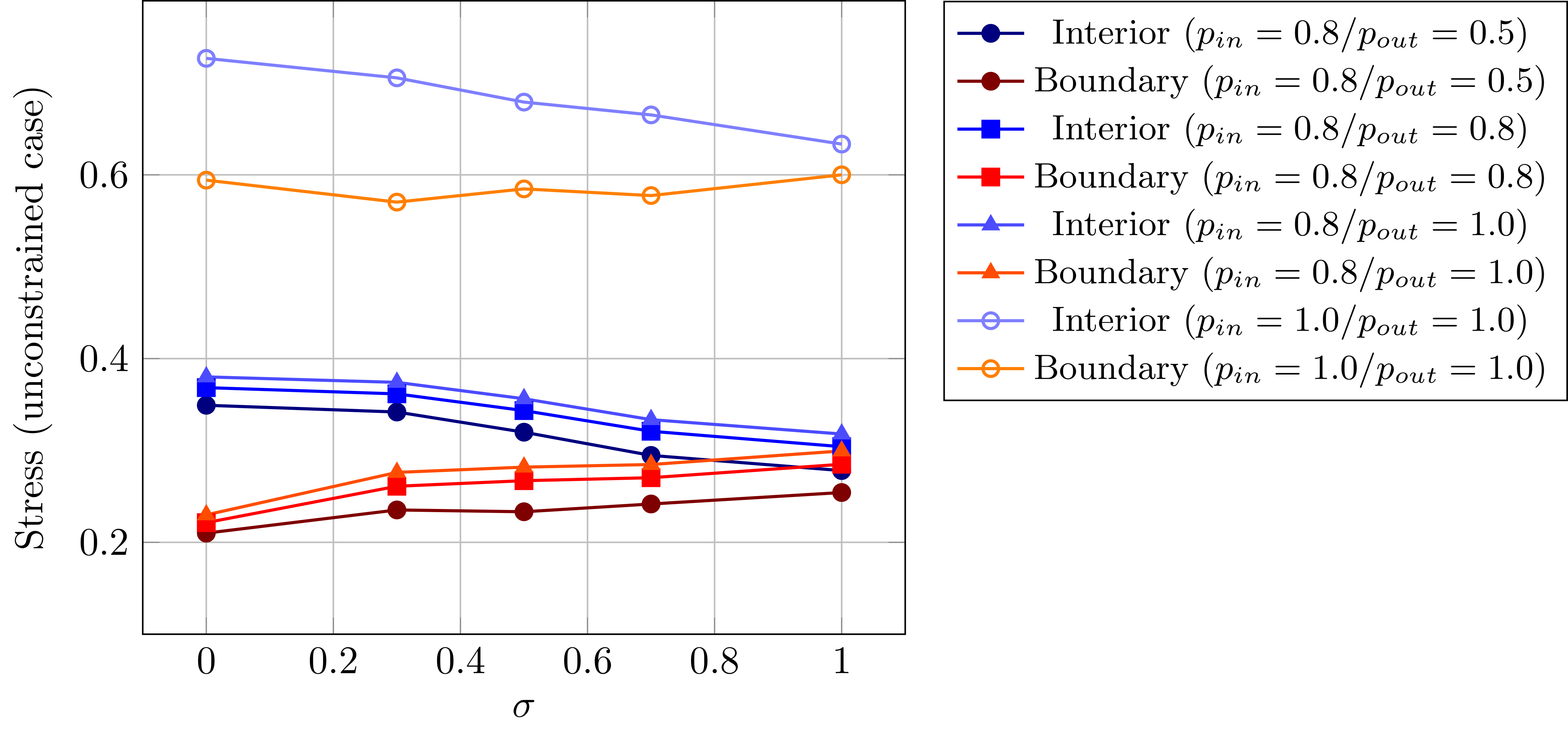} & \\
(e) Stiff interior region &\\
    \end{tabular}
    \caption{{\it Stress after unconstrained energy relaxation, for a range of parameters.}
    (a)~Stress as the interior stiffness $k_\mathrm{in}$ is varied, with the exterior fixed at $k_\mathrm{out}=1$.
    (b)~Comparison between a correlated stiffness field with correlation length $\xi=3$ and an uncorrelated one ($k_\mathrm{in}=10$).
    (c)~Effect of the area fraction occupied by the exterior region, with $p_\text{in}=1$.
    (d)~Comparison across exterior fractions when the interior region is floppy ($k_\mathrm{in}=10$).
    (e)~The same comparison when the interior region is rigid ($k_\mathrm{in}=10$). Stress becomes comparable as $\sigma$ increases. The dependence on boundary shape survives dilution of both regions. Only when the interior becomes floppy does the mechanism change, with prestress released internally rather than transferred outward.}
    \label{fig:unconstrained}
\end{figure*}

\begin{figure*}
    \centering
    \begin{tabular}{c c}
        \includegraphics[width=0.5\linewidth]{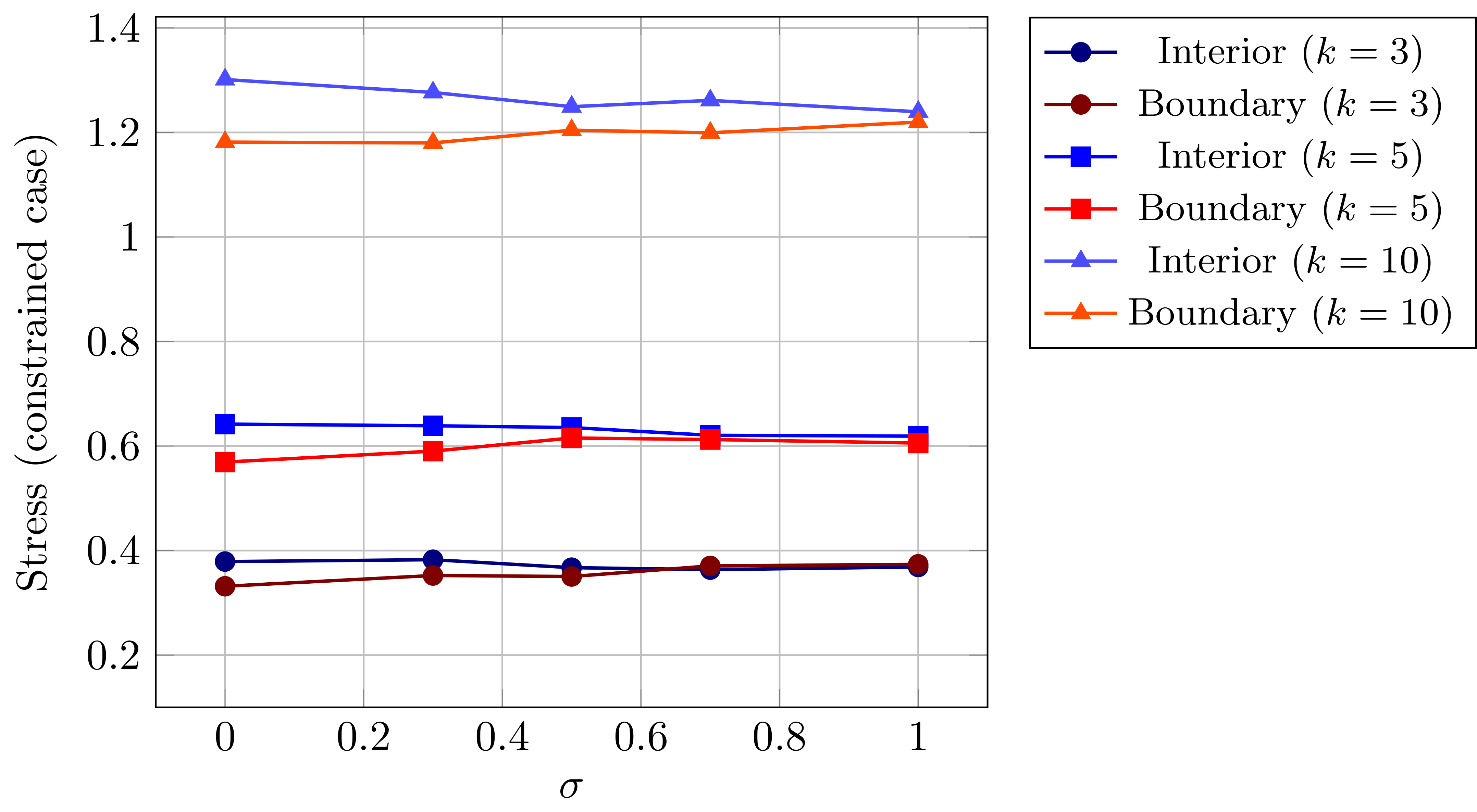} &
        \includegraphics[width=0.5\linewidth]{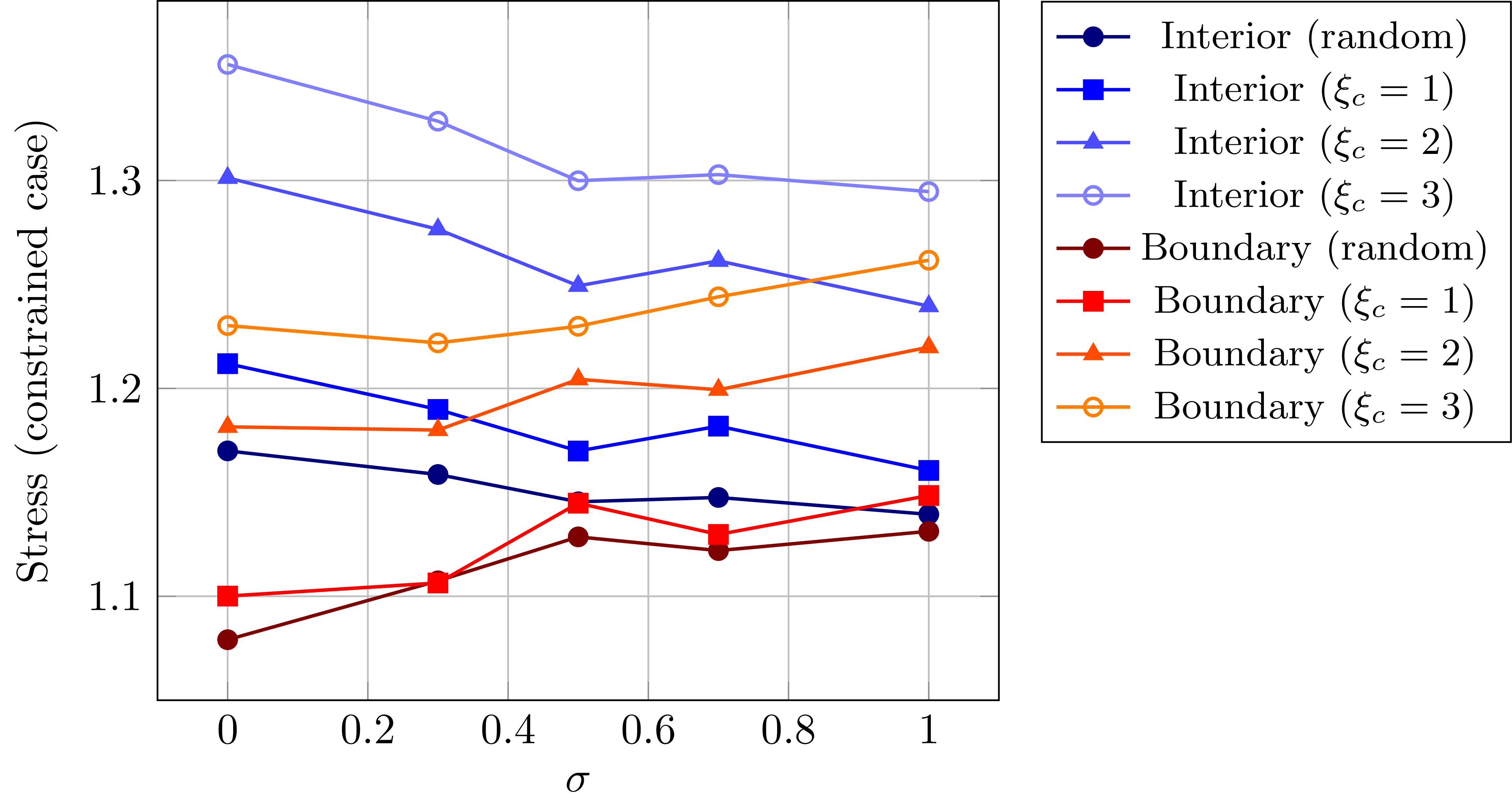} \\
        (a) Interior stiffness ($p_\text{in}=p_\text{out}=1$) &
        (b) Prestress correlation ($p_\text{in}=p_\text{out}=1$) \\
        \includegraphics[width=0.5\linewidth]{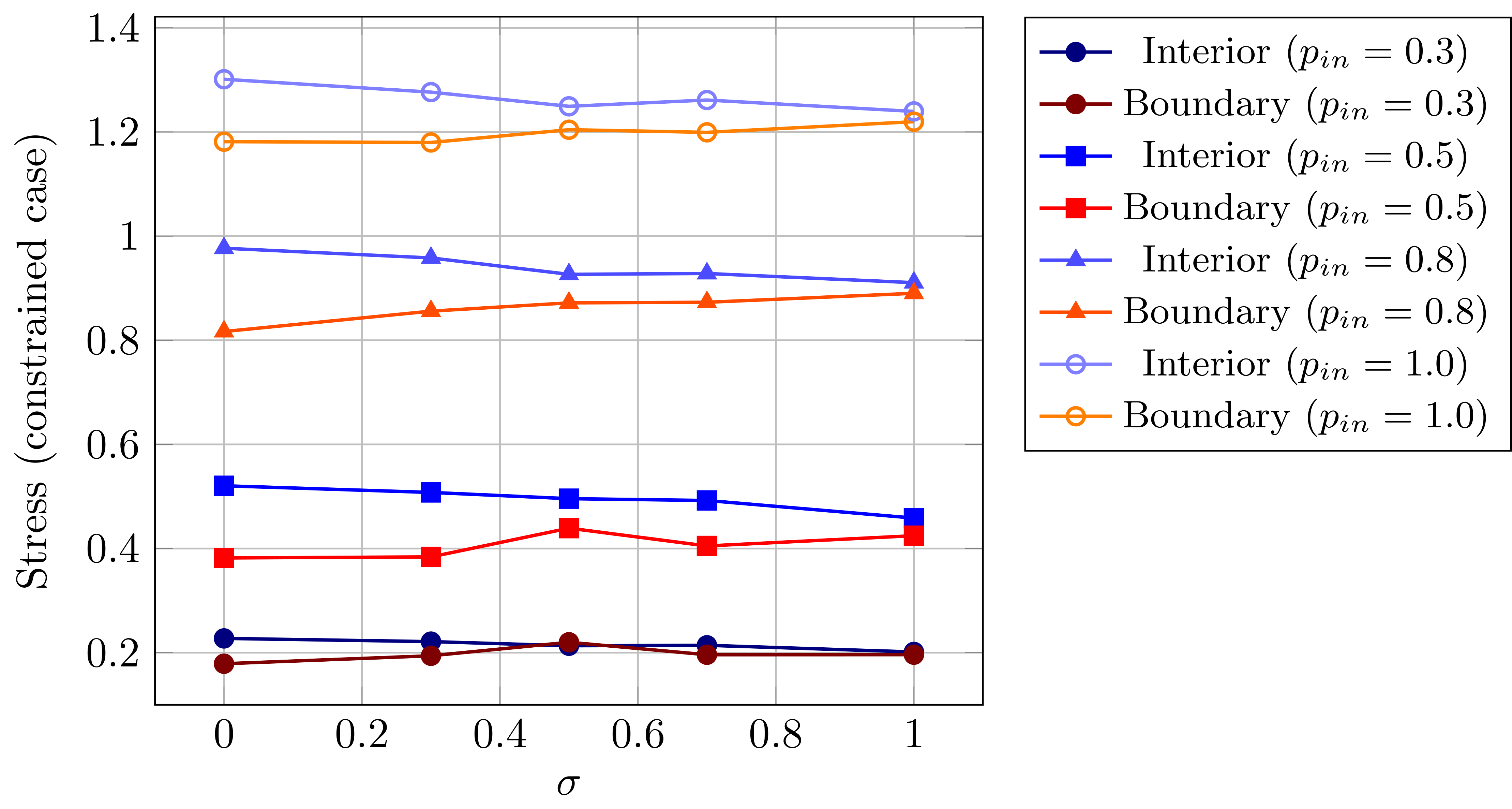} &
        \includegraphics[width=0.5\linewidth]{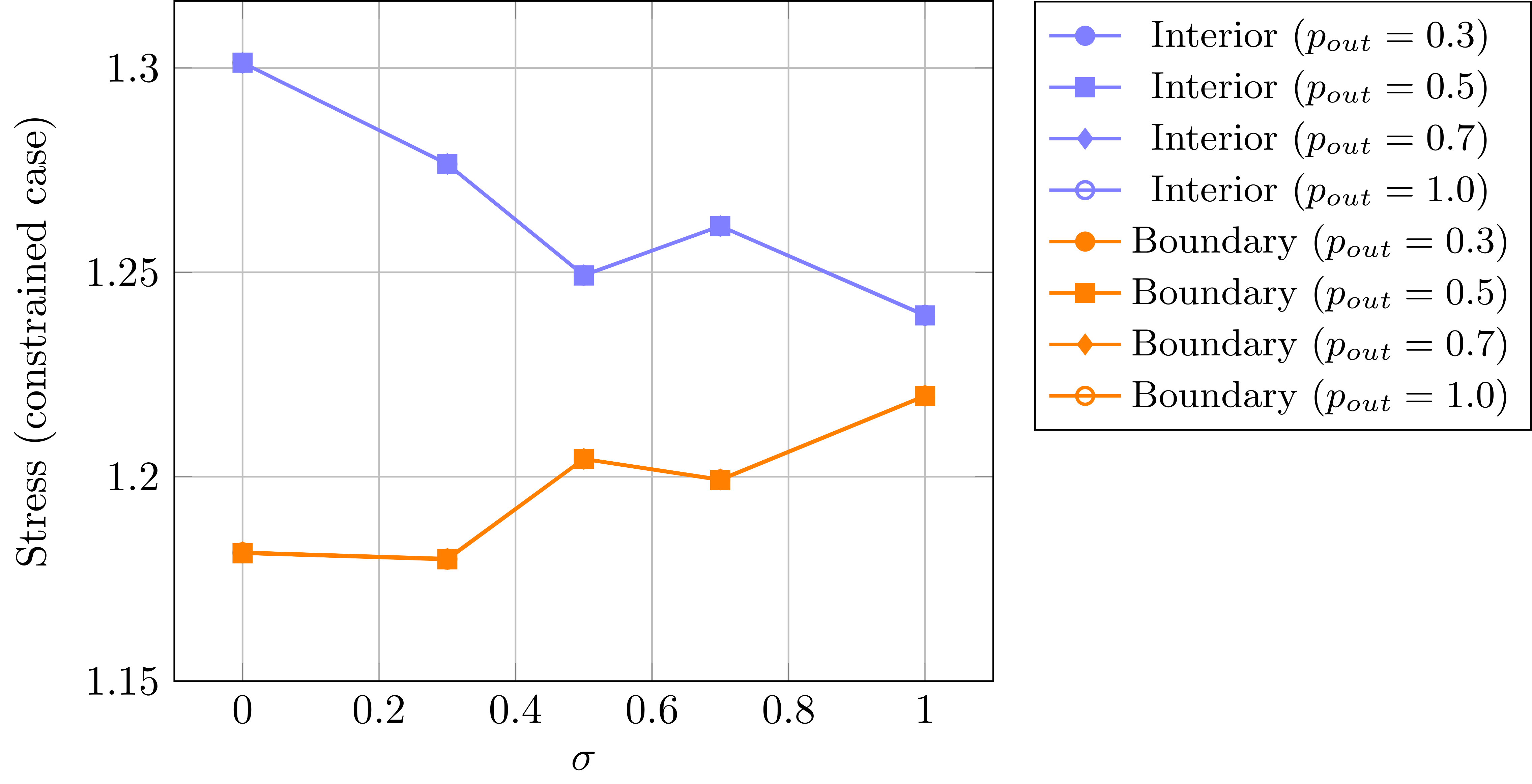} \\
        (c) Floppy interior region ($p_\text{out}=1$) &
        (d) Stiff interior region ($p_\text{in}=1$) \\
    \end{tabular}
    \caption{Stress after constrained energy relaxation across a range of parameters.
    (a)~Interior stress as a function of boundary shape $\sigma$ for varying interior stiffness $k_\text{in}$, with exterior fixed at $k_\text{out}=1$.
    (b)~Comparison between correlated GRF prestress ($\xi_c=3$) and uncorrelated random prestress ($k_\text{in}=10$).
    (c)~Stress measured for various interior bond fractions $p_\text{in}$ to probe the effect of a floppy interior region ($p_\text{out}=1$).
(d)~Stress for a stiff interior region ($p_\text{in}=1$) with varying exterior fraction $p_\text{out}$; all curves overlap, indicating insensitivity to the exterior network fraction. Under constrained relaxation the network responds passively, so varying the interior bond fraction is equivalent to rescaling the interior stiffness, while the exterior fraction has no effect.}

    \label{fig:constrained}
\end{figure*}

First, we measured stress at interior and boundary nodes for both unconstrained (Fig.~\ref{fig:unconstrained}) and constrained (Fig.~\ref{fig:constrained}) cases. As the interior stiffness increases from $k=3$ to $k=10$, both cases show a larger gap between boundary and interior stress (Figs.~\ref{fig:unconstrained} and~\ref{fig:constrained}(a)).

The rigidity transition of a triangular network occurs at bond fraction $p \approx 0.66$ for stretching-only percolation and effective-medium theory, and at $p \approx 0.56$ when nonzero filament bending is included~\cite{Jacobs1996,Mao2013}. To probe the role of network rigidity, we assign separate bond fractions $p$ to the interior and exterior regions, with the remaining fraction $1-p$ assigned a small stiffness $k=0.01$. Since our model includes only stretching, $p=0.5$ lies below the rigid network threshold ($p\approx0.66$) while $p=0.7$ lies just above it. Fig.~\ref{fig:unconstrained}(c) shows results for $p_\text{in}=1$ with varying $p_\text{out}$; the stress trend from $\sigma=0$ to $\sigma=1$ remains consistent across all fractions. 

Fig.~\ref{fig:unconstrained}(d) shows that when the interior network is floppy ($p_\text{in}<0.66$), prestress is released internally, as the additional motional degrees of freedom allow stress redistribution at no extra energy cost, reducing the need to transfer stress to the exterior region (note the $y$-axis is scaled by $10^{-2}$). When the interior is rigid but diluted ($p_\text{in}=0.8$, Fig.~\ref{fig:unconstrained}(e)), lying slightly beyond the rigidity transition, the absolute stress values scale with the rigid fraction, but the trend remains consistent with Fig.~\ref{fig:unconstrained}(c). We also note that decreasing $p_\text{out}$ slightly reduces the stress level, consistent with a softer exterior providing less resistance to stress transfer from the interior.

In contrast, the constrained case is sensitive to the interior network fraction, with stress scaling similarly to the stiffness (Fig.~\ref{fig:constrained}(c)), but insensitive to the exterior network fraction (Fig.~\ref{fig:constrained}(d)), with stress values remaining approximately constant across different rigid/floppy ratios. This suggests that in the constrained case the network behaves passively, such that changing the bond fraction is equivalent to rescaling the total stiffness — as can be seen by comparing Fig.~\ref{fig:constrained}(a) at $k_\text{in}=5$ with Fig.~\ref{fig:constrained}(c) at $p_\text{in}=0.5$, $k_\text{in}=10$. However, boundary fraction takes no effect regardless of fraction $p_\text{out}$ changes.

We also tested different GRF correlation lengths (random, $\xi_c=1,2,3$). Interestingly, the random case shows the largest interior--boundary stress gap for the circular boundary (Fig.~\ref{fig:unconstrained}(b)). Defining $\delta s = s_\text{int} - s_\text{bnd}$, the unconstrained case gives $\delta s_{\sigma=0} - \delta s_{\sigma=1} = 0.187$ for random prestress versus $0.06$ for $\xi_c=3$, indicating that higher input correlation reduces the difference between boundary shapes. The constrained case, however, remains approximately constant at $\delta s_{\sigma=0} - \delta s_{\sigma=1} \approx 0.09$ across all correlation lengths (Fig.~\ref{fig:constrained}(b)).

\begin{figure}
    \centering
    \begin{tabular}{c}
        \includegraphics[width=0.95\linewidth]{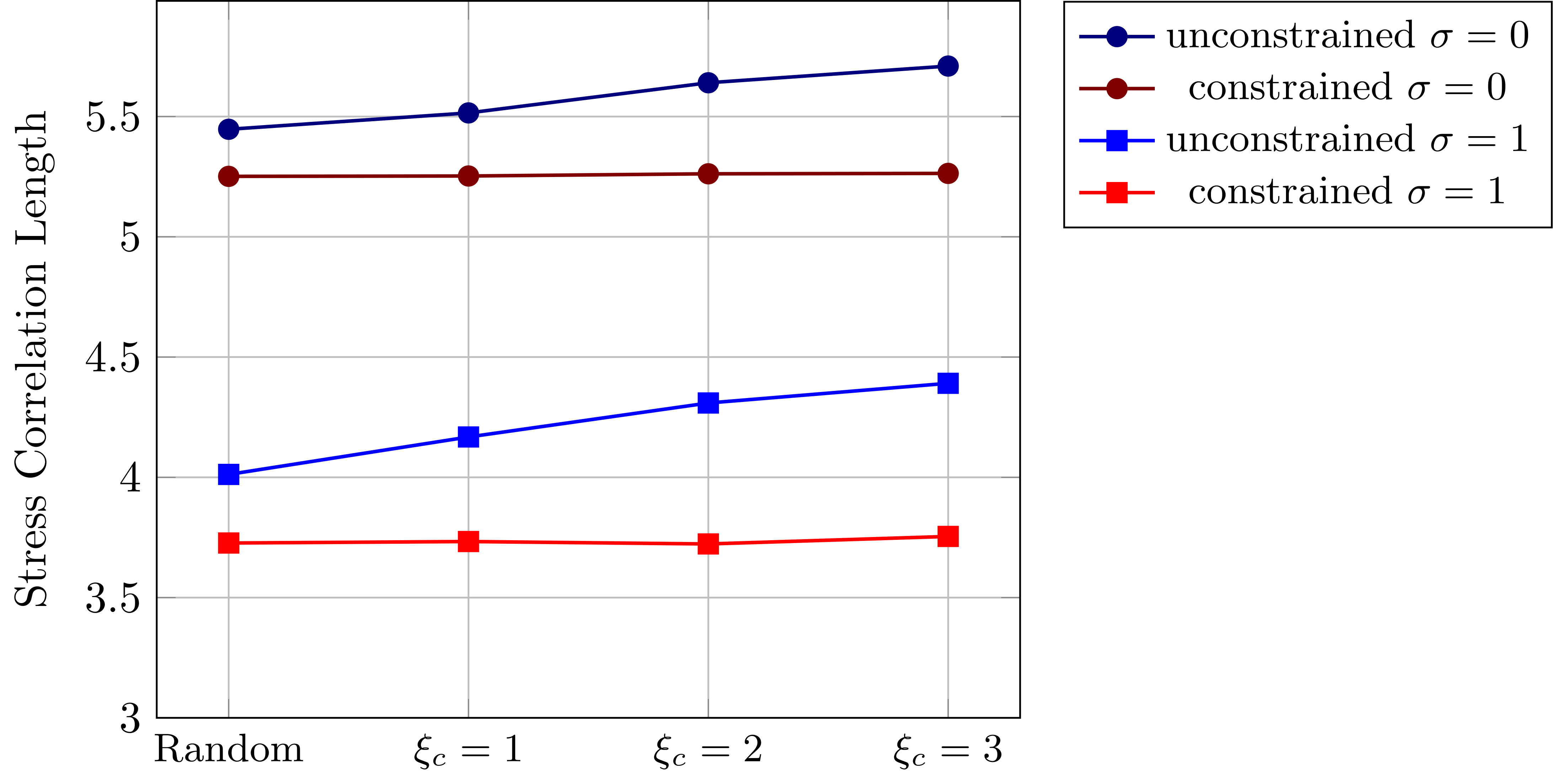}
    \end{tabular}
    \caption{{\it Autocorrelation measurement} ($k_\mathrm{in}=10$, $p_\text{in}=p_\text{out}=1$). In the constrained case $\xi$ is approximately constant, set by the fixed maximum relaxation length; in the unconstrained case $\xi$ grows with $\xi_c$, showing that input correlations are inherited by the output stress field.}
    \label{fig:autocorr}
\end{figure}

To characterize the spatial structure of the resulting stress field, we compute the 2D autocorrelation of bond stresses. Placing bond stresses at their midpoints $R_b$ and interpolating onto a regular grid, the normalized autocorrelation is
\begin{equation}
    C(r) = \frac{\langle s_b\, s_{b'}\rangle_{|R_b - R_{b'}|\approx r}}{\langle s_b^2\rangle},
\end{equation}
which is then radially averaged and fit to $C(r) \sim e^{-r/\xi}$, where $\xi$ is the stress correlation length in units of $L_0$. Figure~\ref{fig:autocorr} shows $\xi$ for each case at $\sigma=0$ and $\sigma=1$. In the constrained case, $\xi$ remains approximately constant, which is expected since the maximum relaxation length is fixed. In the unconstrained case, $\xi$ increases with $\xi_c$, implying that stronger input correlations in the prestress are reflected in the output stress field. Comparing boundary shapes, the circular boundary ($\sigma=0$) yields $\xi \approx 5.5$ while the rough boundary ($\sigma=1$) yields $\xi \approx 4$ for random prestress. Given $R_0=15$, these correspond to roughly $1/4$--$1/3$ of the inclusion radius, indicating that stress correlations extend over a substantial fraction of the inclusion. 

How far does stress imposed in one region penetrate into another? In a continuous medium, Saint-Venant's principle~\cite{HORGAN1983,XU1997} predicts displacement decay $u(x) \sim e^{-x/\lambda}$. For discrete lattices, decay is even more rapid: in an x-braced lattice the dominant decay factor is $\lambda \approx 0.28$ per cell~\cite{STEPHEN1996,Stephen2006}, and for a triangular lattice the Rayleigh spectrum gives a cut-off decay factor $\lambda(q_c) = e^{-2.3} \approx 0.10$ per row~\cite{Karpov2019}. Converting our observed correlation length $\xi \approx 4$--$5.5\,L_0$ to a per-row decay factor via $\lambda_\text{obs} = e^{-d/\xi}$, where $d = L_0\sqrt{3}/2$ is the row spacing, gives $\lambda_\text{obs} \approx 0.80$--$0.84$, closer to unity than the lattice Saint-Venant predictions. This indicates that correlated prestress sustains longer-range stress correlations than a locally applied load would. Furthermore, the circular boundary ($\sigma=0$) yields a larger $\xi$ than the rough boundary ($\sigma=1$), suggesting that boundary roughness suppresses long-range stress correlations and can thereby reduce crack propagation risk (see Appendix \ref{sec:progressive_failure} for a simple crack test).


\subsection{Geometric Partitioning: Size, Number, and Connectivity}
Next, we examine the effect of geometric partitioning by fixing the total number of high-$k$ bonds and dividing the interior region into three configurations: untouching equal-area, untouching unequal-area, and touching unequal-area partitioning. Here ``untouching'' means that multiple interior regions are fully separated by exterior nodes, while ``touching'' means they share boundary nodes (see Fig.~\ref{fig:partitioning_img}).

The total area of the $N=1$ disk is partitioned into inclusions of different sizes ($s_\text{int}=0.581$, $s_\text{bnd}=0.184$), specified by area fractions summing to unity. Each inclusion is placed in a grid cell with radius $R_0\sqrt{f_k}$ for area fraction $f_k$, and the total number of high-$k$ bonds is fixed at $|\mathcal{E}_\text{int}|=2568$. GRF and random prestress is applied and both constrained and unconstrained regimes are evaluated. Parameters: $60\times60$ lattice, $k_\text{in}=10$, $k_\text{out}=1$, $\beta=0.1$, $\xi_c=2$, $100$ realizations (LBFGS 100 iteration for both unconstrained and constrained). The full table is included in Appendix~\ref{sec:partitioning_table}.

Figs.~\ref{fig:partitioning}(a) and (b) show unconstrained results for the touching and untouching cases, respectively. The $x$-axis denotes the partition configuration: $\frac{1}{2}$ indicates a half-size disk, and a prefactor indicates the number of copies, e.g., $2\tfrac{1}{4}$ denotes two disks each of quarter size. Compared to the touching case, which shows only a modest effect, the untouching case exhibits a dramatic stress reduction for more than two partitions; interior and boundary stress become comparable when the number of partitions reaches six. This is related to the number of interior nodes: for equal partitioning, when $N_\text{int}$ and $|\mathcal{E}_\text{bnd}|$ become comparable, the two stress measures approach the same magnitude (see table in the Appendix). For unequal partitioning, $N_\text{int}=821$ and $|\mathcal{E}_\text{bnd}|=476$ for the $2\tfrac{1}{2}+4\tfrac{1}{8}$ configuration (approximately half size compared to equal partitioning). Figs.~\ref{fig:partitioning}(c) and (d) show the constrained results for touching and non-touching regions, respectively. Both exhibit trends similar to those in the unconstrained case, although the absolute magnitudes differ.

The approximately two-order-of-magnitude difference between touching and non-touching arrangements indicates a strong sensitivity to connectivity despite the limited geometric change near the contact. Previous studies of random elastic networks have shown that modifying a small fraction of bonds can substantially alter the mechanical response at distant locations~\cite{Rocks2017}. The large difference between our touching and non-touching arrangements --- where the input (prestress) rather than the network itself is changed --- may similarly reflect the mechanical importance of the bonds connecting the stiff regions, despite their limited spatial extent. Conversely, interrupting this connection allows relative motion to be accommodated through the intervening soft network. The resulting change in the available relaxation modes provides a possible explanation for the large difference in response magnitude. The similar subdivision trends therefore coexist with a strong dependence on whether the stiff regions are directly connected.

\begin{figure*}
    \centering
    \begin{tabular}{c c}
        \includegraphics[width=0.5\linewidth]{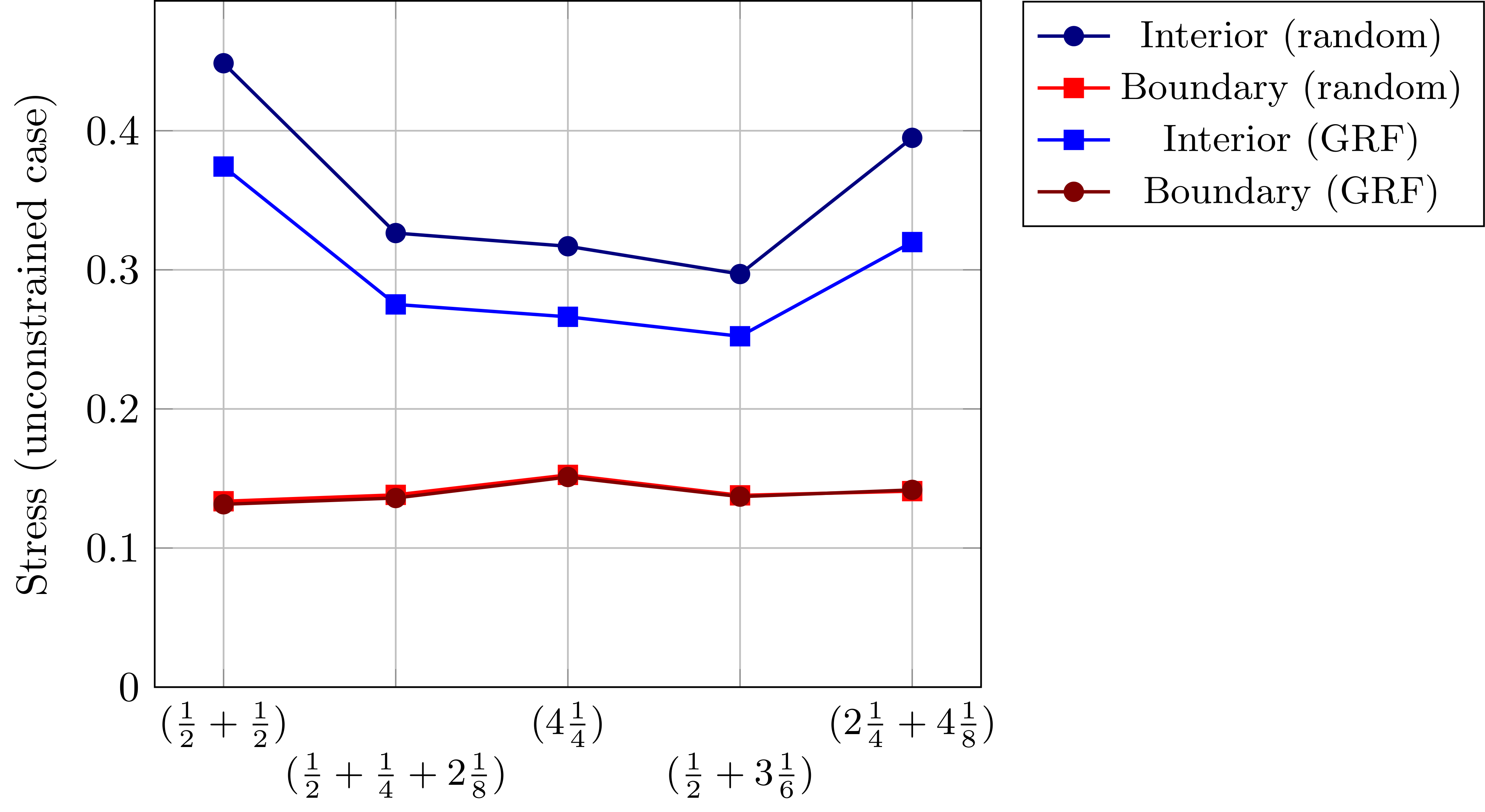} &
        \includegraphics[width=0.5\linewidth]{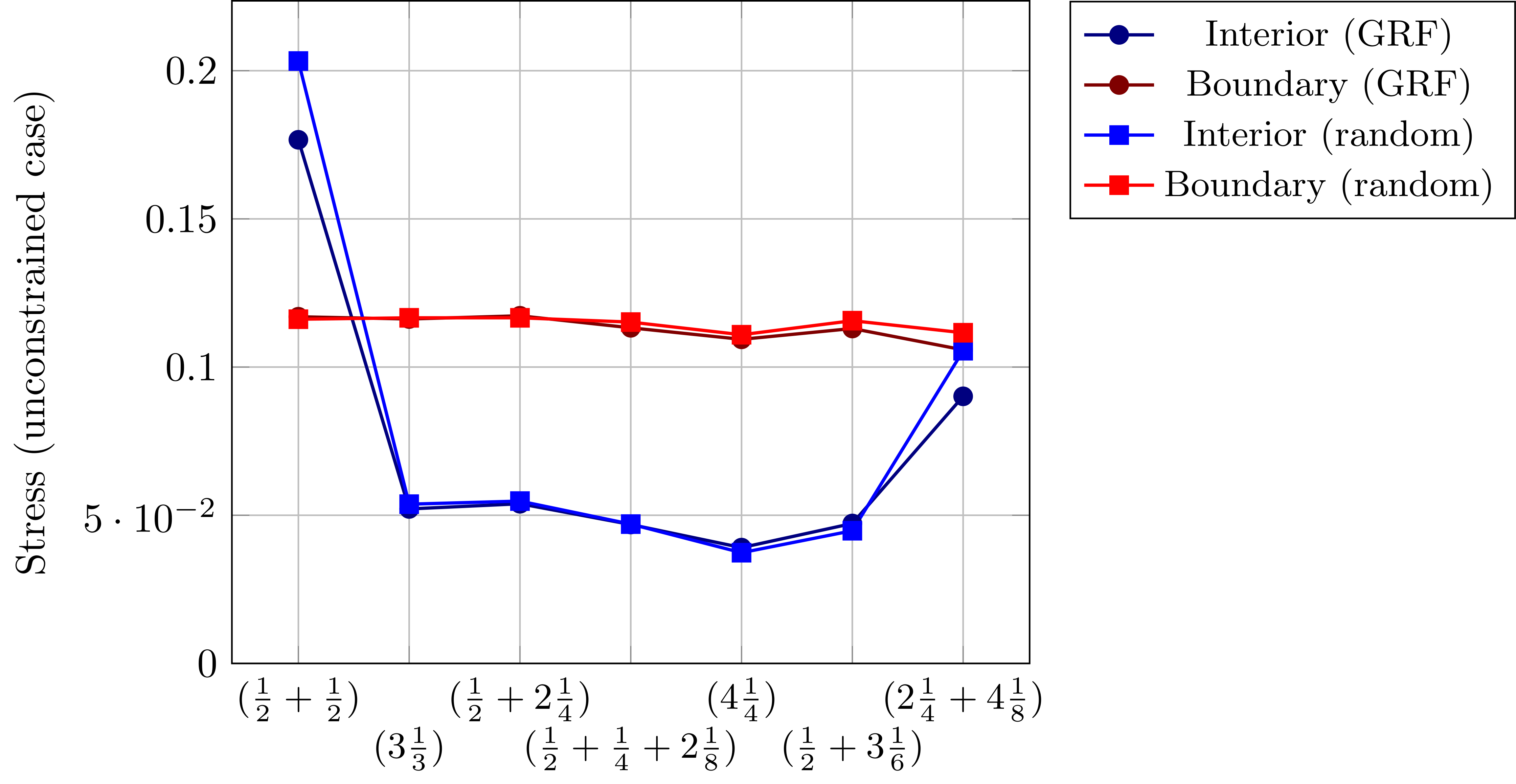} \\
        (a) Unconstrained (touching) & (b) Unconstrained (untouching)\\
        \includegraphics[width=0.5\linewidth]{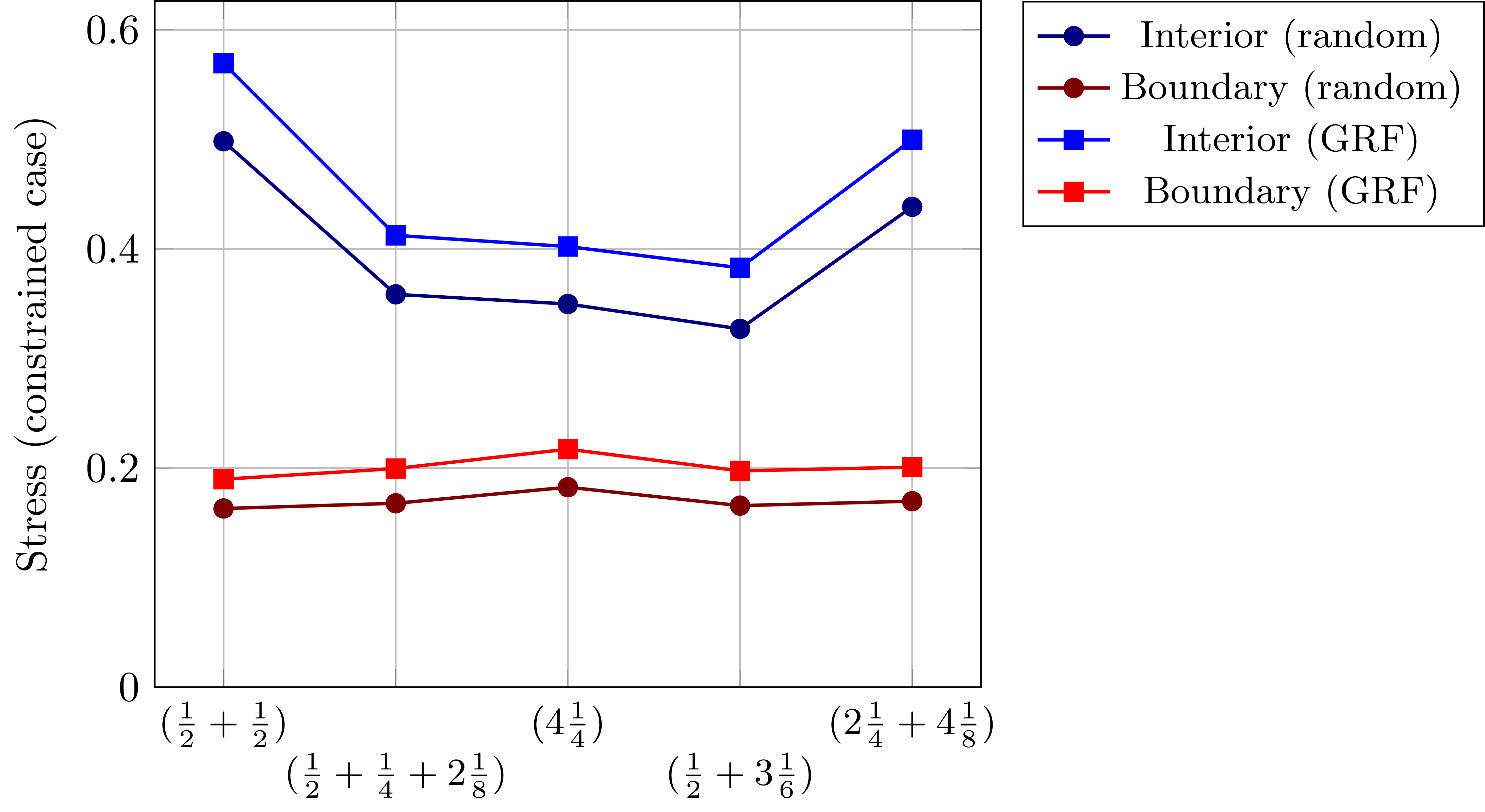} &
        \includegraphics[width=0.5\linewidth]{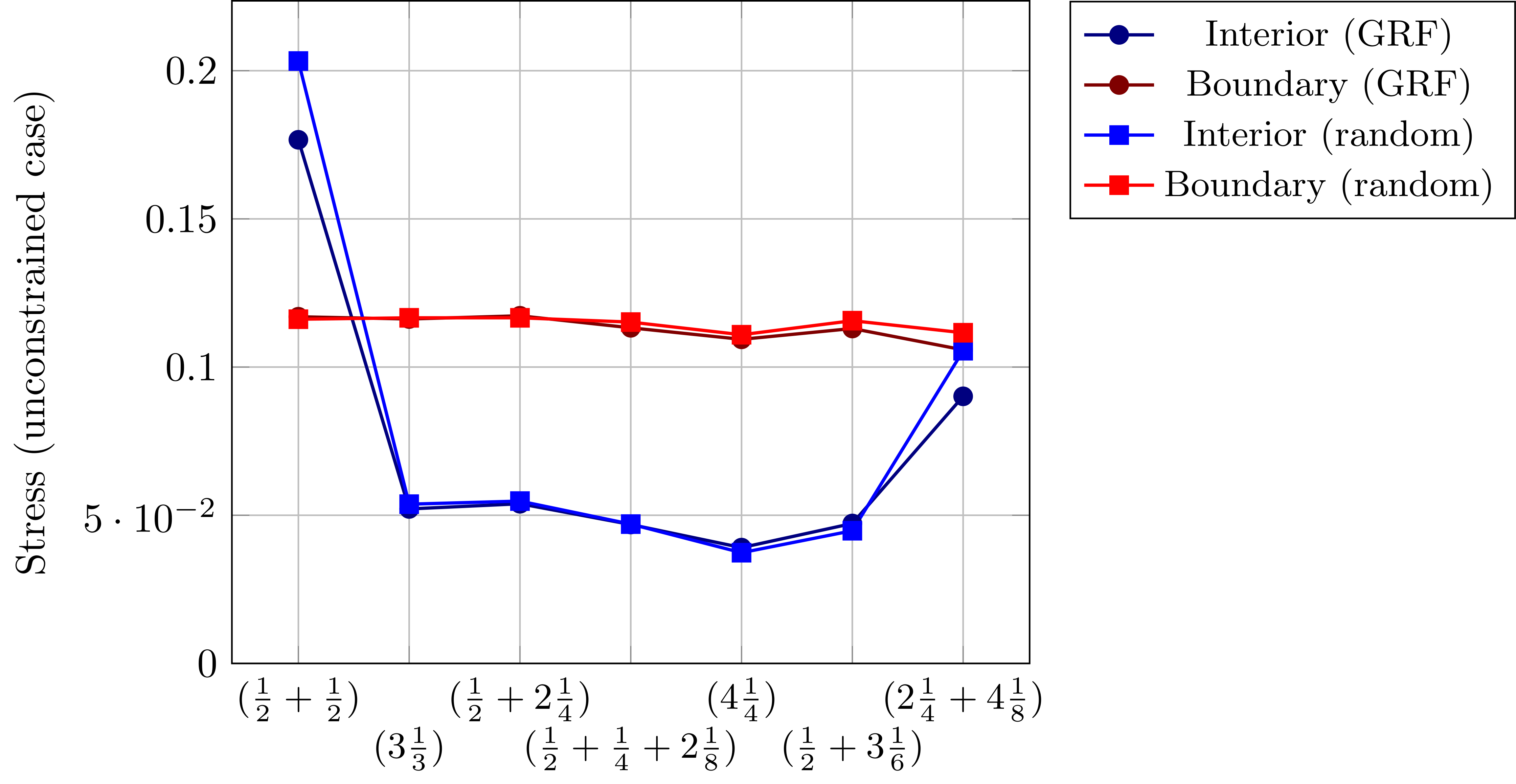} \\
       (c) Constrained (touching) & (b) Constrained (untouching)\\        
    \end{tabular}
\caption{{\it Stress of touching and untouching unequal partitioning.} 
    (a)--(b)~Unconstrained results for the touching and untouching cases,
    respectively.
    (c)--(d)~Constrained results for the touching and untouching cases,
    respectively.
    The $x$-axis denotes the partition configuration: $\tfrac{1}{2}$ indicates
    a half-size disk, and a prefactor gives the number of copies, so that
    $2\tfrac{1}{4}$ denotes two disks each of quarter size.
    The untouching case shows a pronounced stress reduction beyond two
    partitions, while the touching case changes only modestly.}    \label{fig:partitioning}
\end{figure*}
\section{Discussion}
In this study, using a 2D triangular network, we have shown the effect of boundary shape and partitioning in various conditions in terms of stress relaxation. Increasing the number of boundary bonds by a factor of $1.5$ improves stress transfer $(s_\text{int}-s_\text{bnd})$ by approximately 20\% for random stress (13\% for GRF). For the equal partitioning case (touching), $s_\text{int}$, $s_\text{bnd}$, and $s_\text{out}$ become comparable when $|\mathcal{E}_\text{bnd}|=832\approx 821$ ($N_\text{int}$). For unequal partitioning, $N_\text{int}=821$ and $|\mathcal{E}_\text{bnd}|=476$ for the $2\tfrac{1}{2}+4\tfrac{1}{8}$ configuration, approximately half that of equal partitioning, as shown in Fig.~\ref{fig:partitioning_img}(n). The simulation results can be summarized as follows. 
\paragraph{Interface bonds as stress transfer channels.}
Interface bonds serve as the channels through which prestress energy transfers between regions. A larger number of interface bonds provides more parallel transfer paths, reducing the per-bond stress and the displacement gradient between the interior and the boundary. This mechanism offers little benefit in floppy networks or when the interior-to-exterior stiffness ratio is low, but lowers stress at the most highly loaded nodes in both the constrained and unconstrained cases. These results suggest a possible mechanical advantage of subdivided internal structures in plant tissues subjected to growth-generated prestress.

Our results can be interpreted as follows: when the boundary channel is smaller than the stressed region, internal stress remains isolated. If the boundary channel is large enough, stress does not accumulate in the interior region. This finding can be extended to defects. Defects in a crystal can be created by interface mismatch~\cite{MATTHEWS1974}. In our model the interface is imposed by stiffness mismatch rather than nucleated, but in both cases it is the mismatch between regions that sets where stress concentrates. However, defect sites can also act as a boundary layer related to energy transfer. Seung and Nelson~\cite{Seung1988} showed that buckled defects (interfaces) screen or transfer elastic stress: dislocations behave as if they were inextensional for a sufficiently large membrane radius in a 2D elastic sheet. Although screening by defects operates through out-of-plane buckling, which our strictly two-dimensional network does not consider, the principle carries over: an interface relieves stress that would otherwise accumulate in the bulk. This is analogous to our finding related to stress isolation.

Moreover, Zhang and Mao~\cite{Zhang2018} showed that fracturing of topological Maxwell lattices proceeds through domain walls (boundaries). They find that bond-breaking events concentrate near self-stress domain walls, protecting the lattice bulk even deep into the failure process. Although Maxwell lattices sit at the isostatic point while the fully connected triangular lattice is over-coordinated, our diluted networks approach that coordination range, and we find the same qualitative trend there: the interface, not the bulk, determines where stress localizes.

The importance of interface geometry extends to fracture resistance. Bidhendi et al.~\cite{Bidhendi2023} demonstrated that wavy epidermal cell boundaries promote crack deflection and meandering, increasing fracture resistance in plant tissues and plant-inspired synthetic structures. Similarly, Sanner et al.~\cite{Sanner2026} showed that microstructural disorder generated by randomly removing bonds increases the failure fracture energy in a random-walk-like manner. A complementary fracture-mechanics perspective is provided by Ming-Yuan and Hutchinson~\cite{MINGYUAN1989}, who analyzed the competition between crack penetration and deflection at interfaces between dissimilar elastic materials. Together, these studies and our results highlight how interface geometry contributes to mechanical robustness through stress redistribution and crack propagation control.

In our simulations, boundary shape has little effect in floppy networks with a low interior bond fraction under fixed boundary conditions. Under confinement and compressive loading, however, boundary geometry may become mechanically relevant, as reported for animal cells~\cite{Kim2024,Tah2025}, providing a possible parallel to the wavy boundaries found in plant tissues.

\paragraph{Partitioning reduces stress concentration.}
Partitioning creates additional stress-transfer channels and thereby promotes more efficient energy transfer between regions. The effect is strongest for three to four partitions, regardless of whether the prestress is spatially correlated, and diminishes as the number of partitions increases, eventually becoming negligible when the number of interior nodes and the number of boundary edges become comparable. Importantly, the touching (connected) case does not exhibit the same reduction, implying that the independence of inclusions — rather than simply their number — is the key factor, and this holds regardless of stress correlation.

The effect of subdivision depends on the material architecture and loading conditions. In cellular solids, increasing the specimen-to-cell size ratio can increase the apparent Young's
modulus by reducing the relative contribution of compliant surface cells~\cite{Onck2001,Andrews2001}. 

\paragraph{A graded interface relieves stress concentration.}
A graded interface — for example, a transition from high to low fiber density or a combination of differing stiffnesses — reduces the maximum stress. For a fixed applied prestress over a fixed region, a broader buffer zone provides a greater variety of paths along which stress can be redistributed. Beyond plant tissues, a related design principle has been investigated at the tendon-to-bone attachment, where a compliant, functionally graded interface reduces stress concentrations between mechanically dissimilar materials~\cite{Liu2012}.
\begin{figure}
    \centering
    \includegraphics[width=0.9\linewidth]{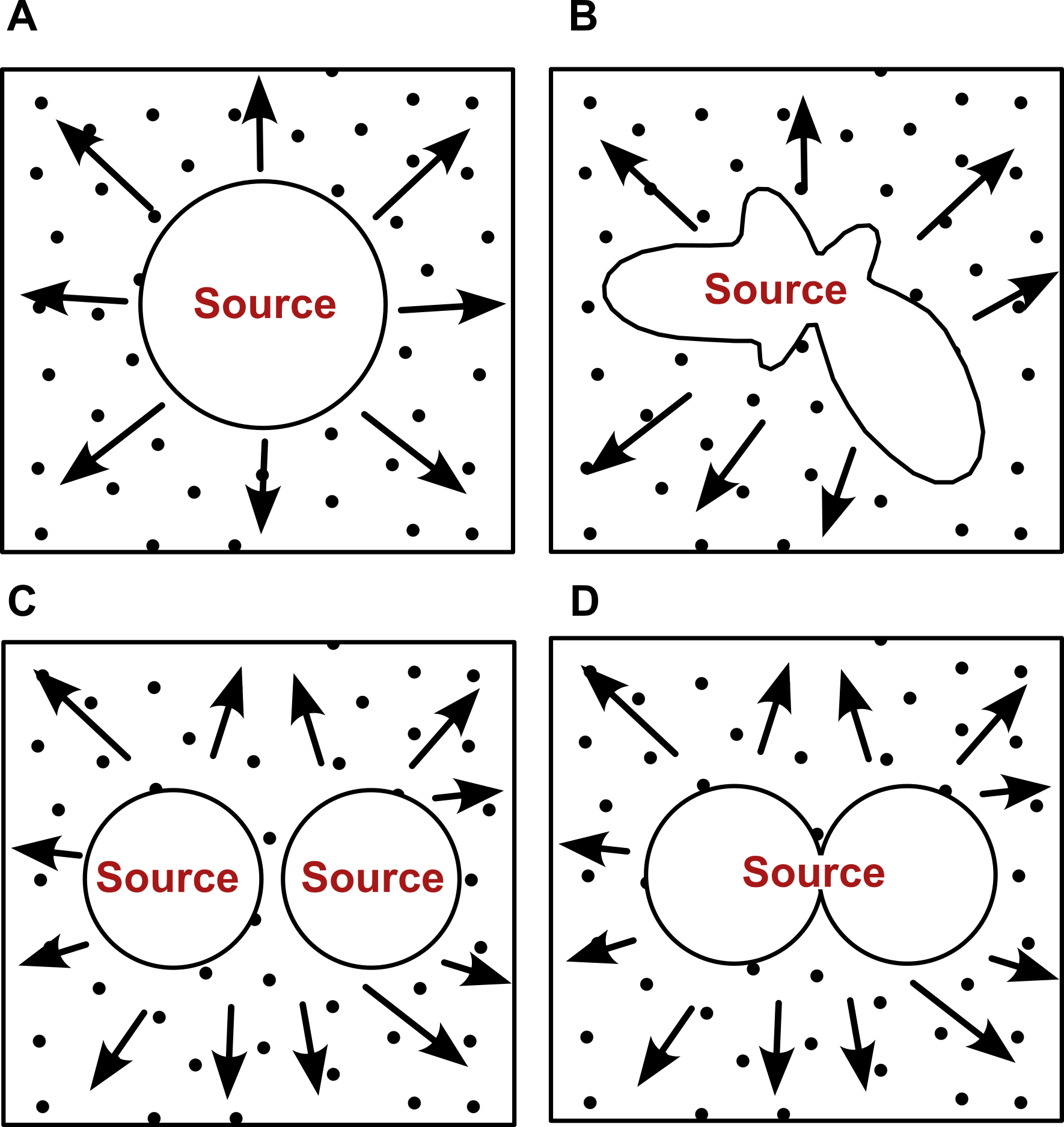}
    \caption{{\it Interpretation of results in terms of energy flow.} Schematic arrows indicate the direction of stress (energy) propagation for (a) disk boundary, (b) rough boundary, (c) two untouching partitions, and (d) two touching partitions.}
    \label{fig-drawing2}
\end{figure}

These results suggest three strategies for controlling stress flow from stiffer to softer regions in elastic networks. First, increasing the interface area between regions provides additional paths for stress transfer (Figs.~\ref{fig-drawing2}(a) and (b)); distributing the stiff phase among multiple subregions increases its contact with the surrounding soft material, reducing the load concentrated along individual paths while preserving the total fraction of stiff material (Figs.~\ref{fig-drawing2}(c) and (d)). Second, subdivision should account for the balance between the number of boundary bonds $|\mathcal{E}_\text{bnd}|$ and interior nodes $N_\text{int}$: interior and boundary stresses become comparable in magnitude at boundary-bond-to-interior-node ratios of approximately $1:1$ for equal-sized regions and $1:2$ for unequal-sized regions. Random and spatially correlated prestress produce broadly similar responses in our simulations, suggesting that the relevant balance depends on how the stiff material and stress sources are distributed throughout the lattice, which together determine the paths available for stress redistribution. Third, grading the stiffness transition facilitates stress redistribution across the interface, although varying the buffer width has little effect on the peak stress within the parameter range examined.
\begin{figure}
    \centering
    \includegraphics[width=0.9\linewidth]{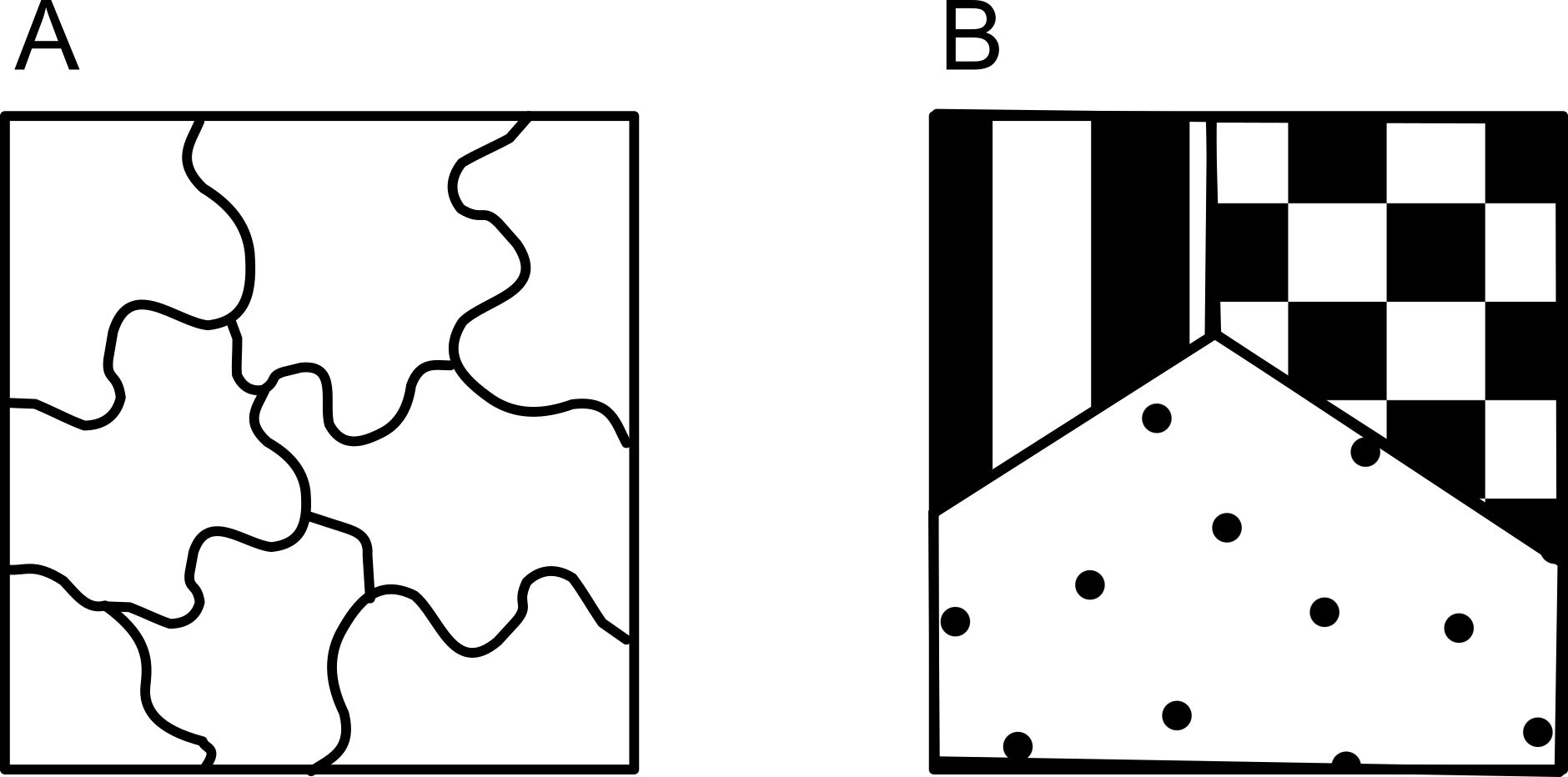}
    \caption{{\it Application of boundary design strategies.} (a) Irregular boundary region. (b) A nonlinear boundary surface realized through a regular lattice structure with varying microstructure orientation.}
    \label{fig-example1}
\end{figure}

\begin{figure}
    \centering
    \includegraphics[width=0.9\linewidth]{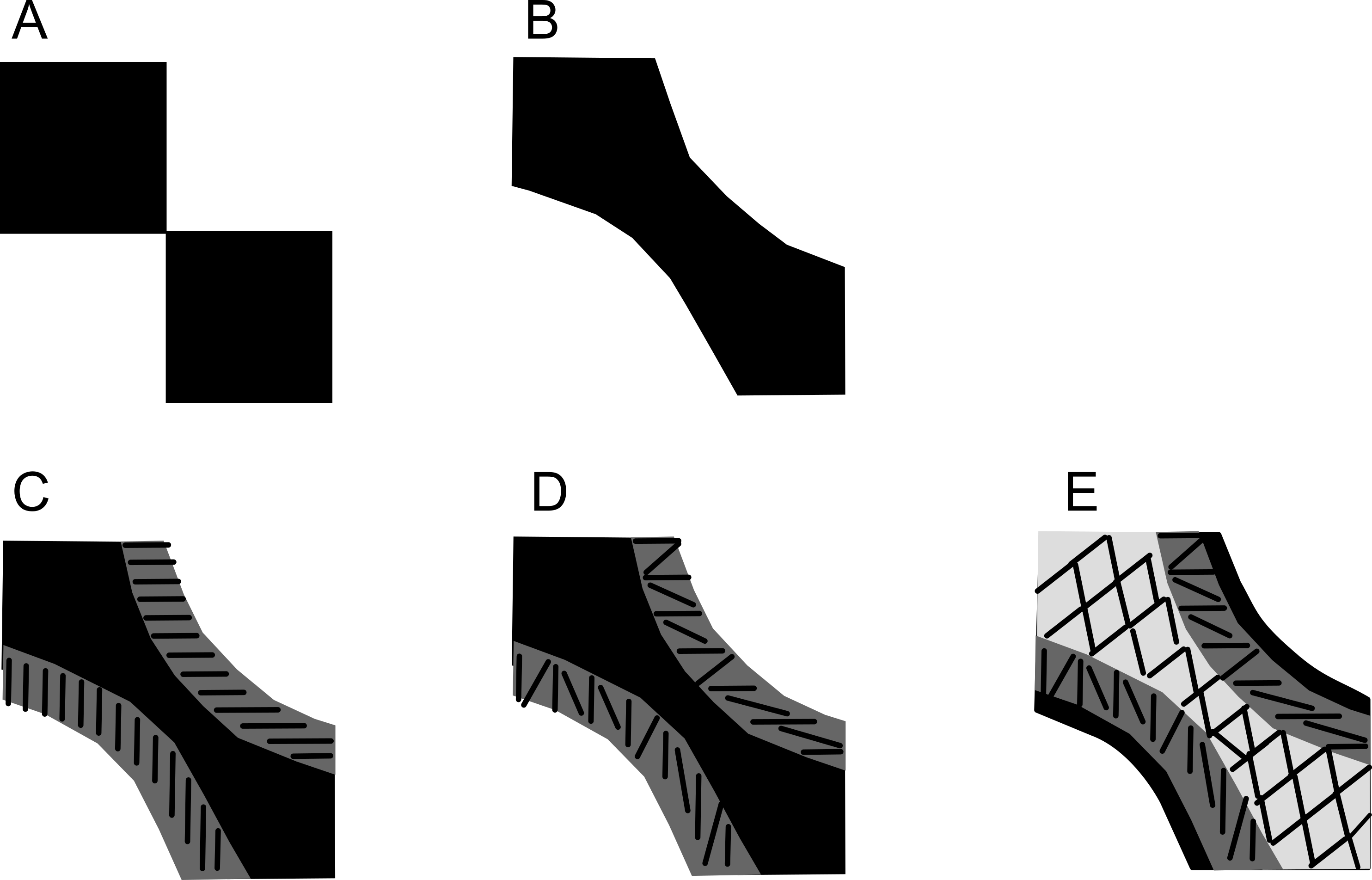}
\caption{{\it Design strategies applied to a checkerboard motif.} (a) Original design: two stiff blocks meeting at a single corner. (b) Smoothing: the corner contact is replaced by a continuous connection, removing the singular junction. (c) Graded interface with regular microstructure or density: a transition region (gray) is added along the boundary. (d) Graded interface with irregular microstructure or density. (e) Graded and layered interface with irregular or regular microstructure and density.}
\label{fig-example2}
\end{figure}

Figs.~\ref{fig-example1} and~\ref{fig-example2} illustrate how these strategies can be realized in practice. Fig.~\ref{fig-example1}(a) shows the irregular boundary of plant pavement cells, which promotes crack meandering through compliant interfaces rather than geometry alone~\cite{Bidhendi2023}; Fig.~\ref{fig-example1}(b) shows a jointed lattice in which the microstructural orientation changes across a straight boundary, arresting shear bands and cracks to yield improved damage tolerance~\cite{Pham2019}. Fig.~\ref{fig-example2} demonstrates the progressive application of these strategies to an engineered design: starting from an initial topology obtained by finite element optimization (a), the sharp corner contact is smoothed (b) to eliminate the stress singularity; graded boundaries with regular or irregular microstructure or density (c)--(d) are then introduced to further reduce stress; and finally (e) combines boundary grading with internal microstructural and density restructuring for maximum stress relief. These strategies are consistent with recent advances in metamaterial design: Liu et al.~\cite{Liu2024} demonstrated that a multilayer strategy combined with topology optimization can generate ultrastiff metamaterials; Vaverka et al.~\cite{Vaverka2024} showed that incorporating various microstructures enables mass savings while maintaining stiffness; and Shu et al.~\cite{Shu2025} reported improvements of approximately 15\% in both thermal and mechanical stiffness through graded hollow lattice microstructures.

These examples motivate the following design principles:
\begin{itemize}
    \item Adding multiple boundary layers reduces stress concentration at designated locations.
    \item Non-uniform regional boundaries guide crack meandering and promote stress redistribution under spatially correlated or directional stress.
\item Arrange the stiff material into multiple disconnected subregions where stress is applied or concentrated to promote stress redistribution and reduce stress concentration.
\end{itemize}
The third design principle is supported here for elastic spring networks representing cellular structures under designated stressed spots, though its generalization requires care. Related studies of metallic cellular solids have shown that increasing the specimen-to-cell size ratio can improve the apparent Young's modulus and compressive strength~\cite{Onck2001,Andrews2001}, although those results concern boundary-induced size effects rather than internal stress redistribution, and their extension to filled two-phase composites requires separate assessment. Su et al.~\cite{Su2020} reported that increasing the number of unit cells at fixed specimen size generally reduced stiffness and yield strength, with architecture-dependent exceptions. Among their architectures, at small unit-cell sizes, the discontinuous-phase design maintains a more uniform shape under compression than its connected counterparts. This observation is qualitatively compatible with our finding that phase separation promotes stress redistribution. Their loading conditions differ from ours: compression is applied at the specimen boundaries, whereas our simulations generate stress within the network through bond rest-length perturbations. Their results therefore provide a qualitative comparison rather than direct evidence for the stress reduction observed here. Moreover, their finding that smaller unit cells generally reduce stiffness indicates that finer subdivision does not necessarily improve mechanical performance. Accordingly, our design principles should guide the arrangement of stress-transfer paths in relation to the stress sources and constraints, rather than prescribe subdivision or phase separation as universally beneficial on their own.

\noindent{\it Acknowledgments.} 
We acknowledge the support of the Government of Canada’s New Frontiers in
Research Fund (NFRF), NFRFE-2023-00723.

\noindent{\it Data Availability.} 
The data and code used in this study will be made publicly available in a repository upon publication.

\setlength{\emergencystretch}{2em}
\bibliographystyle{unsrt}
\bibliography{references}
\appendix
\renewcommand{\thefigure}{A\arabic{figure}}

\setcounter{figure}{0}
\appsec{Supplementary Data of Boundary Region Shape}
\appsubsec{Stress redistribution during progressive failure}
\label{sec:progressive_failure}

For an intact network with quadratic elastic energy, direct matrix solution and energy minimization yield the same equilibrium, provided that the imposed constraints eliminate rigid-body motions and floppy modes. The need for repeated relaxation therefore arises not from a difference between solution methods, but from the evolution of the network during failure: at each step, the bond carrying the highest stress is removed and the remaining network is reoptimized. Each removal changes the connectivity and redistributes the stresses, so the equilibrium problem at step $n$ becomes
\begin{equation}
    \mathbf{u}^{(n)} = \arg\min_{\mathbf{u}} E^{(n)}(\mathbf{u}),
\end{equation}
where $E^{(n)}$ is the elastic energy of the network at step $n$ with the surviving bonds, updated to reflect each removal. A single solution using the initial matrix cannot capture this progressive redistribution.

\begin{figure*}
    \centering
    \begin{tabular}{c c}
        \includegraphics[width=0.35\linewidth]{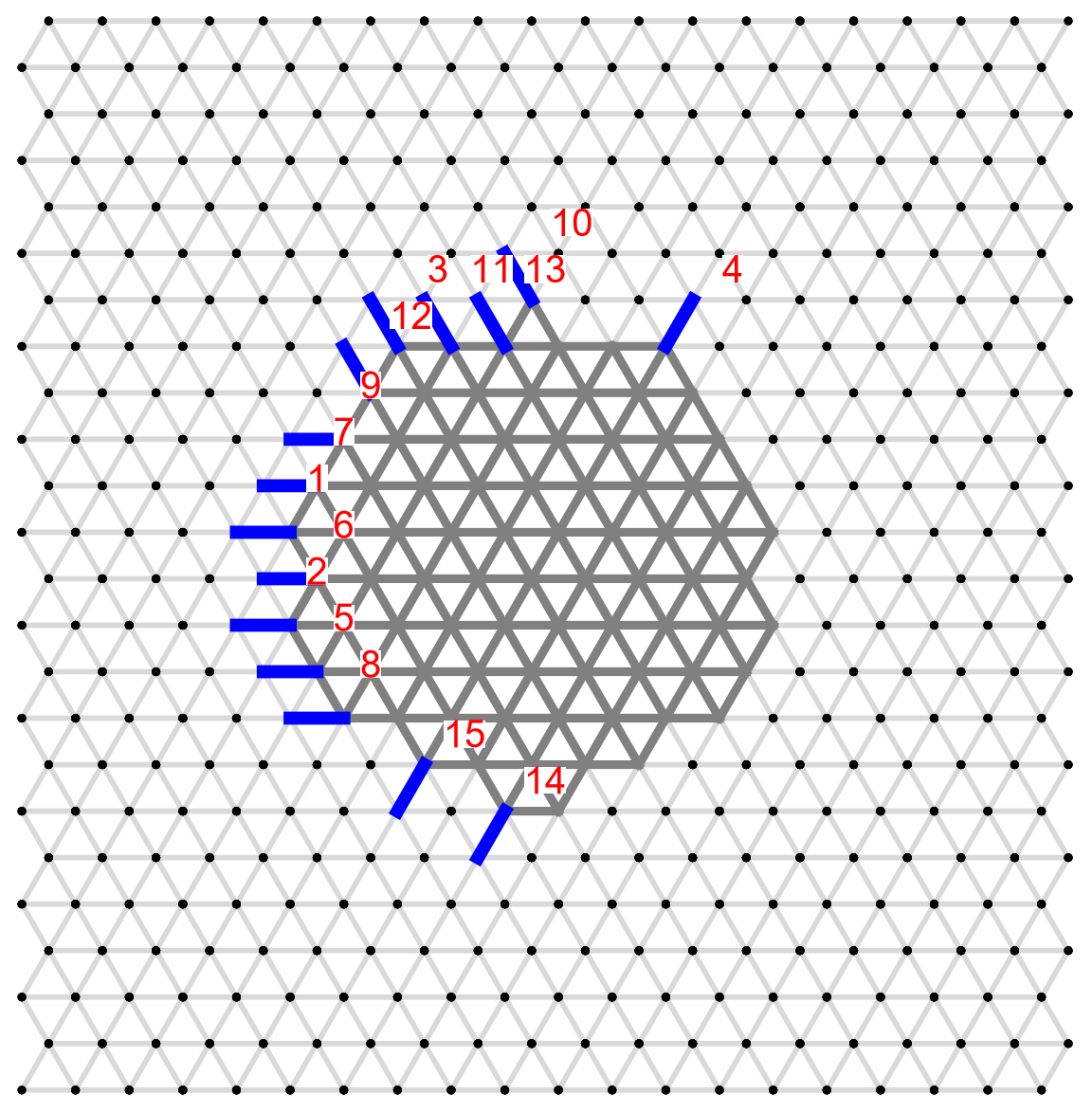}&
        \includegraphics[width=0.35\linewidth]{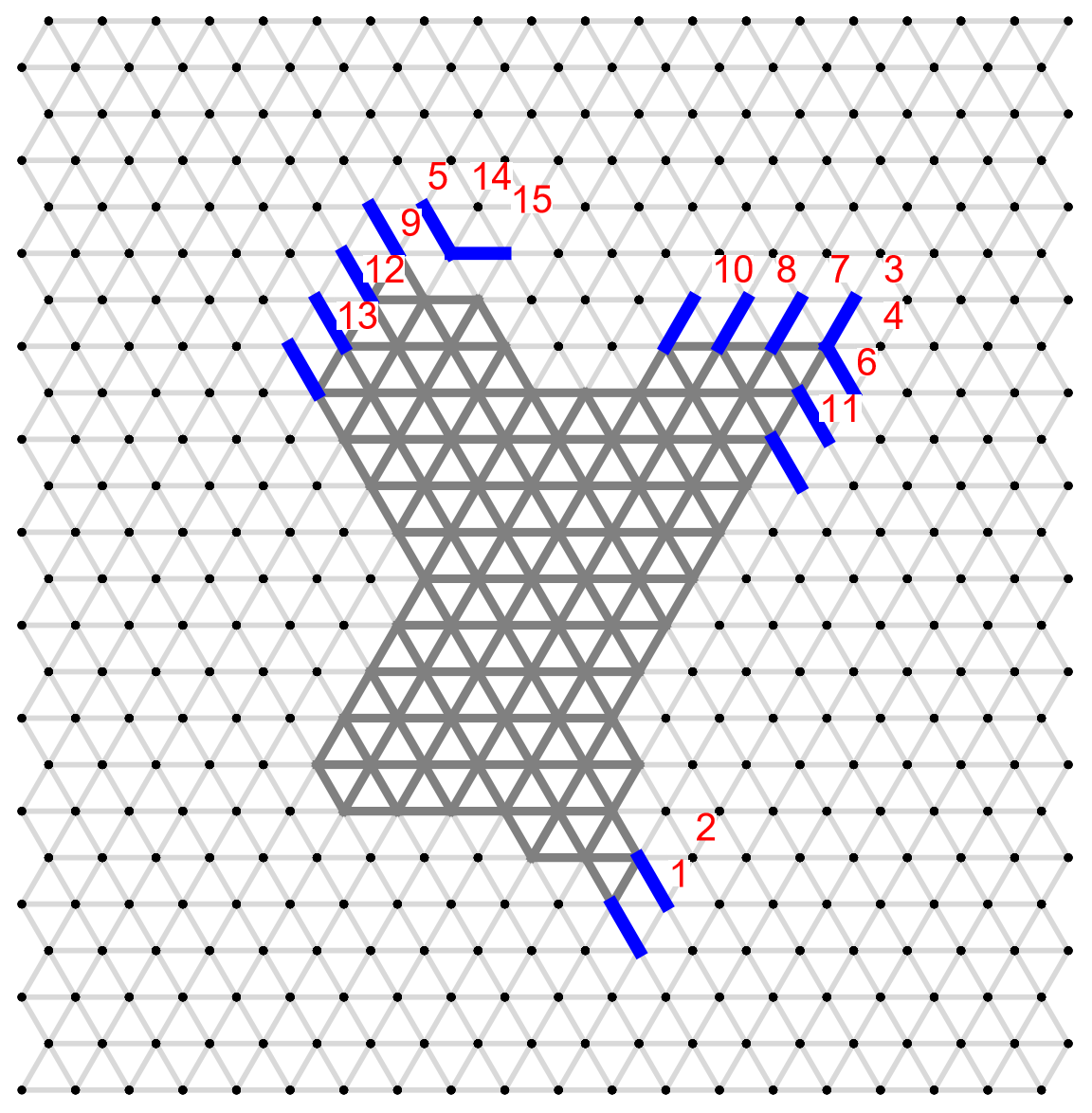} \\
        (a) Disk & (b) Rough boundary \\
    \end{tabular}
    \caption{Crack propagation under uniform prestress for (a) a circular boundary and (b) a rough boundary. Numbers indicate the order of bond removal. At each step, the highest-stress bond is removed and the network is relaxed before the next removal.}
    \label{fig:crack}
\end{figure*}

We compare the two boundary shapes ($|\mathcal{E}_\text{int}|=218$, same geometry as the previous section) under constant prestress. The crack propagation test in Fig.~\ref{fig:crack} provides a direct comparison: the removal sequence reveals the crack path. In the circular case, cracks propagate in a localized and nearly straight path (Fig.~\ref{fig:crack}(a)), accumulating in the interior due to the symmetric geometry, which allows only uniform local rearrangement to reach the energy minimum. In contrast, the rough boundary (Fig.~\ref{fig:crack}(b)) forces cracks to meander and jump between non-adjacent locations, reflecting the geometric mismatch between compression and tension regions that drives more global stress redistribution --- consistent with the tear resistance mechanism observed in plant pavement cells~\cite{Bidhendi2023}.

This distinction is central to assessing the role of interface geometry. The initial stress field identifies the first bond selected for removal, whereas subsequent failures depend on the alternative load paths available. The arrangement of stiff regions and their interfaces therefore influences the failure sequence through repeated stress redistribution. The bond-removal procedure thus extends the intact-network analysis to an evolving failure process: providing a simple test of how structural design affects progressive failure beyond the initial stress distribution.

\appsubsec{Supplementary Data Table of Simple Prestress Test}
\label{sec:supp_table}

Tables~\ref{tab:sigma_stress} and~\ref{tab:sigma_stress_grf} compare the number of nodes and bonds against the stress gap. As noted, $|\mathcal{E}_\text{int}|$ varies by $\sim$6\% across $\sigma$. Increasing $|\mathcal{E}_\text{bnd}|$ by 52\% promotes stress propagation, reducing the gap by 20\% for random prestress and 12.9\% for GRF prestress.

Table~\ref{tab:sigma-data-grf-constrained} shows full data of selected set ($\beta=0.05,0.1,0.2$ for $\xi_c=2$ for GRF). Note that the variance of stress ($\text{Var}^s_\text{int}$, $\text{Var}^s_\text{bnd}$) is larger than its mean, whereas the displacement variance $\text{Var}_\text{int}$ is smaller than $\text{Var}_\text{bnd}$, indicating larger displacement fluctuations at boundary nodes. Fig.~\ref{fig:Cconstrained} shows the full distribution for the constant prestress case, representing the simpler case of a shifting mean value despite large variance.

\begin{table}
\centering
\begin{tabular}{ccccccccc}
\hline
$\sigma$ & $N_\text{bnd}$ & $N_\text{int}$ & $|\mathcal{E}_\text{int}|$ & $|\mathcal{E}_\text{bnd}|$ & $s_\text{int}$ & $s_\text{bnd}$ & Gap (\%) \\
\hline
0.0 & 66  & 295 & 1014 & 138 & 0.8864 & 0.6667 & 24.8 \\
0.3 & 72  & 279 & 978  & 150 & 0.8706 & 0.7413 & 14.9 \\
0.5 & 80  & 271 & 970  & 166 & 0.8441 & 0.7700 & 8.8  \\
0.7 & 94  & 256 & 953  & 194 & 0.8348 & 0.7834 & 6.2  \\
1.0 & 101 & 253 & 957  & 210 & 0.8201 & 0.7873 & 4.0  \\
\hline
\end{tabular}

\caption{Mean interior and boundary bond stress vs.\ boundary roughness $\sigma$ for random prestress. $N_\text{bnd}$: boundary nodes; $N_\text{int}$: interior nodes; $\mathcal{E}_\text{int}$: interior bonds; $\mathcal{E}_\text{bnd}$: boundary bonds. Gap $= (s_\text{int} - s_\text{bnd})/s_\text{int} \times 100\%$.}
\label{tab:sigma_stress}
\end{table}

\begin{table}
\centering
\begin{tabular}{ccccccccc}
\hline
$\sigma$ & $N_\text{bnd}$ & $N_\text{int}$ & $|\mathcal{E}_\text{int}|$ & $|\mathcal{E}_\text{bnd}|$ & $s_\text{int}$ & $s_\text{bnd}$ & Gap (\%) \\
\hline
0.0 & 66  & 295 & 1014 & 138 & 0.7268 & 0.5943 & 18.2 \\
0.3 & 72  & 279 & 978  & 150 & 0.7056 & 0.5703 & 19.2 \\
0.5 & 80  & 271 & 970  & 166 & 0.6792 & 0.5847 & 13.9 \\
0.7 & 94  & 256 & 953  & 194 & 0.6653 & 0.5775 & 13.2 \\
1.0 & 101 & 253 & 957  & 210 & 0.6335 & 0.6001 & 5.3  \\
\hline
\end{tabular}
\caption{Mean interior and boundary bond stress vs.\ boundary roughness $\sigma$ for GRF prestress. $N_\text{bnd}$: boundary nodes; $N_\text{int}$: interior nodes; $\mathcal{E}_\text{int}$: interior bonds; $\mathcal{E}_\text{bnd}$: boundary bonds. Gap $= (s_\text{int} - s_\text{bnd})/s_\text{int} \times 100\%$.}
\label{tab:sigma_stress_grf}
\end{table}

\begin{table*}[p]
\centering
\caption{Boundary roughness data: \texttt{prestress\_type} = \texttt{GRF} ($\xi_c=2$); \texttt{regime} = \texttt{constrained}. Both panels use the original row labels and reported numerical precision.}
\label{tab:sigma-data-grf-constrained}
\footnotesize
\setlength{\tabcolsep}{5pt}
\renewcommand{\arraystretch}{1.05}
\textbf{(a)}\par\smallskip
\begin{tabular*}{\linewidth}{@{\extracolsep{\fill}}lrrrrrrrr@{}}
\toprule
Label & $|\mathcal{E}_\text{bnd}|$ & $N_\text{bnd}$ & $N_\text{int}$ & $|\mathcal{E}_\text{int}|$ & $\xi_\text{int}$ & $\xi_\text{bnd}$ & $\xi_\text{out}$ & $\xi_\text{tot}$ \\
\midrule
$\sigma=0.0$, $\beta=0.05$ & 138 & 66 & 295 & 1014 & 0.11274 & 0.16515 & 0.05806 & 0.07313 \\
$\sigma=0.0$, $\beta=0.10$ & 138 & 66 & 295 & 1014 & 0.23516 & 0.36806 & 0.13515 & 0.16494 \\
$\sigma=0.0$, $\beta=0.20$ & 138 & 66 & 295 & 1014 & 0.44292 & 0.69986 & 0.24000 & 0.29968 \\
\midrule
$\sigma=0.3$, $\beta=0.05$ & 150 & 72 & 279 & 978 & 0.11721 & 0.17434 & 0.06013 & 0.07594 \\
$\sigma=0.3$, $\beta=0.10$ & 150 & 72 & 279 & 978 & 0.25202 & 0.38208 & 0.13536 & 0.16843 \\
$\sigma=0.3$, $\beta=0.20$ & 150 & 72 & 279 & 978 & 0.48357 & 0.72452 & 0.24180 & 0.30869 \\
\midrule
$\sigma=0.5$, $\beta=0.05$ & 166 & 80 & 271 & 970 & 0.12313 & 0.17731 & 0.06406 & 0.08010 \\
$\sigma=0.5$, $\beta=0.10$ & 166 & 80 & 271 & 970 & 0.27418 & 0.38910 & 0.13999 & 0.17596 \\
$\sigma=0.5$, $\beta=0.20$ & 166 & 80 & 271 & 970 & 0.49692 & 0.71035 & 0.24832 & 0.31500 \\
\midrule
$\sigma=0.7$, $\beta=0.05$ & 194 & 94 & 256 & 953 & 0.12157 & 0.17245 & 0.06329 & 0.07896 \\
$\sigma=0.7$, $\beta=0.10$ & 194 & 94 & 256 & 953 & 0.25487 & 0.36346 & 0.13670 & 0.16879 \\
$\sigma=0.7$, $\beta=0.20$ & 194 & 94 & 256 & 953 & 0.50312 & 0.70727 & 0.25539 & 0.32132 \\
\midrule
$\sigma=1.0$, $\beta=0.05$ & 210 & 101 & 253 & 957 & 0.13140 & 0.17100 & 0.06551 & 0.08221 \\
$\sigma=1.0$, $\beta=0.10$ & 210 & 101 & 253 & 957 & 0.26907 & 0.36207 & 0.13892 & 0.17271 \\
$\sigma=1.0$, $\beta=0.20$ & 210 & 101 & 253 & 957 & 0.52496 & 0.69703 & 0.25886 & 0.32699 \\
\bottomrule
\end{tabular*}
\par\medskip
\textbf{(b)}\par\smallskip
\begin{tabular*}{\linewidth}{@{\extracolsep{\fill}}lrrrrrrrr@{}}
\toprule
Label & $s_\text{int}$ & $s_\text{bnd}$ & $s_\text{out}$ & $\text{Var}^s_\text{int}$ & $\text{Var}^s_\text{bnd}$ & $\text{Var}^s_\text{out}$ & $\text{Var}_\text{int}$ & $\text{Var}_\text{bnd}$ \\
\midrule
$\sigma=0.0$, $\beta=0.05$ & 0.18502 & 0.15222 & 0.00766 & 0.01974 & 0.01425 & 0.00011 & 0.00425 & 0.00843 \\
$\sigma=0.0$, $\beta=0.10$ & 0.36681 & 0.29741 & 0.01692 & 0.07804 & 0.05553 & 0.00046 & 0.01736 & 0.03726 \\
$\sigma=0.0$, $\beta=0.20$ & 0.72681 & 0.59425 & 0.03103 & 0.30961 & 0.22554 & 0.00174 & 0.06757 & 0.14796 \\
\midrule
$\sigma=0.3$, $\beta=0.05$ & 0.18023 & 0.14354 & 0.00812 & 0.01930 & 0.01255 & 0.00012 & 0.00475 & 0.00908 \\
$\sigma=0.3$, $\beta=0.10$ & 0.35163 & 0.28863 & 0.01765 & 0.07340 & 0.05079 & 0.00050 & 0.02164 & 0.04066 \\
$\sigma=0.3$, $\beta=0.20$ & 0.70559 & 0.57033 & 0.03245 & 0.29462 & 0.20248 & 0.00189 & 0.08436 & 0.17391 \\
\midrule
$\sigma=0.5$, $\beta=0.05$ & 0.17513 & 0.14620 & 0.00857 & 0.01842 & 0.01274 & 0.00013 & 0.00518 & 0.00880 \\
$\sigma=0.5$, $\beta=0.10$ & 0.34275 & 0.29657 & 0.01861 & 0.07026 & 0.05354 & 0.00056 & 0.02567 & 0.04328 \\
$\sigma=0.5$, $\beta=0.20$ & 0.67919 & 0.58472 & 0.03318 & 0.27750 & 0.20852 & 0.00195 & 0.09535 & 0.16963 \\
\midrule
$\sigma=0.7$, $\beta=0.05$ & 0.16722 & 0.14757 & 0.00892 & 0.01704 & 0.01346 & 0.00014 & 0.00521 & 0.00896 \\
$\sigma=0.7$, $\beta=0.10$ & 0.32952 & 0.29044 & 0.01856 & 0.06688 & 0.05128 & 0.00055 & 0.02059 & 0.03768 \\
$\sigma=0.7$, $\beta=0.20$ & 0.66525 & 0.57747 & 0.03521 & 0.27108 & 0.20774 & 0.00217 & 0.08805 & 0.16238 \\
\midrule
$\sigma=1.0$, $\beta=0.05$ & 0.16163 & 0.15058 & 0.00914 & 0.01609 & 0.01386 & 0.00014 & 0.00552 & 0.00895 \\
$\sigma=1.0$, $\beta=0.10$ & 0.31497 & 0.30474 & 0.01923 & 0.06124 & 0.05697 & 0.00058 & 0.02362 & 0.03552 \\
$\sigma=1.0$, $\beta=0.20$ & 0.63351 & 0.60006 & 0.03596 & 0.24634 & 0.21706 & 0.00216 & 0.09413 & 0.15404 \\
\bottomrule
\end{tabular*}
\end{table*}
\vspace{5mm}

\begin{figure*}
    \centering
    % --- Equal partitioning ---
    \begin{tabular}{c c c c}
        \includegraphics[width=0.22\linewidth]{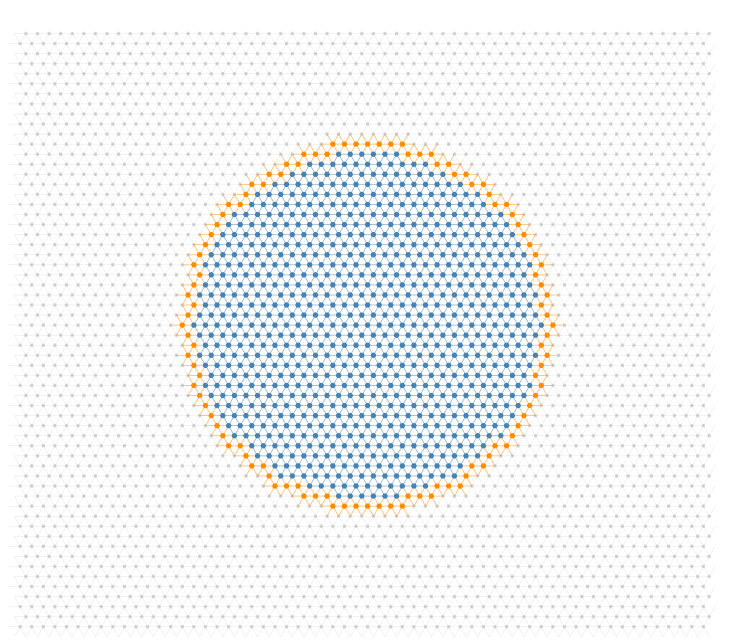}&
        \includegraphics[width=0.22\linewidth]{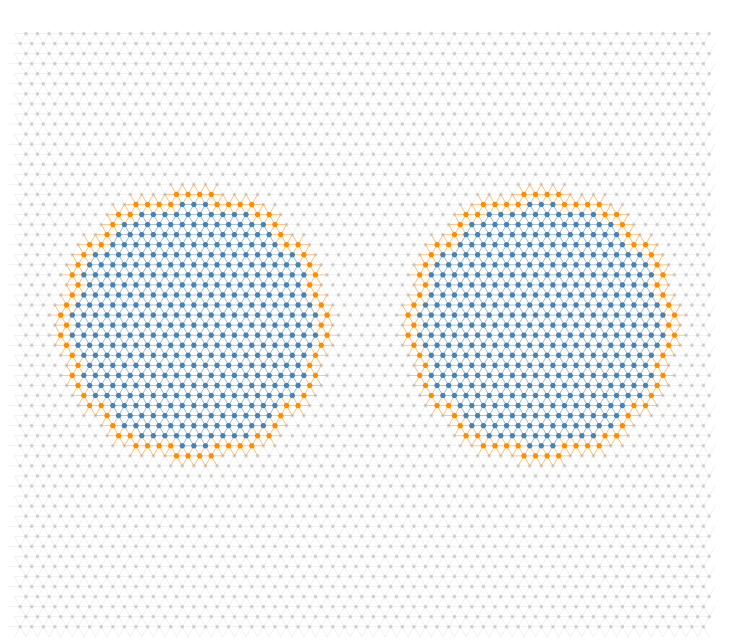}&
        \includegraphics[width=0.22\linewidth]{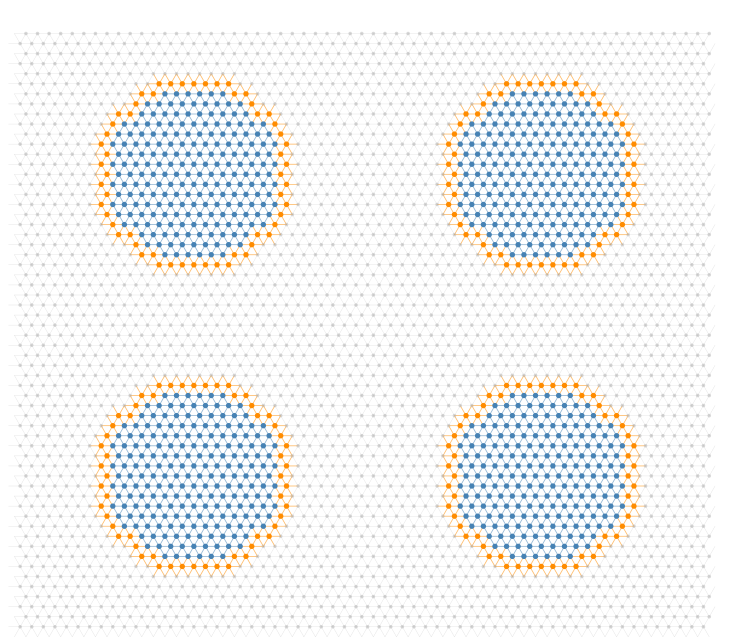}&
        \includegraphics[width=0.22\linewidth]{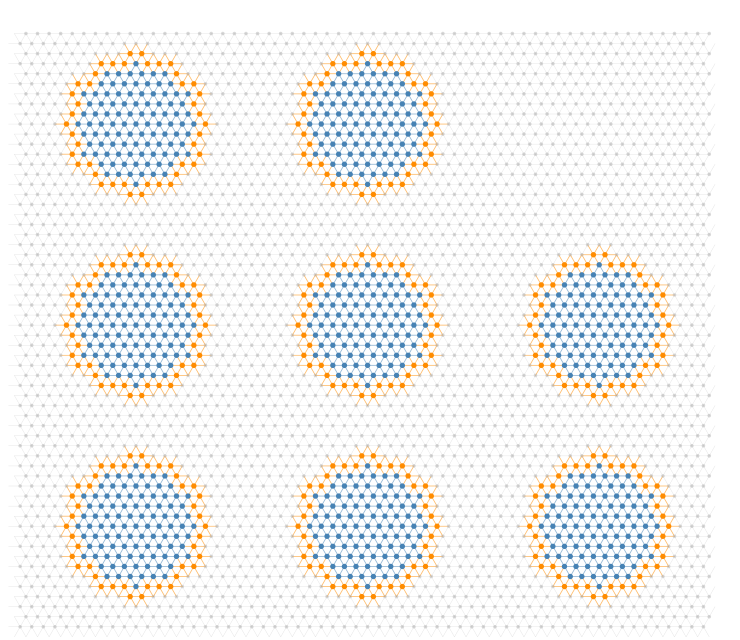}\\
        (a) $N=1$ & (b) $N=2$ & (c) $N=4$ & (d) $N=8$\\[4pt]
        \includegraphics[width=0.22\linewidth]{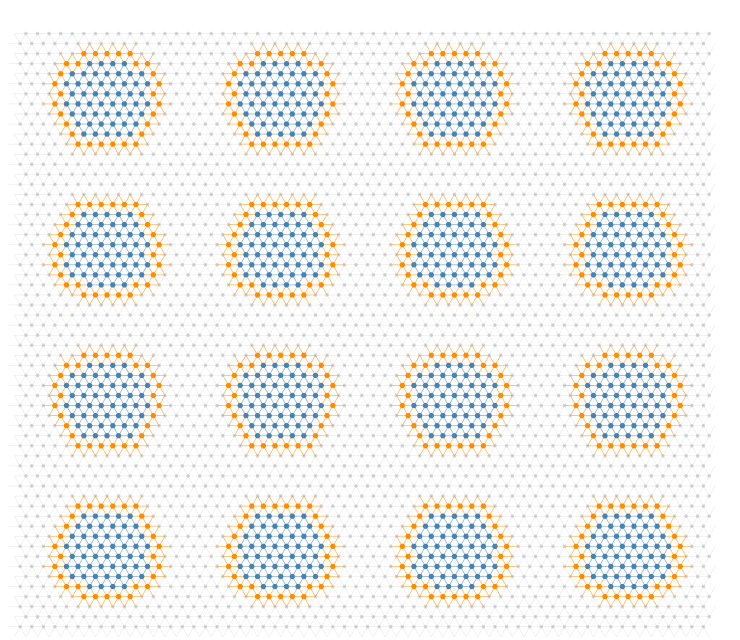}&
        \includegraphics[width=0.22\linewidth]{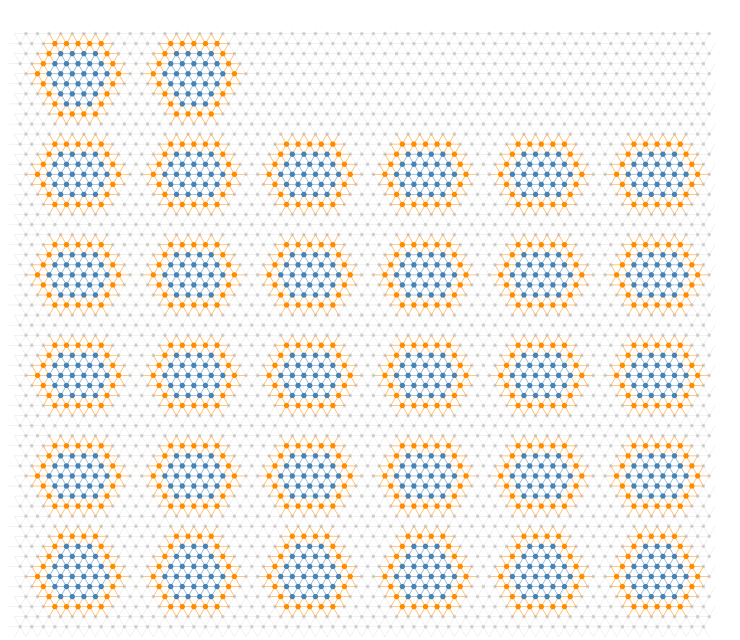}&
        \includegraphics[width=0.22\linewidth]{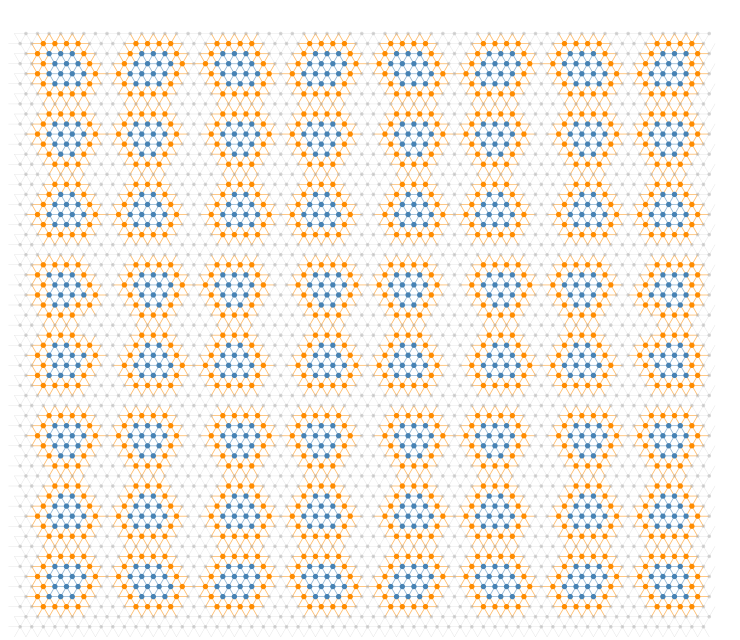}&\\
        (e) $N=16$ & (f) $N=32$ & (g) $N=64$ &\\
    \end{tabular}

    \vspace{6pt}

    % --- Unequal / touching partitioning ---
    \begin{tabular}{c c c c}
        \includegraphics[width=0.22\linewidth]{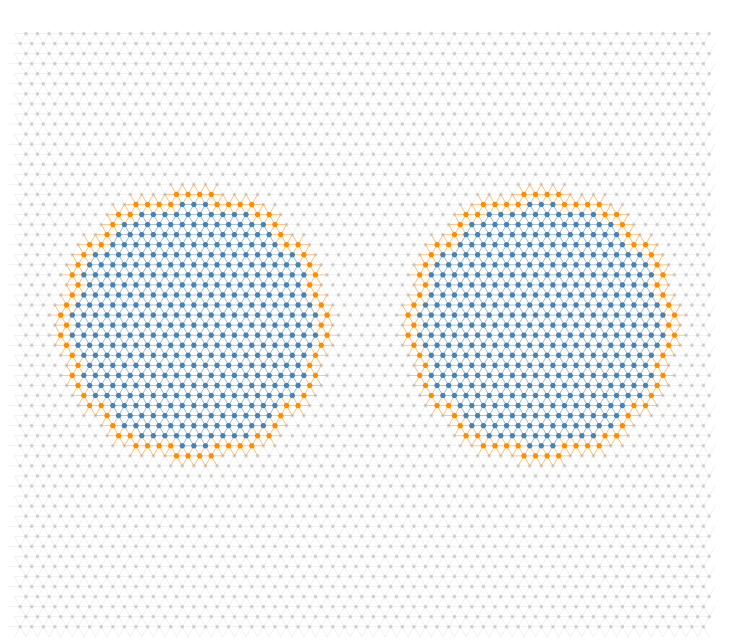}&
        \includegraphics[width=0.22\linewidth]{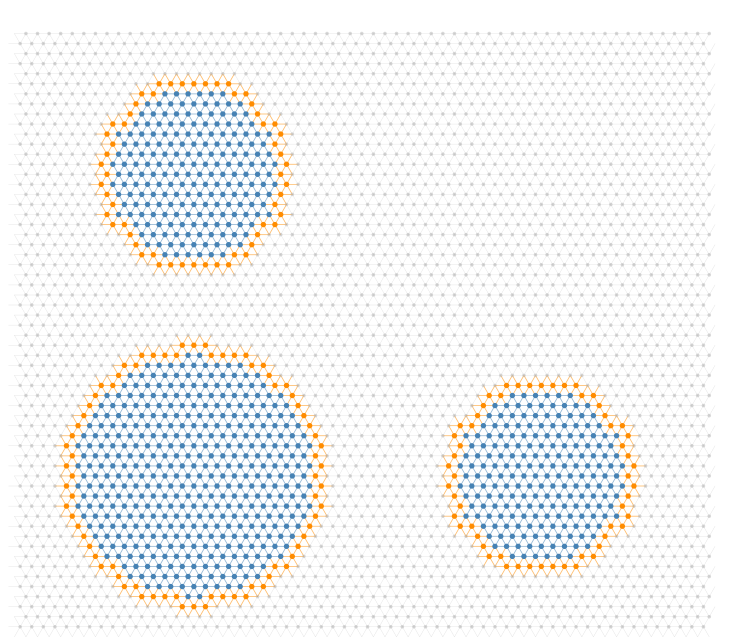}&
        \includegraphics[width=0.22\linewidth]{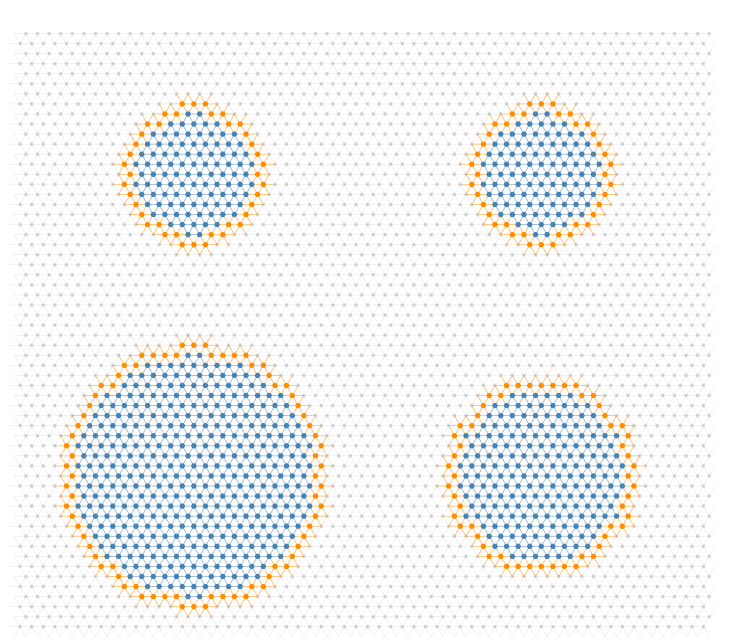}&
        \includegraphics[width=0.22\linewidth]{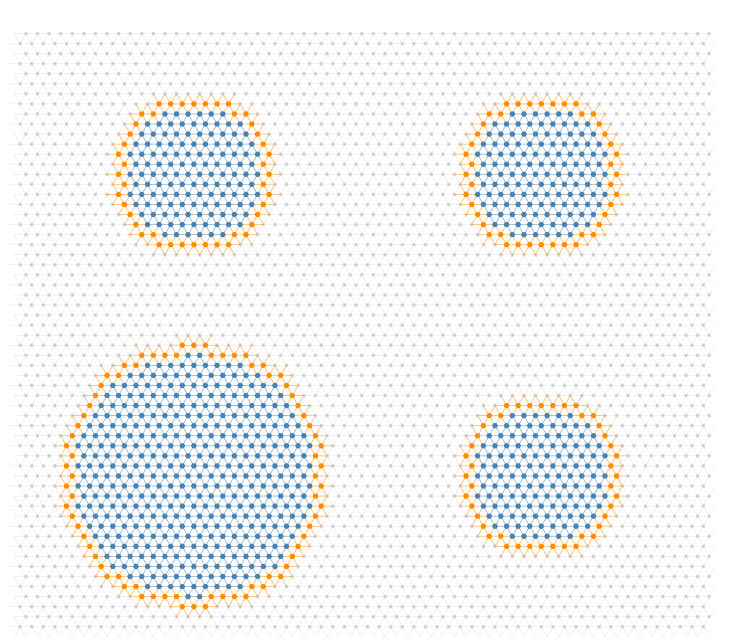}\\
        (h) Untouching $\tfrac{1}{2}+\tfrac{1}{2}$ &
        (i) Untouching $\tfrac{1}{2}+2\tfrac{1}{4}$ &
        (j) Untouching $\tfrac{1}{2}+\tfrac{1}{4}+2\tfrac{1}{8}$ &
        (k) Untouching $\tfrac{1}{2}+3\tfrac{1}{6}$\\[4pt]
        \includegraphics[width=0.22\linewidth]{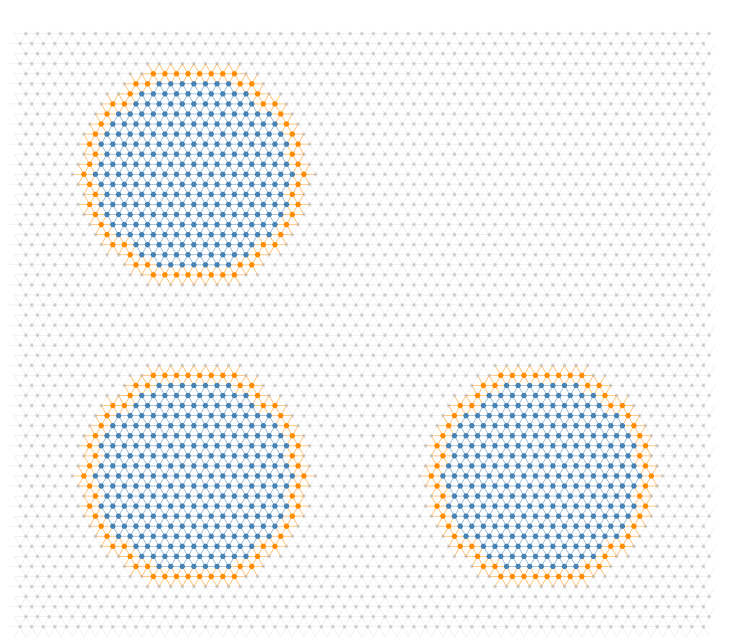}&
        \includegraphics[width=0.22\linewidth]{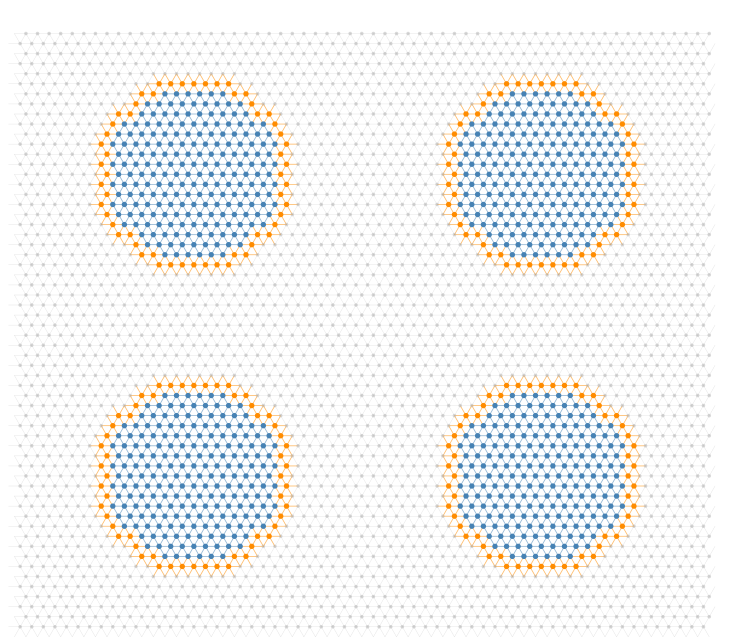}&
        \includegraphics[width=0.22\linewidth]{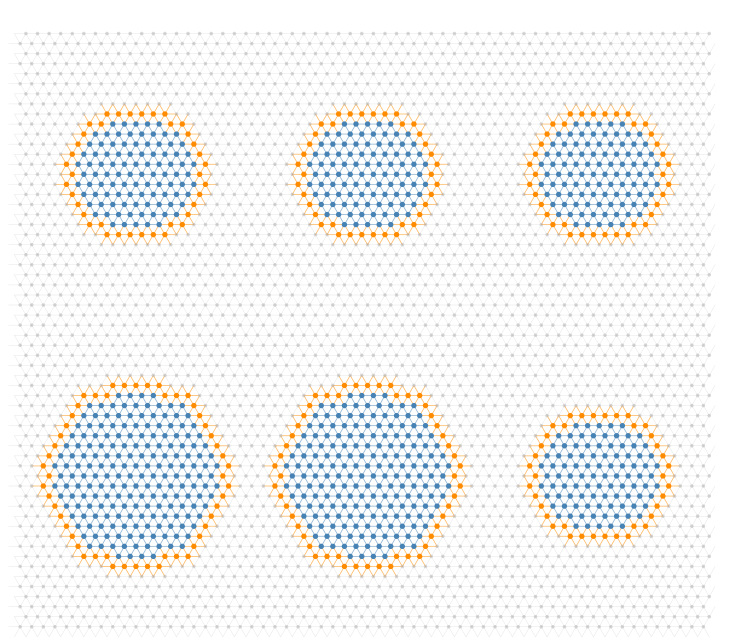}&
        \includegraphics[width=0.22\linewidth]{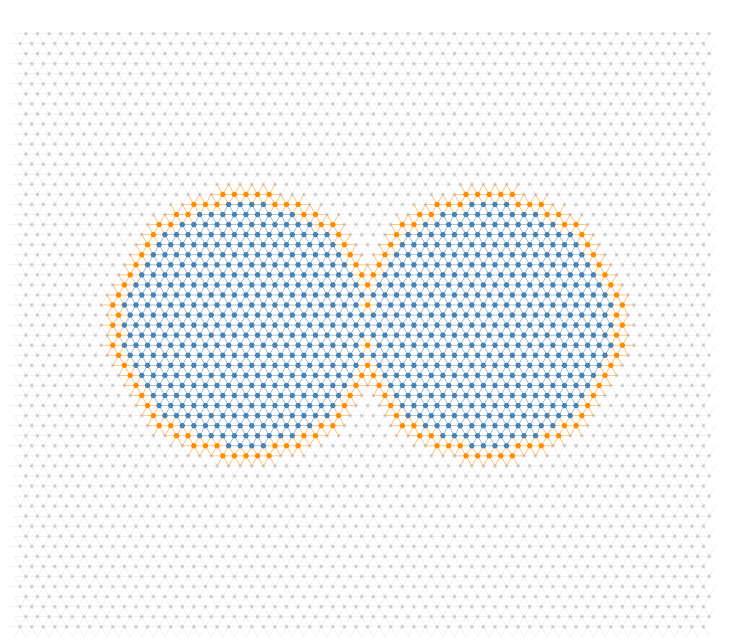}\\
        (l) Untouching $\tfrac{1}{3}+\tfrac{1}{3}+\tfrac{1}{3}$ &
        (m) Untouching $4\tfrac{1}{4}$ &
        (n) Untouching $2\tfrac{1}{4}+4\tfrac{1}{8}$ &
        (o) Touching $\tfrac{1}{2}+\tfrac{1}{2}$\\[4pt]
        \includegraphics[width=0.22\linewidth]{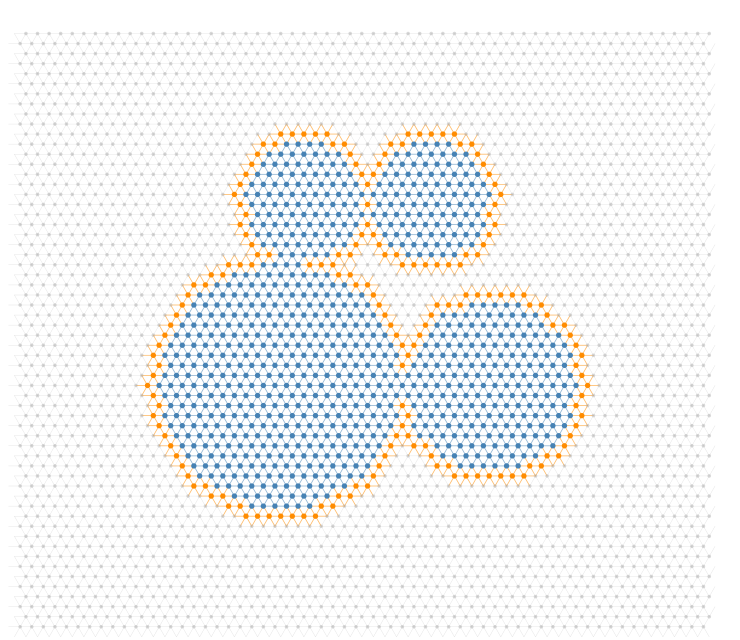}&
        \includegraphics[width=0.22\linewidth]{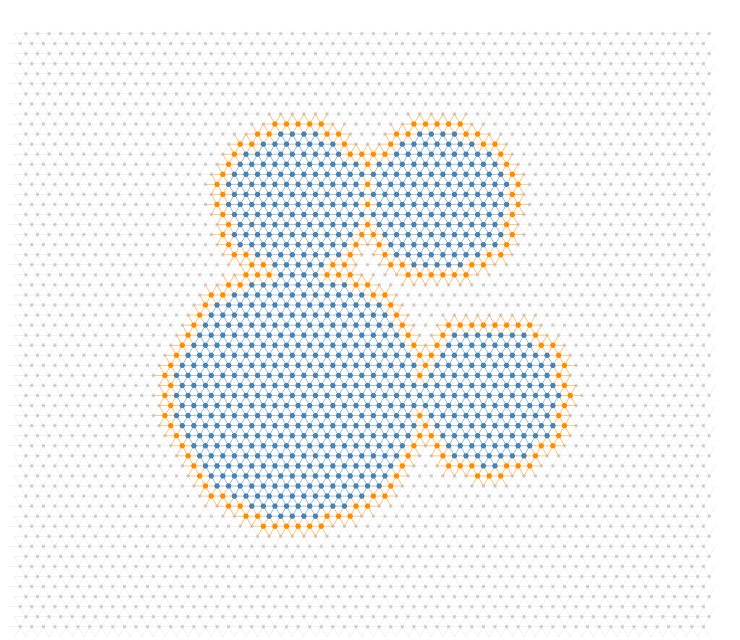}&
        \includegraphics[width=0.22\linewidth]{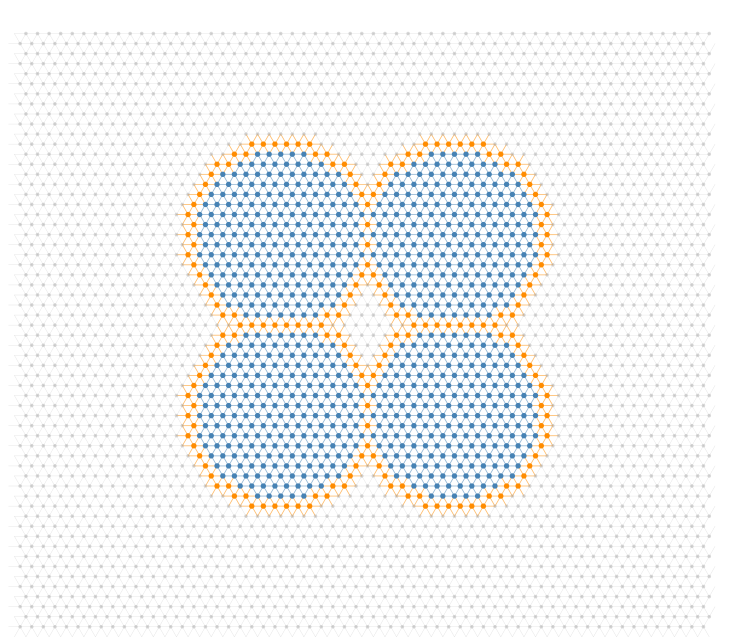}&
        \includegraphics[width=0.22\linewidth]{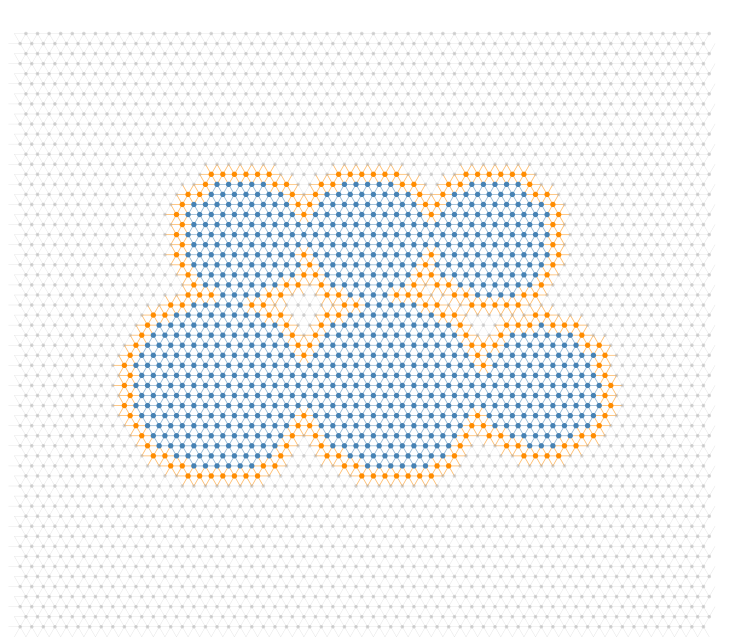}\\
        (p) Touching $\tfrac{1}{2}+\tfrac{1}{4}+2\tfrac{1}{8}$ &
        (q) Touching $\tfrac{1}{2}+3\tfrac{1}{6}$ &
        (r) Touching $4\tfrac{1}{4}$ &
        (s) Touching $2\tfrac{1}{4}+4\tfrac{1}{8}$\\
    \end{tabular}

    \caption{Partitioning configurations. \textit{Top:} Untouching equal partitioning ($N=1$ to $64$). \textit{Bottom:} Untouching and touching unequal partitioning. Orange bonds: boundary bonds. Gray bonds: exterior bonds (low-$k$). Filled circles denote interior nodes ($N_\text{int}$, blue), boundary nodes ($N_\text{bnd}$, orange), and exterior nodes (gray).}
    \label{fig:partitioning_img}
\end{figure*}

\appsec{Supplementary Images and Data of Equal and Unequal Area Partitioning}
\label{sec:partitioning_table}
In this section, we include Fig.~\ref{fig:partitioning_img} and full tables of untouching equal and unequal partitioning statistical data. (The unequal touching case is not included due to its similarity to the simple test results.) In the main text, we briefly introduced the results of equal partitioning. For the partitioning case, $|\mathcal{E}_\text{int}|$ is fixed at $2568$ and the radius of each disk partition is adjusted based on the node count, including switching boundary nodes to interior nodes to maintain a constant. Table~\ref{tab:partitioning-data-grf-constrained} shows the detailed results. An interesting observation is that when $N=4, 8$, the stress is stronger on the boundary and exterior than the interior, indicating faster energy propagation. At $N=16$, comparable stress begins to appear across all regions, which can also be verified through both the ratio $|\mathcal{E}_\text{bnd}|/N_\text{int}$ and the uniform distribution of disks (meaning disks are separated by approximately one diameter). The constrained and random cases show slightly higher stress than the unconstrained and GRF prestress cases, but the overall trends are analogous, as shown in Tables~\ref{tab:partitioning-data-grf-unconstrained}--\ref{tab:partitioning-data-random-unconstrained}.

% BEGIN APPENDIX CONTENT
\begin{table*}[p]
	\centering
	\caption{Partitioning data (untouching): \texttt{prestress\_type} = \texttt{GRF}; \texttt{regime} = \texttt{constrained}. Non-touching arrangement. Both panels use the original row labels and reported numerical precision.}
	\label{tab:partitioning-data-grf-constrained}
	\footnotesize
	\setlength{\tabcolsep}{5pt}
	\renewcommand{\arraystretch}{1.05}
	\textbf{(a)}\par\smallskip
	\begin{tabular*}{\linewidth}{@{\extracolsep{\fill}}lrrrrrrrr@{}}
		\toprule
		Label & $E_\text{bnd}$ & $N_\text{bnd}$ & $N_\text{int}$ & $E_\text{int}$ & $\xi_\text{int}$ & $\xi_\text{bnd}$ & $\xi_\text{out}$ & $\xi_\text{tot}$ \\
		\midrule
		Equal $N=1$  & 210  & 108 & 821 & 2568 & 0.07365 & 0.07593 & 0.00039 & 0.01937 \\
		Equal $N=2$  & 300  & 156 & 821 & 2568 & 0.03747 & 0.07419 & 0.00055 & 0.01215 \\
		Equal $N=4$  & 424  & 224 & 821 & 2568 & 0.01294 & 0.07166 & 0.00076 & 0.00795 \\
		Equal $N=8$  & 624  & 336 & 821 & 2568 & 0.01962 & 0.07031 & 0.00108 & 0.01164 \\
		Equal $N=16$ & 832  & 464 & 821 & 2568 & 0.02780 & 0.06498 & 0.00163 & 0.01581 \\
		Equal $N=32$ & 1104 & 648 & 821 & 2568 & 0.03106 & 0.06473 & 0.00259 & 0.02013 \\
		Equal $N=64$ & 1576 & 980 & 821 & 2568 & 0.03190 & 0.05758 & 0.00374 & 0.02467 \\
		\midrule
		$\tfrac{1}{2}+\tfrac{1}{2}$                         & 300 & 156 & 821 & 2568 & 0.03743 & 0.07436 & 0.00056 & 0.01216 \\
		$\tfrac{1}{3}+\tfrac{1}{3}+\tfrac{1}{3}$            & 354 & 186 & 821 & 2568 & 0.01449 & 0.07466 & 0.00118 & 0.00796 \\
		$\tfrac{1}{2}+\tfrac{1}{4}+\tfrac{1}{4}$            & 358 & 188 & 821 & 2568 & 0.01474 & 0.07391 & 0.00114 & 0.00798 \\
		$\tfrac{1}{2}+\tfrac{1}{4}+\tfrac{1}{8}+\tfrac{1}{8}$ & 400 & 212 & 821 & 2568 & 0.01271 & 0.07322 & 0.00074 & 0.00768 \\
		$\tfrac{1}{4}+\tfrac{1}{4}+\tfrac{1}{4}+\tfrac{1}{4}$ & 424 & 224 & 821 & 2568 & 0.01267 & 0.07140 & 0.00076 & 0.00788 \\
		$\tfrac{1}{2}+\tfrac{1}{6}+\tfrac{1}{6}+\tfrac{1}{6}$ & 392 & 208 & 821 & 2568 & 0.01271 & 0.07431 & 0.00079 & 0.00770 \\
		$\tfrac{1}{4}+\tfrac{1}{4}+\tfrac{1}{8}+\tfrac{1}{8}+\tfrac{1}{8}+\tfrac{1}{8}$ & 476 & 256 & 821 & 2568 & 0.02207 & 0.07452 & 0.00233 & 0.01182 \\
		\bottomrule
	\end{tabular*}
	\par\medskip
	\textbf{(b)}\par\smallskip
	\begin{tabular*}{\linewidth}{@{\extracolsep{\fill}}lrrrrrrrr@{}}
		\toprule
		Label & $s_\text{int}$ & $s_\text{bnd}$ & $s_\text{out}$ & $\text{Var}^s_\text{int}$ & $\text{Var}^s_\text{bnd}$ & $\text{Var}^s_\text{out}$ & $\text{Var}_\text{int}$ & $\text{Var}_\text{bnd}$ \\
		\midrule
		Equal $N=1$  & 0.58065 & 0.18349 & 0.00024 & 0.19462 & 0.08550 & 0       & 0.00073 & 0.00084 \\
		Equal $N=2$  & 0.25998 & 0.18021 & 0.09646 & 0.17133 & 0.08771 & 0.07954 & 0.00169 & 0.00089 \\
		Equal $N=4$  & 0.04267 & 0.16596 & 0.14522 & 0.03489 & 0.07797 & 0.10810 & 0.00087 & 0.00098 \\
		Equal $N=8$  & 0.08708 & 0.16530 & 0.12574 & 0.06598 & 0.07818 & 0.09506 & 0.00117 & 0.00102 \\
		Equal $N=16$ & 0.13642 & 0.13692 & 0.10465 & 0.09556 & 0.06360 & 0.07729 & 0.00142 & 0.00116 \\
		Equal $N=32$ & 0.11327 & 0.12740 & 0.09820 & 0.07662 & 0.05564 & 0.06809 & 0.00141 & 0.00114 \\
		Equal $N=64$ & 0.09101 & 0.10340 & 0.09334 & 0.05685 & 0.04219 & 0.05840 & 0.00134 & 0.00124 \\
		\midrule
		$\tfrac{1}{2}+\tfrac{1}{2}$                         & 0.25588 & 0.17498 & 0.09262 & 0.16103 & 0.07996 & 0.07348 & 0.00168 & 0.00088 \\
		$\tfrac{1}{3}+\tfrac{1}{3}+\tfrac{1}{3}$            & 0.06367 & 0.17927 & 0.14560 & 0.05187 & 0.08479 & 0.11088 & 0.00094 & 0.00088 \\
		$\tfrac{1}{2}+\tfrac{1}{4}+\tfrac{1}{4}$            & 0.06600 & 0.17867 & 0.14335 & 0.05466 & 0.08383 & 0.10791 & 0.00097 & 0.00090 \\
		$\tfrac{1}{2}+\tfrac{1}{4}+\tfrac{1}{8}+\tfrac{1}{8}$ & 0.05438 & 0.17340 & 0.14419 & 0.04551 & 0.08121 & 0.10704 & 0.00086 & 0.00091 \\
		$\tfrac{1}{4}+\tfrac{1}{4}+\tfrac{1}{4}+\tfrac{1}{4}$ & 0.04128 & 0.16556 & 0.14530 & 0.03293 & 0.07706 & 0.10889 & 0.00084 & 0.00098 \\
		$\tfrac{1}{2}+\tfrac{1}{6}+\tfrac{1}{6}+\tfrac{1}{6}$ & 0.05528 & 0.17806 & 0.14775 & 0.04814 & 0.08482 & 0.11187 & 0.00087 & 0.00088 \\
		$\tfrac{1}{4}+\tfrac{1}{4}+\tfrac{1}{8}+\tfrac{1}{8}+\tfrac{1}{8}+\tfrac{1}{8}$ & 0.13297 & 0.17234 & 0.11707 & 0.10223 & 0.07900 & 0.08832 & 0.00130 & 0.00087 \\
		\bottomrule
	\end{tabular*}
\end{table*}
\begin{table*}[p]
	\centering
	\caption{Partitioning data (untouching): \texttt{prestress\_type} = \texttt{GRF}; \texttt{regime} = \texttt{unconstrained}. Both panels use the original row labels and reported numerical precision.}
	\label{tab:partitioning-data-grf-unconstrained}
	\footnotesize
	\setlength{\tabcolsep}{5pt}
	\renewcommand{\arraystretch}{1.05}
	\textbf{(a)}\par\smallskip
	\begin{tabular*}{\linewidth}{@{\extracolsep{\fill}}lrrrrrrrr@{}}
		\toprule
		Label & $E_\text{bnd}$ & $N_\text{bnd}$ & $N_\text{int}$ & $E_\text{int}$ & $\xi_\text{int}$ & $\xi_\text{bnd}$ & $\xi_\text{out}$ & $\xi_\text{tot}$ \\
		\midrule
		Equal $N=1$  & 210  & 108 & 821 & 2568 & 0.30842 & 0.45802 & 0.14004 & 0.18798 \\
		Equal $N=2$  & 300  & 156 & 821 & 2568 & 0.25358 & 0.39559 & 0.14374 & 0.17968 \\
		Equal $N=4$  & 424  & 224 & 821 & 2568 & 0.17963 & 0.32565 & 0.16160 & 0.17593 \\
		Equal $N=8$  & 624  & 336 & 821 & 2568 & 0.18080 & 0.27663 & 0.15533 & 0.17224 \\
		Equal $N=16$ & 832  & 464 & 821 & 2568 & 0.17277 & 0.23392 & 0.15821 & 0.17133 \\
		Equal $N=32$ & 1104 & 648 & 821 & 2568 & 0.16235 & 0.19804 & 0.14167 & 0.15641 \\
		Equal $N=64$ & 1576 & 980 & 821 & 2568 & 0.14236 & 0.16135 & 0.13154 & 0.14205 \\
		\midrule
		$\tfrac{1}{2}+\tfrac{1}{2}$                         & 300 & 156 & 821 & 2568 & 0.24486 & 0.37555 & 0.13500 & 0.17045 \\
		$\tfrac{1}{3}+\tfrac{1}{3}+\tfrac{1}{3}$            & 354 & 186 & 821 & 2568 & 0.19008 & 0.36008 & 0.15509 & 0.17351 \\
		$\tfrac{1}{2}+\tfrac{1}{4}+\tfrac{1}{4}$            & 358 & 188 & 821 & 2568 & 0.18286 & 0.35187 & 0.14698 & 0.16571 \\
		$\tfrac{1}{2}+\tfrac{1}{4}+\tfrac{1}{8}+\tfrac{1}{8}$ & 400 & 212 & 821 & 2568 & 0.17292 & 0.32451 & 0.15007 & 0.16544 \\
		$\tfrac{1}{4}+\tfrac{1}{4}+\tfrac{1}{4}+\tfrac{1}{4}$ & 424 & 224 & 821 & 2568 & 0.17974 & 0.32561 & 0.16102 & 0.17554 \\
		$\tfrac{1}{2}+\tfrac{1}{6}+\tfrac{1}{6}+\tfrac{1}{6}$ & 392 & 208 & 821 & 2568 & 0.19034 & 0.33912 & 0.16143 & 0.17815 \\
		$\tfrac{1}{4}+\tfrac{1}{4}+\tfrac{1}{8}+\tfrac{1}{8}+\tfrac{1}{8}+\tfrac{1}{8}$ & 476 & 256 & 821 & 2568 & 0.17850 & 0.30281 & 0.15452 & 0.17028 \\
		\bottomrule
	\end{tabular*}
	\par\medskip
	\textbf{(b)}\par\smallskip
	\begin{tabular*}{\linewidth}{@{\extracolsep{\fill}}lrrrrrrrr@{}}
		\toprule
		Label & $s_\text{int}$ & $s_\text{bnd}$ & $s_\text{out}$ & $\text{Var}^s_\text{int}$ & $\text{Var}^s_\text{bnd}$ & $\text{Var}^s_\text{out}$ & $\text{Var}_\text{int}$ & $\text{Var}_\text{bnd}$ \\
		\midrule
		Equal $N=1$  & 0.38613 & 0.12497 & 0.01299 & 0.08745 & 0.01783 & 0.00025 & 0.02498 & 0.04787 \\
		Equal $N=2$  & 0.17816 & 0.11766 & 0.07186 & 0.06372 & 0.01504 & 0.03081 & 0.02029 & 0.03830 \\
		Equal $N=4$  & 0.03935 & 0.10935 & 0.09940 & 0.01059 & 0.01358 & 0.03575 & 0.01246 & 0.02911 \\
		Equal $N=8$  & 0.06511 & 0.10572 & 0.08316 & 0.01936 & 0.01290 & 0.02658 & 0.00969 & 0.02202 \\
		Equal $N=16$ & 0.08494 & 0.08285 & 0.06739 & 0.02162 & 0.00734 & 0.01554 & 0.00923 & 0.01682 \\
		Equal $N=32$ & 0.06857 & 0.07288 & 0.06011 & 0.01259 & 0.00519 & 0.01153 & 0.00736 & 0.01185 \\
		Equal $N=64$ & 0.05484 & 0.06010 & 0.05508 & 0.00774 & 0.00384 & 0.00753 & 0.00577 & 0.00805 \\
		\midrule
		$\tfrac{1}{2}+\tfrac{1}{2}$                         & 0.17669 & 0.11698 & 0.07006 & 0.06305 & 0.01578 & 0.02976 & 0.01871 & 0.03459 \\
		$\tfrac{1}{3}+\tfrac{1}{3}+\tfrac{1}{3}$            & 0.05213 & 0.11622 & 0.10149 & 0.01710 & 0.01503 & 0.04050 & 0.01754 & 0.03607 \\
		$\tfrac{1}{2}+\tfrac{1}{4}+\tfrac{1}{4}$            & 0.05388 & 0.11731 & 0.10087 & 0.01952 & 0.01552 & 0.04089 & 0.01713 & 0.03346 \\
		$\tfrac{1}{2}+\tfrac{1}{4}+\tfrac{1}{8}+\tfrac{1}{8}$ & 0.04688 & 0.11325 & 0.10097 & 0.01585 & 0.01488 & 0.03909 & 0.01402 & 0.03059 \\
		$\tfrac{1}{4}+\tfrac{1}{4}+\tfrac{1}{4}+\tfrac{1}{4}$ & 0.03908 & 0.10936 & 0.09968 & 0.01043 & 0.01371 & 0.03607 & 0.01266 & 0.02948 \\
		$\tfrac{1}{2}+\tfrac{1}{6}+\tfrac{1}{6}+\tfrac{1}{6}$ & 0.04722 & 0.11301 & 0.10219 & 0.01586 & 0.01447 & 0.03980 & 0.01665 & 0.03328 \\
		$\tfrac{1}{4}+\tfrac{1}{4}+\tfrac{1}{8}+\tfrac{1}{8}+\tfrac{1}{8}+\tfrac{1}{8}$ & 0.09013 & 0.10581 & 0.08237 & 0.03323 & 0.01164 & 0.02864 & 0.01164 & 0.02691 \\
		\bottomrule
	\end{tabular*}
\end{table*}
\begin{table*}[p]
	\centering
	\caption{Partitioning data (untouching): \texttt{prestress\_type} = \texttt{random}; \texttt{regime} = \texttt{constrained}. Both panels use the original row labels and reported numerical precision.}
	\label{tab:partitioning-data-random-constrained}
	\footnotesize
	\setlength{\tabcolsep}{5pt}
	\renewcommand{\arraystretch}{1.05}
	\textbf{(a)}\par\smallskip
	\begin{tabular*}{\linewidth}{@{\extracolsep{\fill}}lrrrrrrrr@{}}
		\toprule
		Label & $E_\text{bnd}$ & $N_\text{bnd}$ & $N_\text{int}$ & $E_\text{int}$ & $\xi_\text{int}$ & $\xi_\text{bnd}$ & $\xi_\text{out}$ & $\xi_\text{tot}$ \\
		\midrule
		Equal $N=1$  & 210  & 108 & 821 & 2568 & 0.07099 & 0.06824 & 0.00025 & 0.01842 \\
		Equal $N=2$  & 300  & 156 & 821 & 2568 & 0.03595 & 0.06766 & 0.00037 & 0.01139 \\
		Equal $N=4$  & 424  & 224 & 821 & 2568 & 0.01230 & 0.06585 & 0.00052 & 0.00728 \\
		Equal $N=8$  & 624  & 336 & 821 & 2568 & 0.01902 & 0.06531 & 0.00081 & 0.01085 \\
		Equal $N=16$ & 832  & 464 & 821 & 2568 & 0.02730 & 0.06101 & 0.00120 & 0.01490 \\
		Equal $N=32$ & 1104 & 648 & 821 & 2568 & 0.03115 & 0.06250 & 0.00204 & 0.01942 \\
		Equal $N=64$ & 1576 & 980 & 821 & 2568 & 0.03275 & 0.05704 & 0.00333 & 0.02452 \\
		\midrule
		$\tfrac{1}{2}+\tfrac{1}{2}$                         & 300 & 156 & 821 & 2568 & 0.03582 & 0.06815 & 0.00037 & 0.01138 \\
		$\tfrac{1}{3}+\tfrac{1}{3}+\tfrac{1}{3}$            & 354 & 186 & 821 & 2568 & 0.01391 & 0.06879 & 0.00096 & 0.00736 \\
		$\tfrac{1}{2}+\tfrac{1}{4}+\tfrac{1}{4}$            & 358 & 188 & 821 & 2568 & 0.01393 & 0.06772 & 0.00093 & 0.00733 \\
		$\tfrac{1}{2}+\tfrac{1}{4}+\tfrac{1}{8}+\tfrac{1}{8}$ & 400 & 212 & 821 & 2568 & 0.01211 & 0.06749 & 0.00051 & 0.00705 \\
		$\tfrac{1}{4}+\tfrac{1}{4}+\tfrac{1}{4}+\tfrac{1}{4}$ & 424 & 224 & 821 & 2568 & 0.01225 & 0.06587 & 0.00052 & 0.00727 \\
		$\tfrac{1}{2}+\tfrac{1}{6}+\tfrac{1}{6}+\tfrac{1}{6}$ & 392 & 208 & 821 & 2568 & 0.01185 & 0.06833 & 0.00053 & 0.00698 \\
		$\tfrac{1}{4}+\tfrac{1}{4}+\tfrac{1}{8}+\tfrac{1}{8}+\tfrac{1}{8}+\tfrac{1}{8}$ & 476 & 256 & 821 & 2568 & 0.02130 & 0.06873 & 0.00193 & 0.01095 \\
		\bottomrule
	\end{tabular*}
	\par\medskip
	\textbf{(b)}\par\smallskip
	\begin{tabular*}{\linewidth}{@{\extracolsep{\fill}}lrrrrrrrr@{}}
		\toprule
		Label & $s_\text{int}$ & $s_\text{bnd}$ & $s_\text{out}$ & $\text{Var}^s_\text{int}$ & $\text{Var}^s_\text{bnd}$ & $\text{Var}^s_\text{out}$ & $\text{Var}_\text{int}$ & $\text{Var}_\text{bnd}$ \\
		\midrule
		Equal $N=1$  & 0.49934 & 0.15339 & 0.00018 & 0.14297 & 0.05327 & 0       & 0.00075 & 0.00091 \\
		Equal $N=2$  & 0.22499 & 0.15083 & 0.08286 & 0.12465 & 0.05327 & 0.05732 & 0.00162 & 0.00092 \\
		Equal $N=4$  & 0.03815 & 0.14193 & 0.12802 & 0.02653 & 0.04941 & 0.08055 & 0.00082 & 0.00098 \\
		Equal $N=8$  & 0.07946 & 0.14069 & 0.11190 & 0.05373 & 0.04912 & 0.07236 & 0.00114 & 0.00101 \\
		Equal $N=16$ & 0.12454 & 0.11505 & 0.09391 & 0.07445 & 0.03640 & 0.05675 & 0.00141 & 0.00110 \\
		Equal $N=32$ & 0.10606 & 0.10800 & 0.09150 & 0.06043 & 0.03287 & 0.05345 & 0.00145 & 0.00108 \\
		Equal $N=64$ & 0.08732 & 0.09062 & 0.09096 & 0.04464 & 0.02564 & 0.04693 & 0.00140 & 0.00118 \\
		\midrule
		$\tfrac{1}{2}+\tfrac{1}{2}$                         & 0.22486 & 0.14955 & 0.08216 & 0.12379 & 0.05193 & 0.05674 & 0.00161 & 0.00091 \\
		$\tfrac{1}{3}+\tfrac{1}{3}+\tfrac{1}{3}$            & 0.05676 & 0.15144 & 0.12548 & 0.03995 & 0.05353 & 0.08176 & 0.00090 & 0.00091 \\
		$\tfrac{1}{2}+\tfrac{1}{4}+\tfrac{1}{4}$            & 0.05753 & 0.14998 & 0.12593 & 0.04083 & 0.05238 & 0.08226 & 0.00090 & 0.00093 \\
		$\tfrac{1}{2}+\tfrac{1}{4}+\tfrac{1}{8}+\tfrac{1}{8}$ & 0.04769 & 0.14940 & 0.12722 & 0.03433 & 0.05260 & 0.08229 & 0.00081 & 0.00093 \\
		$\tfrac{1}{4}+\tfrac{1}{4}+\tfrac{1}{4}+\tfrac{1}{4}$ & 0.03796 & 0.14288 & 0.12799 & 0.02674 & 0.04936 & 0.08050 & 0.00081 & 0.00098 \\
		$\tfrac{1}{2}+\tfrac{1}{6}+\tfrac{1}{6}+\tfrac{1}{6}$ & 0.04526 & 0.14947 & 0.12888 & 0.03216 & 0.05215 & 0.08390 & 0.00079 & 0.00091 \\
		$\tfrac{1}{4}+\tfrac{1}{4}+\tfrac{1}{8}+\tfrac{1}{8}+\tfrac{1}{8}+\tfrac{1}{8}$ & 0.11538 & 0.14599 & 0.10509 & 0.07602 & 0.04985 & 0.06928 & 0.00124 & 0.00090 \\
		\bottomrule
	\end{tabular*}
\end{table*}

\begin{table*}[p]
	\centering
	\caption{Partitioning data (untouching): \texttt{prestress\_type} = \texttt{random}; \texttt{regime} = \texttt{unconstrained}. Both panels use the original row labels and reported numerical precision.}
	\label{tab:partitioning-data-random-unconstrained}
	\footnotesize
	\setlength{\tabcolsep}{5pt}
	\renewcommand{\arraystretch}{1.05}
	\textbf{(a)}\par\smallskip
	\begin{tabular*}{\linewidth}{@{\extracolsep{\fill}}lrrrrrrrr@{}}
		\toprule
		Label & $E_\text{bnd}$ & $N_\text{bnd}$ & $N_\text{int}$ & $E_\text{int}$ & $\xi_\text{int}$ & $\xi_\text{bnd}$ & $\xi_\text{out}$ & $\xi_\text{tot}$ \\
		\midrule
		Equal $N=1$  & 210  & 108 & 821 & 2568 & 0.13935 & 0.17258 & 0.04559 & 0.07078 \\
		Equal $N=2$  & 300  & 156 & 821 & 2568 & 0.10824 & 0.16256 & 0.04879 & 0.06726 \\
		Equal $N=4$  & 424  & 224 & 821 & 2568 & 0.06749 & 0.14829 & 0.05354 & 0.06263 \\
		Equal $N=8$  & 624  & 336 & 821 & 2568 & 0.07865 & 0.13743 & 0.05839 & 0.07024 \\
		Equal $N=16$ & 832  & 464 & 821 & 2568 & 0.08911 & 0.13022 & 0.06526 & 0.07911 \\
		Equal $N=32$ & 1104 & 648 & 821 & 2568 & 0.09685 & 0.12581 & 0.06887 & 0.08537 \\
		Equal $N=64$ & 1576 & 980 & 821 & 2568 & 0.09472 & 0.11303 & 0.07502 & 0.08976 \\
		\midrule
		$\tfrac{1}{2}+\tfrac{1}{2}$                         & 300 & 156 & 821 & 2568 & 0.10501 & 0.15784 & 0.04704 & 0.06505 \\
		$\tfrac{1}{3}+\tfrac{1}{3}+\tfrac{1}{3}$            & 354 & 186 & 821 & 2568 & 0.06939 & 0.15494 & 0.05125 & 0.06067 \\
		$\tfrac{1}{2}+\tfrac{1}{4}+\tfrac{1}{4}$            & 358 & 188 & 821 & 2568 & 0.06837 & 0.15150 & 0.05018 & 0.05955 \\
		$\tfrac{1}{2}+\tfrac{1}{4}+\tfrac{1}{8}+\tfrac{1}{8}$ & 400 & 212 & 821 & 2568 & 0.06762 & 0.14905 & 0.05373 & 0.06245 \\
		$\tfrac{1}{4}+\tfrac{1}{4}+\tfrac{1}{4}+\tfrac{1}{4}$ & 424 & 224 & 821 & 2568 & 0.06704 & 0.14819 & 0.05413 & 0.06294 \\
		$\tfrac{1}{2}+\tfrac{1}{6}+\tfrac{1}{6}+\tfrac{1}{6}$ & 392 & 208 & 821 & 2568 & 0.06552 & 0.14860 & 0.05184 & 0.06048 \\
		$\tfrac{1}{4}+\tfrac{1}{4}+\tfrac{1}{8}+\tfrac{1}{8}+\tfrac{1}{8}+\tfrac{1}{8}$ & 476 & 256 & 821 & 2568 & 0.07410 & 0.14517 & 0.05393 & 0.06485 \\
		\bottomrule
	\end{tabular*}
	\par\medskip
	\textbf{(b)}\par\smallskip
	\begin{tabular*}{\linewidth}{@{\extracolsep{\fill}}lrrrrrrrr@{}}
		\toprule
		Label & $s_\text{int}$ & $s_\text{bnd}$ & $s_\text{out}$ & $\text{Var}^s_\text{int}$ & $\text{Var}^s_\text{bnd}$ & $\text{Var}^s_\text{out}$ & $\text{Var}_\text{int}$ & $\text{Var}_\text{bnd}$ \\
		\midrule
		Equal $N=1$  & 0.44611 & 0.12048 & 0.00465 & 0.11433 & 0.02173 & 0.00005 & 0.00532 & 0.00826 \\
		Equal $N=2$  & 0.20266 & 0.11666 & 0.07681 & 0.09395 & 0.02014 & 0.04380 & 0.00470 & 0.00740 \\
		Equal $N=4$  & 0.03691 & 0.11029 & 0.11488 & 0.01681 & 0.01840 & 0.05779 & 0.00301 & 0.00637 \\
		Equal $N=8$  & 0.07409 & 0.10915 & 0.10103 & 0.03661 & 0.01962 & 0.05101 & 0.00314 & 0.00564 \\
		Equal $N=16$ & 0.10560 & 0.08682 & 0.07984 & 0.04392 & 0.01039 & 0.03101 & 0.00345 & 0.00524 \\
		Equal $N=32$ & 0.08511 & 0.07891 & 0.07433 & 0.02730 & 0.00723 & 0.02511 & 0.00347 & 0.00481 \\
		Equal $N=64$ & 0.06424 & 0.06487 & 0.06607 & 0.01448 & 0.00492 & 0.01496 & 0.00314 & 0.00405 \\
		\midrule
		$\tfrac{1}{2}+\tfrac{1}{2}$                         & 0.20326 & 0.11615 & 0.07595 & 0.09457 & 0.02002 & 0.04314 & 0.00446 & 0.00704 \\
		$\tfrac{1}{3}+\tfrac{1}{3}+\tfrac{1}{3}$            & 0.05374 & 0.11666 & 0.11504 & 0.02855 & 0.01997 & 0.06261 & 0.00328 & 0.00681 \\
		$\tfrac{1}{2}+\tfrac{1}{4}+\tfrac{1}{4}$            & 0.05478 & 0.11662 & 0.11551 & 0.02984 & 0.02100 & 0.06330 & 0.00324 & 0.00654 \\
		$\tfrac{1}{2}+\tfrac{1}{4}+\tfrac{1}{8}+\tfrac{1}{8}$ & 0.04701 & 0.11518 & 0.11611 & 0.02547 & 0.02073 & 0.06188 & 0.00308 & 0.00648 \\
		$\tfrac{1}{4}+\tfrac{1}{4}+\tfrac{1}{4}+\tfrac{1}{4}$ & 0.03736 & 0.11091 & 0.11476 & 0.01754 & 0.01874 & 0.05772 & 0.00288 & 0.00634 \\
		$\tfrac{1}{2}+\tfrac{1}{6}+\tfrac{1}{6}+\tfrac{1}{6}$ & 0.04478 & 0.11565 & 0.11778 & 0.02363 & 0.02010 & 0.06368 & 0.00287 & 0.00636 \\
		$\tfrac{1}{4}+\tfrac{1}{4}+\tfrac{1}{8}+\tfrac{1}{8}+\tfrac{1}{8}+\tfrac{1}{8}$ & 0.10552 & 0.11158 & 0.09626 & 0.05681 & 0.01711 & 0.05050 & 0.00331 & 0.00608 \\
		\bottomrule
	\end{tabular*}
\end{table*}
% END APPENDIX CONTENT

\end{document}